\documentclass[12pt]{article}
\makeatletter \@addtoreset{equation}{section} \makeatother
\renewcommand{\theequation}{\thesection.\arabic{equation}}
\newcommand{\ba}{\begin{array}}
\newcommand{\ea}{\end{array}}
\newcommand{\beq}{\begin{equation}}
\newcommand{\eeq}{\end{equation}}
\newcommand{\bea}{\begin{eqnarray}}
\newcommand{\eea}{\end{eqnarray}}

\def\bce{\begin{center}}
\def\ece{\end{center}}

\def\nonu{\nonumber}

\def\pa{\partial}

\def\be{\beta}

\def\eps6{{\displaystyle \mathop{\epsilon}^{6}}{}}
\def\g6{{\displaystyle \mathop{g}^{6}}{}}

\def\nab6{{\displaystyle \mathop{\nabla}^{6}}{}}

\def\0{{\sst{(0)}}}
\def\1{{\sst{(1)}}}
\def\2{{\sst{(2)}}}
\def\3{{\sst{(3)}}}
\def\4{{\sst{(4)}}}
\def\5{{\sst{(5)}}}
\def\6{{\sst{(6)}}}
\def\7{{\sst{(7)}}}
\def\8{{\sst{(8)}}}

\def\ba{\begin{array}}
\def\ea{\end{array}}
\def\beq{\begin{equation}}
\def\eeq{\end{equation}}
\def\be{\begin{equation}}
\def\ee{\end{equation}}

\def\eps{\epsilon}

\def\ba{\begin{array}}
\def\ea{\end{array}}
\def\beq{\begin{equation}}
\def\eeq{\end{equation}}
\def\be{\begin{equation}}
\def\ee{\end{equation}}

\def\eps{\epsilon}

\def\eps6{{\displaystyle \mathop{\epsilon}^{6}}{}}

\def\nab6{{\displaystyle \mathop{\nabla}^{6}}{}}

\newcommand{\bean}{\begin{eqnarray*}}
\newcommand{\eean}{\end{eqnarray*}}

\begin{document}
\thispagestyle{empty} \addtocounter{page}{-1}
   \begin{flushright}
\end{flushright}

\vspace*{1.3cm}
 
\centerline{ \Large \bf   
Higher Spin Currents  }
\vspace*{0.5cm}
\centerline{ \Large \bf  
in the ${\cal N}=1$ Stringy  Coset Minimal Model }
\vspace*{1.5cm}
\centerline{{\bf Changhyun Ahn 
}} 
\vspace*{1.0cm} 
\centerline{\it 
Department of Physics, Kyungpook National University, Taegu
702-701, Korea} 
\vspace*{0.8cm} 
\centerline{\tt ahn@knu.ac.kr 
} 
\vskip2cm

\centerline{\bf Abstract}
\vspace*{0.5cm}
This study reconsidered the ${\cal N}=1$ supersymmetric extension of the 
$W_3$ algebra which was studied previously. 
This extension consists of seven 
higher spin supercurrents (fourteen higher spin currents in the 
components) as well as the ${\cal N}=1$ 
stress energy tensor of spins $(\frac{3}{2}, 2)$. 
Thus far, the complete expressions for the higher spin currents have 
not been derived.

This paper constructs them explicitly 
in both 
the $c=4$ eight free fermion model and the supersymmetric coset model based on 
$(A_2^{(1)} \oplus A_2^{(1)}, A_2^{(1)})$ at level $(3,k)$.
By acting with the above spin-$\frac{3}{2}$ current 
on the higher spin-3 Casimir 
current, its fermionic partner, the higher 
spin-$\frac{5}{2}$ current, can be generated and 
combined as a first higher spin 
supercurrent with spins $(\frac{5}{2}, 3)$. 
By calculating the operator product expansions (OPE) 
between  the higher spin supercurrent and itself,
the next two higher spin supercurrents can be generated  with spins 
$(\frac{7}{2}, 4)$ and $(4, \frac{9}{2})$.  
Moreover, the other two higher spin supercurrents with spins $(4, \frac{9}{2})$ and $(\frac{9}{2}, 5)$
can be generated by calculating the OPE between the first higher spin supercurrent 
with spins $(\frac{5}{2}, 3)$ and the second higher spin supercurrent 
with spins $(\frac{7}{2}, 4)$.     
Finally, the higher spin supercurrents, 
$(\frac{11}{2}, 6)$ and $(6, \frac{13}{2})$, can be 
extracted from the right hand side of the OPE between the
higher spin supercurrents, $(\frac{5}{2}, 3)$ and $(4, \frac{9}{2})$.
 
\baselineskip=18pt
\newpage
\renewcommand{\theequation}
{\arabic{section}\mbox{.}\arabic{equation}}

\section{Introduction}

Gaberdiel and Gopakumar proposed the duality between the $W_N(= W A_{N-1})$ 
minimal model in two dimensional 
conformal field theories and 
the higher spin theory of Vasiliev on the $AdS_3$  \cite{GG,GG1}.
The $W_N$ minimal model conformal field theory
is dual, in the 't Hooft $\frac{1}{N}$ expansion, 
to  the higher spin theory coupled to one complex scalar.
One of the levels for the spin-$1$ WZW current in the conformal field theory 
is fixed by $1$ and the other is given by 
the positive integer, $k$.
The motivation of \cite{GG,GG1} is based on the 
work by Klebanov and Polyakov \cite{KP}
in the context of $AdS_4$ bulk theory and three dimensional 
conformal theory: the $O(N)$ vector model.
Therefore, it is 
natural to ask whether similar duality exists in the group
$SO(N)$ rather than $SU(N)=A_{N-1}$. 
In references \cite{Ahn1106,GV}, 
the possible generalization with the $SO(N)$ coset theory
was described.
All these \cite{GG,GG1,Ahn1106,GV} were a truncated version of 
${\cal N}=2$ supergravity theory \cite{PV}.
Furthermore, the full 
${\cal N}=2$ supersymmetric extension 
with higher spin $AdS_3$ supergravity
was studied in \cite{CHR} and in 
references \cite{CG,Reyetal,HP,Ahn1206,CG1,Ahn1208} 
where the dual conformal field theory is given by 
the ${\cal N}=2$ ${\bf CP}^N$ Kazama-Suzuki model  in two dimensions.    
Recently, in \cite{CHR1}, the ${\cal N}=1$ minimal model holography, 
which involves  
truncating the ${\cal N}=2$ theory \cite{PV} to the ${\cal N}=1$ theory 
with $SO(N)$ coset, is also described.

More general coset theories can be derived by generalizing  
level $1$ for the spin-$1$ current to the arbitrary 
positive integer, $l$.
The spin-$3$ current in these general cosets was constructed in reference
\cite{BBSS2}
and the spin-$4$ current was found recently in \cite{Ahn2011}.
The central charge grows like $N^2$ by defining the 't Hooft limit with 
$k, l, N \rightarrow \infty$ with the appropriate relative ratios held finite \cite{GG2}
\footnote{These general cosets are called by ``stringy cosets'' in \cite{GG2}.}. 
A particular case, 
where $k=l=N$, was studied in the context of two dimensional gauge theory 
coupled to the adjoint fermions in reference \cite{Schoutensetal}.

One study \cite{Douglas} reported 
that the cosets can be supersymmetric if one of the levels is equal to $N$(i.e., $l=N$
with arbitrary $k$).
Furthermore, the first model with $k=1$ in this series (with arbitrary $l$ and $N$)
has 
bosonic $W_N$ symmetry.
The supersymmetric version of $W_N$ algebra with $k=1, l=N$ 
was studied \cite{SS}.
On the other hand, 
supersymmetric extended $W_3$ symmetry can be expected by 
taking $N=3=l$ with arbitrary $k$.

This paper reconsiders 
the following coset minimal model studied previously \cite{ASS}
previously, 
with arbitrary $k$ and 
$l=N=3$, 
\bea
\frac{\widehat{SU}(3)_k \oplus \widehat{SU}(3)_3}
{\widehat{SU}(3)_{k+3}}.
\label{coset}
\eea
The central charge can be obtained and  is given by
\bea
c = 4 \left[ 1- \frac{18}{(k+3)(k+6)} \right],  \qquad k=1, 2, \cdots.
\label{centralc1}
\eea
For $k=1(c=\frac{10}{7})$ with (\ref{centralc1}), 
the explicit higher spin current as well as the superconformal generators
were constructed  \cite{HR} and  given in terms of 
the WZW currents of above coset (\ref{coset}). 
This extended algebra coincides with the ``minimal'' super $W_3$ algebra found in reference \cite{IMY},
where there are  
two extra higher spin currents of spin-$\frac{5}{2}$ and spin-$3$.
In reference \cite{ASS}, 
some of the higher spin currents were found in the $c=4$ free fermion model 
which is equivalent to 
$k \rightarrow 
\infty $ limit in (\ref{centralc1}). 
Moreover, those in 
the generic $c < 4$ supersymmetric coset models based on (\ref{coset})  
were found.
The existence of eight supercurrents 
via the character technique was described.
In reference \cite{Ahn1106}, the possible application 
in the context of minimal model holography \cite{GG,GG1} was suggested 
\footnote{In the first example  in reference \cite{Ahn1106}, the levels of the 
$SO(N)$ coset minimal model are given by $(k,2)$. The central charge behaves 
as $N$ in the large $N$ limit. A couple of conformal dimensions for the fields, in the large $(N,k)$ 't Hooft limit, are calculated. 
In the second example, where the levels are given by 
$(k,3)$ in the $SU(N)$ coset minimal model, 
the higher spin current of spin $\frac{5}{2}$
in the large $(N,k)$ limit was found.    
Finding the corresponding dual theories in the $AdS_3$ higher spin gravity
is an open problem.   }.
If the discrete series for the ${\cal N}=2$ 
superconformal minimal models with $c=\frac{3m}{(m+2)}$ is considered,
$c=\frac{8}{3}$ (i.e., $k=3$) corresponding to $m=16$.

The main item in the Casimir construction  \cite{BBSS1,BBSS2} 
for the level $1$ WZW models for
simply laced Lie algebras is to identify the
complete set of independent generating currents. 
One of the aims of this study  was 
to determine 
the complete set of currents proposed in reference 
\cite{ASS}  and some of the 
algebra they satisfy. The starting point is 
the bosonic spin-$3$ current found in \cite{BBSS1,BBSS2} with fixed $N=3$ and 
finite or infinite $k$. 
Of course, the ${\cal N}=1$ stress energy tensor consisting of 
spin-$2$ and spin-$\frac{3}{2}$ currents can be obtained from the usual Sugawara construction.
All the higher spin currents 
can then 
be obtained from the spin-$\frac{3}{2}$ current and spin-$3$ current.
These two currents generate  higher spin currents \footnote{These higher spin currents are primary under the 
stress energy tensor and they  transform similarly under the spin-$\frac{3}{2}$ current. Combining these two, 
the higher spin currents are superprimary under the ${\cal N}=1$ stress energy tensor. }.

That is, by acting with the above spin-$\frac{3}{2}$ current 
on the higher spin-3 current (by calculating the operator product expansion (OPE) between them),  its fermionic partner, the higher 
spin-$\frac{5}{2}$ current, can be generated in the right hand side of OPE, 
which can be combined as a first higher spin 
${\cal N}=1$ supercurrent with spins $(\frac{5}{2}, 3)$ \footnote{Although 
the notation will be explained in 
section $2$, let us describe the convention used here. The fermionic current of spin $\frac{5}{2}$ and its 
superpartner bosonic current of spin $3$ can be written in terms of a single ${\cal N}=1$ supercurrent, which is 
denoted simply by specifying the spins in this way. Similarly, one denotes 
the ${\cal N}=1$ superconformal 
supercurrent by $(\frac{3}{2}, 2)$. }.
By construction, the OPE between the composite fields can be calculated through 
the basic fundamental OPEs between the bosonic and fermionic WZW currents.
By calculating the OPE between  these two higher spin currents of spin-$\frac{5}{2}$ and spin-$3$
(the three nontrivial OPEs should be calculated in the component approach),
the next two higher spin ${\cal N}=1$ supercurrents can be generated 
 with spins 
$(\frac{7}{2}, 4)$ and $(4, \frac{9}{2})$ that appear in the singular terms of the OPE.  
The explicit forms of these 
in terms of WZW currents for the $c=4$($k \rightarrow \infty$) model 
were already presented in reference 
\cite{ASS}. For the arbitrary central charge (i.e. 
finite $k$), 
the explicit forms are not known.
Moreover, the other two higher spin ${\cal N}=1$ 
supercurrents with spins, $(4, \frac{9}{2})$ and $(\frac{9}{2}, 5)$,
are generated by calculating the OPE between the first higher spin supercurrent 
with spins $(\frac{5}{2}, 3)$ and the second higher spin supercurrent 
with spins $(\frac{7}{2}, 4)$ by repeating similar procedures.     
A range of quasi-primary field currents  can be 
written in terms of the 
known higher spin currents and the ${\cal N}=1$ superconformal currents.
Finally, the higher spin supercurrents, 
$(\frac{11}{2}, 6)$ and $(6, \frac{13}{2})$, can be 
extracted from the right hand side of the OPE between the
higher spin supercurrents, $(\frac{5}{2}, 3)$ and $(4, \frac{9}{2})$.
This will produce the most complicated calculations.

How does one calculate the OPE explicitly and extract the correct 
primary currents in the right hand side of OPE?
In general,
the OPE \cite{Bowcock,BFKNRV,BS}
(See also \cite{Nahm1,Nahm2}) 
between two quasi-primary fields $\Phi_i(z)$ with spin-$h_i$ and $\Phi_j(w)$ with spin-$h_j$(the spins
are positive integer or half-integer)
has the following form 
\footnote{The OPE of the stress energy tensor with the quasi-primary field does not have a third-order pole. 
The primary field is also a quasi-primary field because it satisfies with this condition for the quasi-primary field. In general, 
the OPE between the stress energy tensor and 
quasi-primary field can have a nonzero singular term(s) with the order $n > 3$. 
Of course, for the primary field, 
the trivial vanishing 
higher order singular terms with the order $n \geq 3$ occur.
This paper is restricted to the definition of a quasi-primary field as follows:
what is meant by a quasi-primary field is the one 
that does not contain the primary field. When refering to a 
quasi-primary field, 
it is considered that the nonvanishing higher order (greater than $3$) 
singular terms exist.}
\bea
&& \Phi_i(z) \; \Phi_j(w)  =  \frac{1}{(z-w)^{h_i+h_j}} \, \gamma_{ij} \nonu \\
&& +  \sum_k C_{ijk} \sum_{n=0}^{\infty} \frac{1}{(z-w)^{h_i+h_j-h_k-n}}
\left[\frac{1}{n!} \frac{\Gamma(h_i-h_j+h_k+n)}{\Gamma(h_i-h_j+h_k)} \frac{\Gamma(2h_k)}{\Gamma(2h_k+n)}\right] 
\pa^n \Phi_k(w).
\label{PhiPhi}
\eea
$\gamma_{ij}$ corresponds to a metric on the space of quasi-primary fields.
The quantity, $C_{ijk}$, 
appears in the three-point function between the quasi-primary fields, 
$\Phi_i(z), \Phi_j(z)$ and $\Phi_k(z)$. 
The index $k$ specifies all the quasi-primary fields occurring 
in the right hand side.
The descendant fields for the quasi-primary field 
$\Phi_k(w)$ are (multiple) derivatives of $\Phi_k(w)$:$\pa^n \Phi_k(w)$. 
Furthermore, the coefficient functions are written in terms of various 
Gamma functions  and depend on the spins 
and number of derivatives. 
Provided $n \geq h_i+h_j-h_k$, regular terms can be obtained. 
Otherwise, singular terms exist when $n \leq h_i+h_j-h_k$. 
By noting that  the  Gamma function has the following property, 
$\frac{\Gamma(h_i-h_j+h_k+n)}{\Gamma(h_i-h_j+h_k)}=(h_i-h_j+h_k)(h_i-h_j+h_k+1) \cdots (h_i-h_j+h_k+n-1)$, which is 
a Pochhammer function, 
for $h_i-h_j+h_k \leq 0$ (i.e. the spin of $\Phi_j(z)$ is greater than the sum of the spin of $\Phi_i(z)$ and the spin of 
$\Phi_k(z)$), the summation over $n$ terminates to a finite summation.
For example,  when $(h_i-h_j+h_k+n-1)$ is vanishing for particular $n=n_0$, then
the coefficient for $n > n_0$ always contains this vanishing value $(h_i-h_j+h_k+n_0-1)$.   
For the factor, $ \frac{\Gamma(2h_k)}{\Gamma(2h_k+n)}$, there is no zero for the positive $h_k$ and $n$.
Note that for different $k$' values 
(different quasi-primary fields), the  $h_k$ values of the corresponding spin
can be equal to each other.   
That is, for given singular terms, several different quasi-primary 
fields of the same spin can coexist. 
This feature  will be shown in the next sections.

Determining 
the possible quasi-primary or primary fields, $\Phi_k(w)$, in the 
right hand side is a nontrivial task.  
For lower higher spin quasi-primary or primary 
fields, the number of quasi-primary or primary fields is limited 
but the number of quasi-primary or primary fields  increases with increasing 
spin.  
As mentioned before, in general, the quasi-primary fields in (\ref{PhiPhi})
are given in terms of the 
composite fields between the WZW currents of the integer or half-integer spins.
By construction, the OPEs can be obtained 
from the basic fundamental OPEs between the WZW currents.
All the singular terms for given spins, $h_i$ and $h_j$, can then 
be obtained.
The most singular term should be analyzed first 
followed by the next lower singular terms because
once the lowest quasi-primary field (or primary field) is found, 
then its descendant structure 
is fixed completely  according to (\ref{PhiPhi}).  
After the nontrivial highest singular term is analyzed 
(the quasi-primary field or primary field appears in the
right hand side), 
the next singular term contains the descendant field for the previous quasi-primary or primary fields, and 
the remaining terms contains 
the new quasi-primary fields or primary fields according to (\ref{PhiPhi}).   

In general, finding these new fields is nontrivial.
On the other hand, 
they should transform as quasi-primary or primary fields under the spin-$2$ 
current, as  explained 
before.
Therefore, the OPEs between the spin-$2$ and ''the remaining terms'' 
in the given singular terms should be calculated. 
In general, this will include all the higher singular terms where the order 
is greater than $3$.
Therefore it is important to determine 
what kind of quasi-primary or primary fields occur.
By taking the possible quasi-primary fields with the 
undetermined coefficient functions, 
the remaining items should be expressed in terms of these quasi-primary or primary fields.
The unknown coefficient functions can be fixed  using 
the condition that the third-order pole 
should vanish.
Furthermore, the OPEs between 
the spin-$\frac{3}{2}$ current and ``the remaining terms'' 
should be calculated  to 
observe 
the complete structure of the possible quasi-primary fields. For the quasi-primary fields, there is no 
constraint on the singular terms but for the primary fields, the OPE between the spin-$\frac{3}{2}$ current 
and  primary current should contain either the first-order pole or the second- and first-order poles.  
This suggests that at least the OPE of spin-$\frac{3}{2}$ and the primary field 
should not contain  higher 
order singular terms where the order is greater than $2$. 
The details will be seen in the next sections.

Section $2$ reviews
the ${\cal N}=1$ superconformal algebra 
and the ``minimal'' ${\cal N}=1$ super $W_3$ algebra \cite{IMY} 
and describes  
the higher spin currents in the $c=4$ free fermion model.

In section $3$, the higher spin currents in $c< 4$ coset model are constructed explicitly.

Section $4$ summarizes the results in this paper and discusses 
the future directions.

In the Appendices, some OPEs relevant to the sections $2$ and $3$ are 
presented.

All the relevant works along the line of references 
\cite{GG,GG1} are presented 
in the recent review papers \cite{GG2,AGKP}
and also further works are reported in \cite{PPR}-\cite{Campoleonietal}.

The mathematica package \cite{Thielemans} is used all the time.
       
\section{The higher spin currents in the $c=4$ eight free fermion model }

\subsection{The ${\cal N}=1$ superconformal algebra: review}

Let us describe the ${\cal N}=1$ supersymmetric extension of Virasoro algebra.
The ${\cal N}=1$ superconformal algebra is generated by the 
${\cal N}=1$ super stress energy tensor of spin-$\frac{3}{2}$ \cite{FMS},
\bea
\hat{T}(Z) = \frac{1}{2} G(z) + \theta \; T(z),
\label{that}
\eea
where $Z=(z, \theta)$ is a complex supercoordinate, $T(z)$ is the usual bosonic 
stress energy tensor of spin-$2$ and $G(z)$ is its fermionic superpartner of spin-$\frac{3}{2}$.
The superconformal algebra, in components, is summarized by 
the three OPEs as follows. The OPE between the fermionic field of spin-$\frac{3}{2}$ and itself can be expressed  as
\bea
G(z)\; G(w) = \frac{1}{(z-w)^3} \, 
\frac{2}{3} c + \frac{1}{(z-w)} \, 2 T(w) +\cdots,
\label{gg}
\eea
where the right hand side of this OPE contains 
the bosonic stress energy tensor and the central term.
The equation contains 
no second-order singular term because there is no spin-$1$ field. 
The standard OPE between the bosonic stress energy tensor and itself (i.e., Virasoro algebra) 
is given by
\bea
T(z) \; T(w) = \frac{1}{(z-w)^4} \, \frac{c}{2} 
+\frac{1}{(z-w)^2} \, 2 T(w) +\frac{1}{(z-w)} \, \pa T(w) +\cdots,
\label{tt}
\eea
and finally, the fermionic field is primary with respect to $T(w)$, i.e.,
\bea
T(z) \; G(w) = \frac{1}{(z-w)^2} \, \frac{3}{2} 
G(w) +\frac{1}{(z-w)} \, \pa G(w) +\cdots.
\label{tg}
\eea
Of course, the OPE $G(z) \; T(w)$ can be obtained from 
(\ref{tg}) in the standard way.
The ${\cal N}=1$ superconformal algebra is represented by 
(\ref{gg}), (\ref{tt}) and (\ref{tg}) or its ${\cal  N}=1$ single OPE with (\ref{that}). 
These OPEs can also be expressed 
using the (anti)commutator relations for the 
modes of $T(z)$ and $G(z)$, as usual. Ramond algebra
is used for the integer mode of $G(z)$ and Neveu-Schwarz algebra is used for
the half-integer mode of $G(z)$.
According to the definition of quasi-primary field, 
the stress energy tensor, $T(z)$, is 
a quasi-primary field.
The coefficients $2$ in (\ref{gg}), $2$ in (\ref{tt}) and $\frac{3}{2}$ in (\ref{tg}) play the role of $C_{ijk}$ 
in (\ref{PhiPhi}) and the central terms in (\ref{gg}) and (\ref{tt}) correspond to $\gamma_{ij}$ in (\ref{PhiPhi}). 
Moreover, the relative coefficients,  $\frac{1}{2}$ in (\ref{tt}) 
and $\frac{2}{3}$ in (\ref{tg}),
appearing in the first-order singular term
coincide with the general expression given in (\ref{PhiPhi}). 
Appendix $A$ presents more details of the coefficient functions.

\subsection{The ``minimal'' ${\cal N}=1$ super $W_3$ algebra where $c=\frac{10}{7}$: review}

The simplest extension of the previous ${\cal N}=1$ superconformal algebra is to 
add a single higher spin superprimary current of spin-$\frac{5}{2}$
\bea
\hat{W}(Z) = \frac{1}{\sqrt{6}} U(z) +\theta \; W(z),
\label{what}
\eea
where $W(z)$ is a bosonic spin-$3$ current
and $U(z)$ is a fermionic spin-$\frac{5}{2}$ current. 
These are primary fields with respect to the 
stress energy tensor, $T(z)$, such as (\ref{tg}).
Furthermore, the spin-$\frac{3}{2}$ current $G(z)$ transforms $U(w)$ into $W(w)$ and 
vice versa (fermion goes to the boson and the boson goes to the fermion).
\bea
G(z) \; U(w) = \frac{1}{(z-w)} \, \sqrt{6} \, W(w) +\cdots,
\label{gu}
\eea
and 
\bea
G(z) \; W(w) = \frac{1}{(z-w)^2} \, \frac{5}{\sqrt{6}} \,U(w) +\frac{1}{(z-w)} \, \frac{1}{\sqrt{6}} \,
\pa U(w) +\cdots.
\label{gw}
\eea
This suggests that once any component field of (\ref{what}) is found,  
its superpartner can be determined automatically from (\ref{gu}) or (\ref{gw}).
Again, the relative coefficient $\frac{1}{5}$ showing 
the first-order singular term in (\ref{gw})
comes from the general expression in (\ref{PhiPhi}). 
The role of the spin-$\frac{3}{2}$ current $G(z)$ 
is very important and 
this property will be used continually. 

By assuming that the OPE between the additional supercurrent (\ref{what}) and itself
does not generate any new superprimary current (i.e. the ``minimal'' extension),
the possible structures in the right hand side of the OPE can be written.
The unknown coefficient functions can be determined using 
the so-called Jacobi identities
for normal ordered graded commutators of the supercurrents $\hat{T}(Z)$ and $\hat{W}(Z)$.
The  ``minimal'' ${\cal N}=1$ super $W_3$ algebra is 
associative for $c=\frac{10}{7}$, where the unitary 
representation exists and 
$c=-\frac{5}{2}$ with a nonunitary representation \cite{IMY}.
For $c=\frac{10}{7}$, the three OPEs in the 
components are summarized as follows.
The OPE between the bosonic spin-$3$ current and itself leads to
the following
\bea
W(z) \; W(w) & = & \frac{1}{(z-w)^6} \, \frac{10}{21} +\frac{1}{(z-w)^4} \, 2 T(w)
+\frac{1}{(z-w)^3} \, \pa T(w) \nonu \\
& + & \frac{1}{(z-w)^2} \left[ \frac{3}{10} \pa^2 T +\frac{56}{51} \left( T^2 -\frac{3}{10} \pa^2 T 
\right)  \right](w)
\nonu \\
& + & \frac{1}{(z-w)} \left[ \frac{1}{15} \pa^3 T + (\frac{1}{2}) \frac{56}{51}  
\pa \left( T^2 -\frac{3}{10} \pa^2 T 
\right) \right](w) +\cdots,
\label{ww}
\eea
which is precisely 
the same as the Zamolodchikov's $W_3$ algebra for $c=\frac{10}{7}$, as 
expected.
The coefficient $\frac{1}{2}$ for the descendant field with spin-$5$ associated with the 
quasi-primary field of spin $4$ 
in (\ref{ww}) is derived 
from the general expression in (\ref{PhiPhi}).
The coefficients, $\frac{1}{2}, \frac{3}{20}$ and $\frac{1}{30}$, 
appearing in the descendant fields of the
stress energy tensor can be obtained similarly \footnote{The quasi-primary field has the following OPE with the stress 
energy tensor, 
$T(z) \; (TT -\frac{3}{10} \pa^2 T)(w) = \frac{1}{(z-w)^4} \, (\frac{22}{5} +c) \, T(w) + 
{\cal O}((z-w)^{-2})$, where there is no third-order singular term.}.  

The OPE between the spin-$3$ and spin-$\frac{5}{2}$ can be summarized as
\bea
W(z) \; U(w) 
 & = & \frac{1}{(z-w)^4} \, \frac{3}{\sqrt{6}} G(w)  +
\frac{1}{(z-w)^3} \, (\frac{2}{3}) \frac{3}{\sqrt{6}} \pa G(w) \nonu \\
& + & \frac{1}{(z-w)^2}
\left[ (\frac{1}{4}) \frac{3}{\sqrt{6}} \pa^2 G +
\frac{77\sqrt{6}}{187} \left( G T -\frac{1}{8} \pa^2 G \right)   \right](w)
\nonu \\
& + & \frac{1}{(z-w)} \left[ 
 (\frac{1}{15}) \frac{3}{\sqrt{6}} \pa^3 G + (\frac{4}{7})
\frac{77\sqrt{6}}{187} \pa \left( G T -\frac{1}{8} \pa^2 G \right)  \right. \nonu \\
& + &  \left.  \frac{4\sqrt{6}}{17} \left( \frac{4}{3} T \pa G - G \pa T -\frac{4}{15} \pa^3 G 
\right) 
\right](w) +\cdots.
\label{wu}
\eea
The coefficients $\frac{2}{3}, \frac{1}{4}$ and $\frac{1}{15}$, 
for the descendant 
fields for the spin-$\frac{3}{2}$ field were written down intentionally 
in the right hand side (\ref{wu}) to emphasize that they can be determined 
from (\ref{PhiPhi}).
The spin-$\frac{7}{2}$ and spin-$\frac{9}{2}$ has two quasi-primary fields.
The coefficient $\frac{4}{7}$ appearing in the descendant field for the former 
can be obtained from the general formula.
The first-order singular term consists of the descendant field,
 $\pa^3 G(w)$ for $G(w)$, the descendant field 
$\pa (G T -\frac{1}{8} \pa^2 G)(w)$ for 
spin-$\frac{7}{2}$ quasi-primary field $ (G T -\frac{1}{8} \pa^2 G)(w)$ 
and the quasi-primary field $ (\frac{4}{3} T \pa G - G \pa T -\frac{4}{15} \pa^3 G)(w)$ of spin-$\frac{9}{2}$.
In other words, three independent terms, which are characterized by 
$\pa^3 G(w), \pa G T(w)$ and $G \pa T(w)$ 
can be 
rewritten in terms of the 
above three terms \footnote{The
 following OPEs can be obtained easily  to determine if they are
really quasi-primary fields. They are
$T(z) \; (G T -\frac{1}{8} \pa^2 G)(w)=\frac{1}{(z-w)^4} \, \frac{37}{8} \, G(w) + 
{\cal O}((z-w)^{-2})$ and $T(z) \; (\frac{4}{3} T \pa G - G \pa T -\frac{4}{15} \pa^3 G)(w)=- \frac{1}{(z-w)^5}
\, \frac{33}{5} G(w) -\frac{1}{(z-w)^4} \, (-\frac{1}{3}) \, \frac{33}{5} \pa G(w) +{\cal O}((z-w)^{-2})$. The coefficient, 
$-\frac{1}{3}$, coincides with the general expression (\ref{PhiPhi}) by 
substituting $h_i=2, h_j=\frac{9}{2}, h_k=\frac{3}{2}$,
and $n=1$. The minus sign is because $h_i-h_j+h_k =-1$.
Appendices $B$ and $C$ presents 
other quasi-primary fields and their OPEs between 
$T(z)$ or $G(z)$.}. 

Finally, the spin-$\frac{5}{2}$ and spin-$\frac{5}{2}$ OPE can be 
expressed as
\bea
U(z) \; U(w)  & = & \frac{1}{(z-w)^5} \, \frac{4}{7}  + \frac{1}{(z-w)^3} \, 2T(w) +
\frac{1}{(z-w)^2} \, \pa T(w) \nonu \\
&+&\frac{1}{(z-w)} \, \left[ \frac{3}{10} \pa^2 T +\frac{63}{68} \left( T^2 -\frac{3}{10} \pa^2 T 
\right) +  P_4^{uu} \right](w) +\cdots.
\label{uu}
\eea
The relative coefficients, $\frac{1}{2}$ and 
$\frac{3}{20}$, appearing in the descendant fields for the stress energy 
tensor, $T(w)$,
can be analyzed previously and provide the correct values. 
In the first-order singular term, 
there is a quasi-primary field 
$(T^2 -\frac{3}{10} \pa^2 T)(w)$
and primary field of spin-$4$ \cite{HR} given as
\bea
 P_4^{uu}(w) = \frac{21}{17} \left[ -\frac{7}{10} \pa^2 T + 
\frac{7}{12} \left( T^2 -\frac{3}{10} \pa^2 T 
\right) + G \pa G \right](w),
\label{p4uu}
\eea
where the central charge, $c=\frac{10}{7}$, is used.
For the general $c$, the coefficient, $\frac{7}{12}$, should be replaced by 
$\frac{17}{(22+5c)}$.
Note that this primary field is not an additional field because this can be obtained from the currents, 
$T(w)$ and $G(w)$.
In this example, at the first-order singular term of (\ref{uu}), 
there are two types of 
quasi-primary fields. In the language of (\ref{PhiPhi}), for a fixed 
$h_k$, there are two degeneracies.
More precisely, 
one is a spin-$4$ quasi-primary field and the other is a 
spin-$4$ primary field
(\ref{p4uu}). 
This primary field (\ref{p4uu}) has an unusual OPE with an above 
spin-$\frac{3}{2}$ current $G(z)$ in the next subsection.

\subsection{The $c=4$ free fermion model}

A consistent generalization of 
the above minimal extension of $W_3$ algebra for an 
arbitrary central charge is needed.
This subsection 
will consider the particular supersymmetric coset model 
introduced in the introduction. Before going into detail,
this section
first  describes 
its particular limit where all the algebraic structures are observed.

Consider the eight fermion fields $\psi^a$ of spin-$\frac{1}{2}$ where 
the $SU(3)$ adjoint index $a$ runs from $1$ to $8(=3^2-1)$. 
In this paper, $N$ is fixed by $3$.
The OPE of this fermion field is given by
\bea
\psi^a(z) \; \psi^b(w) =-\frac{1}{(z-w)} \, \frac{1}{2} \delta^{ab} +\cdots.
\label{psipsi}
\eea
Define the spin-$1$ Kac-Moody current $J^a(z)$ as
\bea
J^a(z) = f^{abc} \psi^b \psi^c(z).
\label{Jdef}
\eea
Then it is easy to calculate the OPE 
\footnote{More precisely, one can start by writing $\tt
<< OPEdefs.m$ in the mathematica notebook.
The operators
$\tt Fermionic[psi[1],psi[2],\cdots,psi[8]]$ are defined 
for the model in this section
(and $\tt Bosonic[K[1],K[2],\cdots,K[8]]$
for the coset model in section $3$).
The singular OPEs between the basic operators are then given by
$\tt OPE[psi[m_{-}],psi[n_{-}]]=
MakeOPE[\{-\frac{1}{2} One*Delta[m,n]\}];$
$\tt OPE[K[m_{-}],K[n_{-}]]=
MakeOPE[\{-\frac{k}{2} One*Delta[m,n],Sum[f[m,n,p]*K[p],\{p,1,8\}]\}];
$ Finally, the structure constants
$d$ and $f$ symbols
$\tt f[1, 2, 3] = 1; f[1, 4, 7] = \frac{1}{2}; \cdots 
d[1, 1, 8] = \frac{1}{\sqrt{3}}; d[2, 2, 8] = \frac{1}{\sqrt{3}}; 
\cdots$ should be defined.
The command, $\tt OPESimplify[OPE[A,B],Factor]$ or 
$\tt OPESimplify[OPEPole[3][A,B],Factor]$
can be used
 for any operators $A$ and $B$. 
All the detailed descriptions are given in 
reference \cite{Thielemans}. }
between this spin-$1$ current and itself, which leads to
\bea
J^a (z) \; J^b(w) = -\frac{1}{(z-w)^2} \frac{3}{2} \delta^{ab} + 
\frac{1}{(z-w)} f^{abc} J^c(w) +\cdots,
\label{JJ}
\eea
where the level is given by $3$.
Similarly, 
\bea
\psi^a(z) \; J^b(w) = \frac{1}{(z-w)} \, f^{abc} \psi^c(w) +\cdots.
\label{psiJ}
\eea
In the context of (\ref{PhiPhi}), the above OPEs (\ref{psipsi}), (\ref{JJ}) and 
(\ref{psiJ}) can be described easily.

For a given ${\cal N}=1$ super Kac-Moody algebra where 
$\hat{Q}^a(Z) = \sqrt{3}\, \psi^a(z) +\theta \; J^a(z)$, characterized by 
(\ref{psipsi}), (\ref{JJ}) and (\ref{psiJ}),
the ${\cal N}=1$ superconformal algebra
is realized by the spin-$2$ current
\bea
T(z)= \psi^a \pa \psi^a (z)
\label{c4t}
\eea
and the spin-$\frac{3}{2}$ current
\bea
G(z)= - \frac{2}{3\sqrt{3}} \, \psi^a J^a (z).
\label{c4g}
\eea
They satisfy (\ref{gg}), (\ref{tt}) and (\ref{tg}) for $c=4$.
The normalizations in (\ref{c4t}) and (\ref{c4g}) are fixed automatically.

Now we are ready to construct the higher spin currents.
In reference \cite{BBSS1}, 
the spin-$3$ current is described by the third order Casimir operator 
for $A_2^{(1)}=\widehat{SU}(3)$,
\bea
W(z) = \sqrt{\frac{2}{1215}} \, i \, d^{abc} J^a J^b J^c (z),
\label{c4w}
\eea
where $d^{abc}$ is a completely symmetric traceless $SU(3)$ invariant tensor of rank $3$ and the spin-$1$ current is defined as (\ref{Jdef}).
As mentioned previously, 
its fermionic superpartner of spin-$\frac{5}{2}$ can be determined from the 
relation (\ref{gw}) with (\ref{c4g}) and (\ref{c4w}) 
\bea
U(z)= \sqrt{\frac{2}{375}} \, i \, d^{abc} \psi^a J^b J^c (z).
\label{c4u}
\eea
Thus far, the currents are given by the 
${\cal N}=1$ superconformal generators and 
${\cal N}=1$ $W_3$ supercurrent.
On the other hand, 
additional higher spin currents 
can be observed
once the OPEs between these currents are calculated.

In reference \cite{ASS}, extra higher spin currents were given.
Let us present them in ${\cal N}=1$ superspace with 
their components. Their spins are denoted  in the subscript 
and the prime notation is  
used to describe the different field content with the same spin
\bea
\hat{W} (Z)  & = & \frac{1}{\sqrt{6}} \, U(z) + \theta \; W(z), 
\nonu \\ 
\hat{O}_{\frac{7}{2}}(Z) & = &  O_{\frac{7}{2}} (z) +\theta \; O_4 (z),
\nonu \\
\hat{O}_4 (Z) & = &  O_{4'} (z) +\theta \; O_{\frac{9}{2}} (z),
\nonu \\
\hat{O}_{4'}(Z) & = &  O_{4''} (z) +\theta \; O_{\frac{9}{2}'} (z),
\nonu \\
\hat{O}_{\frac{9}{2}}(Z) & = &  O_{\frac{9}{2}''} (z) +\theta \; O_5 (z),
\nonu \\
\hat{O}_{\frac{11}{2}}(Z) & = &  O_{\frac{11}{2}} (z) +\theta \; O_6 (z),
\nonu \\
\hat{O}_6 (Z) & = &  O_{6'} (z) +\theta \; O_{\frac{13}{2}} (z).
\label{higherspincurrents}
\eea
The normalization factor $\frac{1}{\sqrt{2 h_{\hat{O}}+1}}$ can also be 
introduced in front of 
the $\theta$-independent term, like $\hat{W}(Z)$.
Some of the currents were  explicitly found  in reference \cite{ASS}.
The higher spin currents will be calculated  
in terms of eight fermion fields $\psi^a(z)$.
In particular, some OPEs between 
$\hat{W}(Z)$, $\hat{O}_{\frac{7}{2}}(Z)$ and $\hat{O}_{4'}(Z)$
will be calculated.
There is no reason why $\hat{O}_{4'}(Z)$ was considered instead of 
$\hat{O}_4(Z)$.
In these computations, unknown higher spin currents arise 
in the right hand side of the OPE.
Twelve higher spin currents
in terms of $\psi^a(z)$ were constructed explicitly.

Consider the lower higher spin currents first.

$\bullet$ Construction of higher spin supercurrents, 
$\hat{O}_{\frac{7}{2}}(Z)$ and $\hat{O}_4(Z)$

First consider the OPE between the spin-$\frac{5}{2}$ currents.
All the singular terms can be obtained 
using the defining equation (\ref{psipsi}), (\ref{psiJ}) and (\ref{c4u}).
The OPE can be expressed as   
\bea
U(z) \; U(w) 
 & = & \frac{1}{(z-w)^5} \, \frac{8}{5}  + \frac{1}{(z-w)^3} \, 2T(w) +
\frac{1}{(z-w)^2} \, \pa T(w) \nonu \\
&+&\frac{1}{(z-w)} \, \left[ \frac{3}{10} \pa^2 T +\frac{9}{14} \left( T^2 -\frac{3}{10} \pa^2 T 
\right) +  P_4^{uu} + P_{4'}^{uu} \right](w) +\cdots,
\label{c4uu}
\eea
where the spin-$4$ primary field is given by
\bea
 P_4^{uu}(z) = 
\frac{75}{407} \left[ -\frac{7}{10} \pa^2 T + 
\frac{17}{42} \left( T^2 -\frac{3}{10} \pa^2 T 
\right) + G \pa G \right](z)
\label{p4-uu}
\eea
which is identical to the one in (\ref{p4uu}) with $c=4$. 
Note that the new spin-$4$ primary field, compared to the OPE (\ref{uu}), 
arises, as reported elsewhere \cite{HR}
\bea
P_{4'}^{uu}(z) = -\frac{18}{407} \psi^a 
\pa^3 \psi^a(z) +\mbox{other lower order derivative terms},
\label{c4'uu}
\eea
where only eight out of   156 terms are presented.
In the context of (\ref{PhiPhi}), at the first-order singular term,
there are triplet degeneracies for given spin-$4$ (quasi)primary fields.
One way to see this extra new primary field (\ref{c4'uu}) is to subtract the
$\frac{3}{10} \pa^2 T(w)$-term, $(TT -\frac{3}{10} \pa^2 T)(w)$-term
with an arbitrary coefficient, and the spin-$4$ term (\ref{p4-uu}) 
with an undetermined coefficient,  
from the first-order singular term.
These two unknown coefficients ($\frac{9}{14}$ and $\frac{75}{407}$) 
were fixed by the (quasi)primary condition. 
This spin-$4$ field (\ref{c4'uu}) will provide some component field of ${\cal N}=1$ 
superprimary field $\hat{O}_{\frac{7}{2}}(Z)$ or $\hat{O}_4(Z)$.

Let us move on to the next OPE between the spin-$3$ current (\ref{c4w}) 
and  spin-$\frac{5}{2}$ current (\ref{c4u})
\bea
W(z) \; U(w) 
 & = & \frac{1}{(z-w)^4} \, \frac{3}{\sqrt{6}} G(w)  +
\frac{1}{(z-w)^3} \, (\frac{2}{3}) \frac{3}{\sqrt{6}} \pa G(w) \nonu \\
& + & \frac{1}{(z-w)^2}
\left[ (\frac{1}{4}) \frac{3}{\sqrt{6}} \pa^2 G +
\frac{11\sqrt{6}}{37} \left( G T -\frac{1}{8} \pa^2 G \right) +O_{\frac{7}{2}}  \right](w)
\nonu \\
& + & \frac{1}{(z-w)} \left[ 
 (\frac{1}{15}) \frac{3}{\sqrt{6}} \pa^3 G + (\frac{4}{7})
\frac{11\sqrt{6}}{37} \pa \left( G T -\frac{1}{8} \pa^2 G \right) + 
(\frac{4}{7}) \pa O_{\frac{7}{2}} \right. \nonu \\
& + &  \left.  
\frac{4\sqrt{6}}{77} \left( \frac{4}{3} T \pa G - G \pa T -\frac{4}{15} \pa^3 G 
\right)  + O_{\frac{9}{2}} 
\right](w) +\cdots.
\label{c4wu}
\eea
Compared to the minimal extension in previous subsection, 
there are two new primary fields \cite{HR}.
One is the spin-$\frac{7}{2}$ field
\bea
O_{\frac{7}{2}}(z) = 
-\frac{\sqrt{2}}{37} \, f^{abc} \, \psi^a \psi^b \pa^2 \psi^c(z) +
\mbox{other first-order derivative terms}
\label{c4o72}
\eea
where only the highest derivative terms of 99 terms is presented.
The other is a spin-$\frac{9}{2}$ field
\bea
O_{\frac{9}{2}}(z) = -\frac{4\sqrt{2}}{231} \, f^{abc} \, 
\psi^a \psi^b \pa^3 \psi^c(z) 
 +\mbox{other lower order derivative terms}
\label{c4o92}
\eea
which consists of 270 terms.
The structure constant, $\frac{11\sqrt{6}}{37}$ and 
$\frac{4\sqrt{6}}{77}$, in (\ref{c4wu}) 
are determined by the (quasi)primary condition as 
before. Generally, these are given in terms of 
the central charge, which are different from those in (\ref{wu})
because the $c=4$ model is considered.
The advantage of the 
$c=4$ model is that because the explicit form for the WZW 
currents is known,
one can always calculate the OPE and determine the singular terms.
On the other hand, the construction in a minimal extension is based on 
the assumption that there are some extended generators whose 
realizations are  unknown. Therefore, 
the possible structures with unknown coefficient 
functions should be expressed in 
the right hand side of the OPE 
and the Jacobi identities should be used to fix them.
This subsection and next section 
arranges the known singular terms 
and extracts all the possible (quasi)primary fields 
using (\ref{PhiPhi}) or the expression in Appendix $A$, (\ref{Coeff1}) and 
(\ref{Acoefficient}).

For the spin-$3$ current (\ref{c4w}) and  spin-$3$ current OPE,  
\bea
W(z) \; W(w) =
 & = & \frac{1}{(z-w)^6} \, \frac{4}{3} +\frac{1}{(z-w)^4} \, 2 T(w)
+\frac{1}{(z-w)^3} \, \pa T(w) \nonu \\
& + & \frac{1}{(z-w)^2} \left[ (\frac{3}{10}) \pa^2 T +\frac{16}{21} 
\left( T^2 -\frac{3}{10} \pa^2 T 
\right) +P_4^{ww} +P_{4'}^{ww} \right](w)
\nonu \\
& + & \frac{1}{(z-w)} \left[ (\frac{1}{15}) 
\pa^3 T +(\frac{1}{2}) \frac{16}{21}  
\pa \left( T^2 -\frac{3}{10} \pa^2 T 
\right) +(\frac{1}{2}) 
\pa P_4^{ww} + (\frac{1}{2}) \pa P_{4'}^{ww} \right](w) \nonu \\
& + & \cdots,
\label{c4ww}
\eea
where the primary field of spin-$4$ is
\bea
 P_4^{ww}(z) = -\frac{48}{407} \left[ -\frac{7}{10} \pa^2 T + \frac{17}{42} \left( T^2 -\frac{3}{10} \pa^2 T 
\right) + G \pa G \right](z)
\label{p4-ww}
\eea
and the new primary spin-$4$ field \cite{HR} is 
\bea
P_{4'}^{ww}(z) =
\frac{2}{1221} \psi^a \pa^3 \psi^a(z) 
+\mbox{other lower order derivative terms}
\label{c4'ww}
\eea
whose number of terms is the same as that for spin-$4$ field in 
(\ref{c4'uu}).
Note that these two 
currents do not appear in (\ref{ww}). 
The spin-$4$ field (\ref{c4'ww}) will provide 
some component of the 
superprimary field $\hat{O}_{\frac{7}{2}}(Z)$ or $\hat{O}_4(Z)$.
Although the spin-$4$ field (\ref{p4-uu}) or (\ref{p4-ww}) 
is primary under the 
stress energy tensor, the OPE with a spin-$\frac{3}{2}$ current exhibits 
unusual behavior as mentioned previously.  
If the primary field is one of the components in the superprimary field 
with a given spin, it should transform like  (\ref{gu}) or (\ref{gw}).
On the other hand, the OPE of this spin-$4$ field 
and the spin-$\frac{3}{2}$ current 
has third-order and fourth-order singular terms.
This suggests 
that there is no superpartner for this spin-$4$ field
\footnote{\label{gt} More explicitly, the OPE 
is obtained $G(z) \; (TT +\frac{42}{17} 
G \pa G -\frac{69}{34} \pa^2 T)(w) = -\frac{1}{(z-w)^4} \, \frac{1221}{68}
\, G(w) -\frac{1}{(z-w)^3} \, (-\frac{1}{3}) \, \frac{1221}{68} \pa G(w) +
{\cal O}((z-w)^{-2})$.}.

Thus far, the spin-$3$ current 
and spin-$\frac{5}{2}$ current are given in (\ref{c4w}) and (\ref{c4u}),
whereas 
the spin-$\frac{7}{2}$ field and  spin-$\frac{9}{2}$ field 
are found in (\ref{c4o72}) 
and (\ref{c4o92}), respectively.
These are located in the first three supercurrents in the list 
(\ref{higherspincurrents}).
Other higher spin currents, the superpartners of 
$O_{\frac{7}{2}}(z)$ and $O_{\frac{9}{2}}(z)$,  
should be found in terms of eight fermion fields.

How should the other higher 
spin currents corresponding to $O_4(z)$ or $O_{4'}(z)$ be determined?
Consider the OPE between the spin-$\frac{3}{2}$ current $G(z)$ 
and  spin-$\frac{7}{2}$
current $O_{\frac{7}{2}}(w)$.
This OPE was calculated 
because there are 
explicit forms in (\ref{c4g}) and (\ref{c4o72}), respectively. 
The results showed that 
this OPE leads to the following first-order singular term with
$ \left(-\frac{1}{\sqrt{6}} P_{4'}^{uu} +\sqrt{6} P_{4'}^{ww} \right)(w)$, 
where the spin-$4$ currents are given in (\ref{c4'uu}) 
and (\ref{c4'ww}) as before.
This suggests that
the following current of spin-$4$, due to  
${\cal N}=1$ supersymmetry, can be constructed
\bea
O_4(z) = \left( -\frac{1}{\sqrt{6}} P_{4'}^{uu} +\sqrt{6} P_{4'}^{ww} \right)(z).
\label{c4o4}
\eea
That is, 
\bea
G(z) \; O_{\frac{7}{2}}(w) =\frac{1}{(z-w)} \, O_4 (w) +\cdots.
\label{c4go72}
\eea
Steps should be taken to ensure 
that the OPE $G(z)$ with $O_4(w)$ leads to 
expected singular terms with $O_{\frac{7}{2}}(w)$ by ${\cal N}=1$ supersymmetry.  
\bea
G(z) \; O_4(w) =\frac{1}{(z-w)^2} \, 7 O_{\frac{7}{2}}(w) +\frac{1}{(z-w)} \, \pa O_{\frac{7}{2}}(w) +\cdots.
\label{c4go4}
\eea

Furthermore, the OPE between the spin-$\frac{3}{2}$ current $G(z)$ and spin-$\frac{9}{2}$ current 
$O_{\frac{9}{2}}(w)$ (\ref{c4o92}) 
can be calculated  to determine the superpartner with a spin-$4$ field.
Similar to (\ref{gw}), 
\bea
G(z) \; O_{\frac{9}{2}}(w) =\frac{1}{(z-w)^2} \, 8 O_{4'}(w) +
\frac{1}{(z-w)} \, \pa O_{4'}(w) +\cdots,
\label{OPEgo92}
\eea
where the superpartner of $O_{\frac{9}{2}}(z)$ is given by
\bea
O_{4'}(z) = \frac{1}{8} \left( \frac{16}{7} \sqrt{\frac{2}{3}} P_{4'}^{uu} 
-\frac{4}{7} \sqrt{6} P_{4'}^{ww}\right)(z). 
\label{c4o4'}
\eea
In general, the coefficient in the second-order pole in (\ref{OPEgo92})
is equal to $2$ times the spin of the current appearing in that singular 
term. In (\ref{OPEgo92}), $ 8 = 2 \times 4$ whereas 
in (\ref{c4go4}), $7= 2 \times \frac{7}{2}$.
Similar to (\ref{gu}),  
\bea
G(z) \; O_{4'}(w) =\frac{1}{(z-w)} \,  O_{\frac{9}{2}}(w) +\cdots.
\label{OPEgo4}
\eea
Therefore, the supercurrents, $\hat{O}_{\frac{7}{2}}(z)$ and $\hat{O_4}(z)$,
in (\ref{higherspincurrents}) are determined completely \footnote{The 
OPEs can be written conveniently as $G(z) \; P_{4'}^{uu}(w) =
\frac{1}{(z-w)^2} \, \sqrt{6} O_{\frac{7}{2}}(w)
+\frac{1}{(z-w)} \left( \frac{\sqrt{6}}{6} \pa O_{\frac{7}{2}} +
2\sqrt{6} O_{\frac{9}{2}} \right)(w) +\cdots $ and $G(z) \; P_{4'}^{ww}(w) =
\frac{1}{(z-w)^2} \, 4\sqrt{\frac{2}{3}} O_{\frac{7}{2}}(w)
+\frac{1}{(z-w)} \left( \frac{4}{7} \sqrt{6} \pa O_{\frac{7}{2}} +
\sqrt{\frac{2}{3}} O_{\frac{9}{2}} \right)(w) +\cdots $. 
This is  why the primary fields 
$ P_{4'}^{uu}(z)$ and $ P_{4'}^{ww}(w)$ are unsuitable for the ${\cal N}=1$ 
supercurrents. \label{unusual} }.

The supercurrent $\hat{O}_{\frac{7}{2}}(Z)$ was constructed 
with (\ref{c4go72}) and 
(\ref{c4go4}) 
and $\hat{O}_4(Z)$ with (\ref{OPEgo92}) and (\ref{OPEgo4}) in the list of 
(\ref{higherspincurrents}). Indeed, these were reported in 
reference \cite{ASS}.
Their component fields are given by (\ref{c4o72}), (\ref{c4o4}), (\ref{c4o4'}) 
and (\ref{c4o92}).
In the 
${\cal N}=1$ notation, the above superfusion rule  between the supercurrent 
$\hat{W}(Z)$ and itself can be rewritten as
$\left[\hat{W} \right] \left[\hat{W} \right]  =   
\left[\hat{I} \right] + \left[\hat{O}_{\frac{7}{2}}
\right] +\left[\hat{O}_4 \right]$. The right hand side of this OPE contains 
two higher spin superprimary fields.

The OPE between $\hat{W}(Z_1)$ and 
$\hat{O}_{\frac{7}{2}}(Z_2)$ which is the next lower 
higher spin supercurrent could be considered.

$\bullet$ The construction of higher spin supercurrents 
$\hat{O}_{4'}(Z)$ and $\hat{O}_{\frac{9}{2}}(Z)$

The next lower 
higher spin currents in (\ref{higherspincurrents})
can be obtained.
The OPE between the spin-$\frac{5}{2}$ current $U(z)$ given 
in (\ref{c4u}) and the spin-$\frac{7}{2}$
current $O_{\frac{7}{2}}(w)$ given in (\ref{c4o72}) can be calculated.
The results showed that
\bea
&& U(z) \; O_{\frac{7}{2}} (w)  =   \frac{1}{(z-w)^3}\, \frac{36}{185} W(w) \nonu \\
&&  + 
\frac{1}{(z-w)^2} \left[ (\frac{1}{3}) \frac{36}{185} \pa W -\frac{6\sqrt{6}}{481} \left( 
G U -\frac{\sqrt{6}}{3} \pa W \right) + O_{4''} \right](w) 
\nonu \\
&& +  \frac{1}{(z-w)} \left[ (\frac{1}{14}) \frac{36}{185} \pa^2 W -(\frac{3}{8}) 
\frac{6\sqrt{6}}{481} \pa \left( 
G U -\frac{\sqrt{6}}{3} \pa W \right) + (\frac{3}{8}) \pa O_{4''} \right. \nonu \\
&& +  \left. \frac{764}{8325} \left( T W -\frac{3}{14} \pa^2 W  \right) +
\frac{287}{5550\sqrt{6}} \left( G \pa U -\frac{5}{3} \pa G U -\frac{\sqrt{6}}{7} \pa^2 W 
\right) +
O_5 \right](w) \nonu \\
&& +\cdots.
\label{uo72}
\eea
For the primary field $W(w)$ with the structure constant, 
$\frac{36}{185}$, in the right hand side, 
the relative coefficients for its descendant fields appearing in various singular terms
can be read off from (\ref{PhiPhi}). See also  Appendix $A$ for the detailed 
coefficients in the structure 
constants.
Consider the second-order singular terms.
The first term originating from $W(w)$ is fixed.  
Therefore, how  can the next quasi-primary or primary field be 
observed?
From the second-order pole, the OPE between $T(z)$ and 
the second order-pole subtracted by $(\frac{1}{3}) \frac{36}{185} \pa W(w)$
can be calculated to extract the possible 
quasi-primary fields (i.e. the exact expressions and number of 
quasi-primary fields).
$T(z) \left(\{U O_{\frac{7}{2}}\}_{-2} -(\frac{1}{3}) \frac{36}{185} \pa W\right)(w) =
+ {\cal O}((z-w)^{-2})$ can then be obtained. 
In other words, it transforms as a primary field. On the other hand, 
$G(z) \left(\{U O_{\frac{7}{2}}\}_{-2} -(\frac{1}{3}) \frac{36}{185} \pa W\right)(w) =
\frac{1}{(z-w)^3} \, \frac{2\sqrt{6}}{37} \, U(w) + {\cal O}((z-w)^{-2})$
can be calculated \footnote{
A simplified notation here $ \{U O_{\frac{7}{2}}\}_{-2}(w)$ 
was used for the second order pole of 
(\ref{uo72}) in the spirit of \cite{BBSS1,BBSS2,BS}.}.
This suggests that the remaining terms in the second-order pole contain a 
primary field with unusual behavior with $G(z)$. This  can be obtained 
explicitly. By subtracting 
$G U(w)$ plus the other derivative term with unknown coefficient 
into the above second-order singular terms, 
the unwanted 
third-order pole, which is proportional to $U(w)$, can be removed 
by choosing 
the correct coefficient.  
This is because when the OPE $G(z)$ with $G U(w)$ is calculated, 
the $U(w)$ term in the third-order
singular term can be derived via (\ref{gg}).
Therefore, $(G U -\frac{\sqrt{6}}{4} \pa W)(w)$ might be 
a possible candidate for the primary field that needs to 
be subtracted 
\footnote{
One obtains the OPE $G(z) \; (G U -\frac{\sqrt{6}}{4} \pa W)(w)= \frac{1}{(z-w)^3} \, \frac{13}{3} U(w) + 
{\cal O}((z-w)^{-2})$. See also Appendix $C$. In the ${\cal N}=1$ supercurrent,
this primary field originates from $\hat{T} \hat{W}(Z_2)$ \cite{ASS}.}.
The coefficient 
$-\frac{6\sqrt{6}}{481}$ in (\ref{uo72}) in front of this field was fixed by 
requiring that  there should be no third-order pole from the 
superprimary condition. 
Then we are left with the following primary field.
The spin-$4$ primary field can be expressed as
\bea
O_{4''}(z)= \frac{384}{13} \sqrt{\frac{2}{5}} \, i \, \psi^1 \psi^2 \psi^3 \psi^4 \psi^5 \psi^6 \psi^7 \psi^8(z) +
\mbox{other derivative terms}.
\label{c4o4''}
\eea

Consider the last first-order singular terms.
The first line of these terms in (\ref{uo72}) describes 
the descendant field for the spin-$3$ current and two 
descendant fields for the spin-$4$ primary fields.
As stated before, 
the difference between the whole first-order singular terms
and those three descendant terms were calculated
to ensure 
that there are two quasi-primary fields presented in the second line of
the first order-singular terms in (\ref{uo72}).
The OPE $T(z)$ with $ \left[ \{U O_{\frac{7}{2}}\}_{-1} -
\mbox{first line}
\right](w)$ leads to 
$\frac{1}{(z-w)^4} \, \frac{162}{259} W(w) + {\cal O}((z-w)^{-2})
$.
This suggests that 
the extra quasi-primary field, $T W(w)$, plus derivative terms,
should be considered to cancel 
the fourth-order term  $\frac{162}{259} W(w)$ for the superprimary condition.
Furthermore, the OPE between $G(z)$ and 
 $ \left[ \{U O_{\frac{7}{2}}\}_{-1} -
\mbox{first line} \right](w)$ should be also calculated, which  leads to
$\frac{1}{(z-w)^4} \, \frac{33}{518} \sqrt{\frac{3}{2}} \, U(w) +
\frac{1}{(z-w)^3} \, (-\frac{1}{5})  
\frac{33}{518} \sqrt{\frac{3}{2}} \, \pa U(w) + {\cal O}((z-w)^{-2})$.
This suggests that 
the extra quasi-primary field consisting of 
$G \pa U(w), \pa G U(w)$ and other derivative 
terms should be considered to remove the higher singular terms above.  
Finally, the consistent coefficients, 
$\frac{764}{8325}$ and 
$\frac{287}{5550\sqrt{6}}$, in the second line of the first-order pole in 
(\ref{uo72})
are 
fixed from the above analysis (i.e. superprimary condition) 
and the following spin-$5$ primary current   
remains
\bea
O_{5}(z) = -\frac{8}{225} \, \sqrt{\frac{2}{15}} \, i \,
f^{abc} d^{ade} \, \psi^b \psi^c \psi^d \pa^3 \psi^e (z) 
+\mbox{other lower order derivative terms}.
\label{c4o5}
\eea
Therefore, two primary currents (\ref{c4o4''}) and 
(\ref{c4o5}) were obtained where the former is the $\theta$-independent 
component field  of $\hat{O}_{4'}(Z)$ 
and the latter is the $\theta$-dependent component field of 
$\hat{O}_{\frac{9}{2}}(Z)$.

Let us calculate the OPE of spin-$\frac{5}{2}$ current (\ref{c4u}) and spin-$4$ current (\ref{c4o4}) to determine the superpartners corresponding to the 
above 
two primary fields.
Each singular term, starting from fourth-order singular term
can be obtained.
The final result is presented first, which explains how 
this result can be obtained explicitly  
\bea
&& U(z) \; O_4 (w) = \frac{1}{(z-w)^4} \, \frac{6\sqrt{6}}{37} U(w) +
\frac{1}{(z-w)^3} \, (\frac{1}{5}) \frac{6\sqrt{6}}{37} \pa U(w)  
\nonu \\
&& +\frac{1}{(z-w)^2} \, \left[(\frac{1}{30}) \frac{6\sqrt{6}}{37} \pa^2 U +
\frac{1596}{12025} \left( G W -\frac{1}{6\sqrt{6}} \pa^2 U \right) +
\frac{232 \sqrt{6}}{7215} \left( T U -\frac{1}{4} \pa^2 U
\right) + P_{\frac{9}{2}}\right](w) 
\nonu \\
&& + \frac{1}{(z-w)} \left[ (\frac{1}{210}) \frac{6\sqrt{6}}{37} \pa^3 U +
(\frac{1}{3})  \frac{1596}{12025} \pa \left( G W -\frac{1}{6\sqrt{6}} \pa^2 U \right) +
(\frac{1}{3}) \frac{232 \sqrt{6}}{7215} \pa \left( T U -\frac{1}{4} \pa^2 U
\right) \right. \nonu \\
&& \left. + (\frac{1}{3}) \pa P_{\frac{9}{2}} + Q_{\frac{11}{2}} \right](w) +\cdots.
\label{c4uo4}
\eea
The structure of the right hand side appears 
similar to (\ref{uo72}) by changing 
$W(w)$ to $U(w)$ and vice versa.
Once the structure constant, $\frac{6\sqrt{6}}{37}$ 
in the highest order singular term
is found, the other relative coefficients in the lower singular terms, 
which are associated with the spin-$\frac{5}{2}$
current $U(w)$, are determined automatically  using the formula (\ref{PhiPhi}).
Therefore, the other terms in the right hand side of (\ref{c4uo4})
should be determined.
Consider the nontrivial second-order singular terms.
One should ensure 
that there are 
three extra quasi-primary fields including the last primary field with 
the right structure constants.
As performed before, the OPE between $T(z)$ and 
$ \left(\{U O_{4}\}_{-2} -(\frac{1}{30}) 
\frac{6\sqrt{6}}{37} \pa^2 U\right)(w)$ was calculated.
This OPE has $\frac{18\sqrt{6}}{37} U(w)$ in the fourth-order singular term.
This shows that the extra quasi-primary field should contain a $T U(w)$ term.
Moreover, the OPE between $G(z)$ and 
 $ \left(\{U O_{4}\}_{-2} -(\frac{1}{30}) 
\frac{6\sqrt{6}}{37} \pa^2 U\right)(w)$ should be calculated. 
This OPE has 
the following third-order singular term,
$\frac{1}{(z-w)^3} \, \frac{276}{185} W(w)$. 
Then $G W(w)$ can be considered 
 a quasi-primary field with some derivative term. 
The final new spin-$\frac{9}{2}$ primary field can be obtained 
by subtracting these two candidates from the second-order singular terms,
\bea
P_{\frac{9}{2}}(z) =
\frac{1}{13} \, \sqrt{\frac{2}{5}} \, i \,
d^{abc} f^{bde} f^{cfg} \psi^a \psi^d \psi^e \psi^f \pa^2 \psi^g(z) +
\mbox{other first-order derivative terms}, 
\label{c4p92}
\eea
which will play the role of the component field of some superprimary field. 

Let us focus on the next first-order singular term.
Now all possible descendant fields originating from 
the quasi-primary fields of spin-$\frac{5}{2}$ 
and of spin-$\frac{9}{2}$ can be written with the correct 
coefficient functions.
Then the following quasi-primary field of spin-$\frac{11}{2}$
remains
\bea
Q_{\frac{11}{2}}(z) & = & -\frac{584}{7215} \, \sqrt{\frac{2}{3}}\, 
\left( T \pa U -\frac{5}{4} \pa T U -\frac{1}{7} \pa^3 U \right)(z) \nonu \\
& + & 
\frac{10}{481}\,
\left( G \pa W -2 \pa G W -\frac{1}{21}\, \sqrt{\frac{2}{3}} 
\, \pa^3 U \right)(z)
+\frac{3}{4} \left( G O_{4''} -\frac{2}{9} \pa O_{\frac{9}{2}'} \right)(z).
\label{q112}
\eea
This expression is expressed in terms of the 
various WZW currents and it is difficult to express this
in terms of (\ref{q112}).
Note that there is 
a new primary field, $O_{\frac{9}{2}'}(z)$, 
in the right hand side of (\ref{q112}).
From the newly obtained spin-$4$ primary field in (\ref{c4o4''}),  
the OPE of $G(z)$ with $O_{4''}(w)$ can be calculated, which 
leads to a nonzero first-order singular term that is 
equal to the above $O_{\frac{9}{2}'}(w)$ where
\bea
O_{\frac{9}{2}'}(z) = -\frac{8}{65} \, \sqrt{\frac{2}{5}} \, i \,
d^{abc} f^{bde} f^{cfg} \psi^a \psi^d \psi^e \psi^f \pa^2 \psi^g(z) +
\mbox{other first-order derivative terms}. 
\label{c4o92'}
\eea
The OPE between  $G(z)$ and the field $O_{\frac{9}{2}'}(w)$
should be calculated for a double check.
This provides a consistent result. 
That is, the second-order singular term has $8 O_{4''}(w)$,
whereas the first-order singular term is given by $\pa O_{4''}(w)$.
These are combined into a single superprimary field, 
$\hat{O}_{4'}(Z)$, as shown in (\ref{higherspincurrents}).
The remaining current, $O_{\frac{9}{2}''}(z)$, should be determined.
The quasi-primary property of (\ref{q112}) 
can be checked by the following OPE with $T(z)$
\bea
T(z) \; Q_{\frac{11}{2}} (w) & = & \frac{1}{(z-w)^5} \, \frac{72}{259} \sqrt{6} \, U(w) +
\frac{1}{(z-w)^4} \, (-\frac{1}{5})  \frac{72}{259} \sqrt{6} \, \pa U(w)
\nonu \\
& + & {\cal O}((z-w)^{-2}).
\label{c4tq112}
\eea
No third-order pole exists.
Furthermore, the following OPE can be obtained 
with the spin-$\frac{3}{2}$ current
\bea
G(z) \; Q_{\frac{11}{2}} (w) & = &
\frac{1}{(z-w)^4} \, \frac{48}{1295} \, W(w) \nonu \\
& + & \frac{1}{(z-w)^3} \left[ 
(-\frac{1}{6}) \frac{48}{1295} \pa W  +\frac{16}{3} O_{4''} +\frac{32}{481} \sqrt{6} \left( 
G U -\frac{\sqrt{6}}{3} \pa W  \right) \right](w) 
\nonu \\ 
& + &  {\cal O}((z-w)^{-2}).
\label{c4gq112}
\eea
Note that in (\ref{q112}), three quantities specified 
by the bracket can be checked to ensure that 
they are quasi-primary fields.
Appendices $B$ and $C$ 
present some properties of the various quasi-primary fields
described in this paper.
According to the definition of quasi-primary 
field introduced in the introduction, 
any linear combination of quasi-primary fields leads to 
another quasi-primary field that can be written in terms of the 
known currents.

Consider the next OPE between the spin-$3$ current (\ref{c4w}) 
and the spin-$\frac{7}{2}$ current (\ref{c4o72})  to complete
the OPE between $\hat{W}(Z_1)$ and $\hat{O}_{\frac{7}{2}}(Z_2)$.
The result is as follows: 
\bea
&& W(z) \; O_{\frac{7}{2}} (w) = \frac{1}{(z-w)^4} \,\frac{6}{37} U(w) +
\frac{1}{(z-w)^3} \, (\frac{2}{5})  \frac{6}{37} \pa U(w)
\nonu \\
&& + \frac{1}{(z-w)^2} \left[  (\frac{1}{10})  \frac{6}{37} \pa^2 U  +\frac{412}{7215} \left(
T U -\frac{1}{4} \pa^2 U \right) +
\frac{116\sqrt{6}}{12025} \left( G W -\frac{1}{6\sqrt{6}} \pa^2 U \right)+ \frac{1}{\sqrt{6}} P_{\frac{9}{2}}
\right. \nonu \\
&&  \left. + \frac{1}{\sqrt{6}} O_{\frac{9}{2}'} \right](w)
+\frac{1}{(z-w)} \left[ 
 (\frac{2}{105})  \frac{6}{37} \pa^3 U  +(\frac{4}{9}) \frac{412}{7215} \pa \left(
T U -\frac{1}{4} \pa^2 U \right) \right. \nonu \\
&& \left. +
(\frac{4}{9}) 
\frac{116\sqrt{6}}{12025} \pa \left( G W -\frac{1}{6\sqrt{6}} \pa^2 U \right)
+ (\frac{4}{9}) \frac{1}{\sqrt{6}} \pa P_{\frac{9}{2}} 
+  (\frac{4}{9}) \frac{1}{\sqrt{6}} \pa O_{\frac{9}{2}'} +Q_{\frac{11}{2}'} \right](w) +\cdots.
\label{c4wo72}
\eea
This appears similar to the previous OPE (\ref{c4uo4}).
As stated 
before, once the structure constant appearing in front of $U(w)$ in the right hand side of
(\ref{c4wo72}) is found from the corresponding singular term with the 
WZW currents, 
the relevant coefficients
associated with its descendant fields in the second-order and first-order poles 
are known from (\ref{PhiPhi}).
The  next step is to look at the next nontrivial 
lower singular terms to determine if there is 
a  new primary field or not. If not, 
the singular terms should be expressed 
in terms of the known quasi-primary fields or 
new unknown quasi-primary fields.   
The calculated OPE between $T(z)$ and 
$ \left(\{W O_{\frac{7}{2}}\}_{-2} -(\frac{1}{10}) 
\frac{6}{37} \pa^2 U\right)(w)$  becomes
$-\frac{1}{(z-w)^4} \, \frac{9}{37} U(w) -\frac{1}{(z-w)^3} \, (\frac{3}{5}) \frac{9}{37} 
\pa U(w) + {\cal O}((z-w)^{-2})$.
On the other hand, the OPE with $G(z)$ leads to 
$-\frac{6\sqrt{6}}{185} W(w)$ in the third-order singular term.
This suggests that 
the two quasi-primary fields can be extracted 
in the second-order singular terms
and the structure constants are fixed by the primary condition.
The remaining terms are characterized by two independent 
spin-$\frac{9}{2}$ currents, 
$P_{\frac{9}{2}}(w)$ (\ref{c4p92}) and 
$O_{\frac{9}{2}'}(w)$ (\ref{c4o92'}), which were considered previously.
The first-order singular terms are described as follows.
Because two quasi-primary fields and two 
spin-$\frac{9}{2}$  primary fields were found  at 
the second-order singular terms,
their coefficient functions 
are determined without ambiguities. 
Therefore, any quasi-primary field should be identified  after 
extracting those known field contents from the first-order pole.
The remaining field is then spin-$\frac{11}{2}$ quasi-primary field
\bea
Q_{\frac{11}{2}'}(z)  & = &  -\frac{146}{7215} \sqrt{\frac{2}{3}} 
\left( G \pa W -2 \pa G W -\frac{1}{21} \sqrt{\frac{2}{3}} \pa^3 U 
\right)(z) \nonu \\
& + & \frac{16}{64935} \left( T \pa U -\frac{5}{4} \pa T U -\frac{1}{7} 
\pa^3 U \right)(z)
+  \frac{1}{4} \sqrt{\frac{3}{2}} \left( G O_{4''} -\frac{2}{9} \pa O_{\frac{9}{2}'}\right)(z).
\label{q112'}
\eea
All the field contents for this quasi-primary field 
are given in terms of the previously determined quasi-primary fields. 
Each quasi-primary field in (\ref{q112'}) also appears in (\ref{q112}).
The only difference is the relative coefficients between them. 
As shown in (\ref{c4tq112}) and (\ref{c4gq112}),
the following OPE can be derived
\bea
T(z) \; Q_{\frac{11}{2}'}(w) & = &  \frac{1}{(z-w)^5} \, \frac{80}{777} U(w)+
\frac{1}{(z-w)^4} \, (-\frac{1}{5}) \frac{80}{777} \pa U (w) \nonu \\
& + & {\cal O}((z-w)^{-2}).
\label{c4tq112'}
\eea
No third-order pole exists.
Similarly, 
\bea
G(z) \;  Q_{\frac{11}{2}'}(w) & = & \frac{1}{(z-w)^4} \, \frac{144}{1295} \sqrt{6} W(w) \nonu \\
& + &
\frac{1}{(z-w)^3} \left[ (-\frac{1}{6}) \frac{144}{1295} \pa W +
\frac{8}{3} \sqrt{\frac{2}{3}} O_{4''} +\frac{32}{481} \left( G U-\frac{\sqrt{6}}{3} \pa W 
\right)   \right](w)  \nonu \\
& + & {\cal O}((z-w)^{-2}). 
\label{c4gq112'}
\eea
Of course, the field contents in the right hand side of (\ref{c4tq112'}) and (\ref{c4gq112'})
are the same as those in (\ref{c4tq112}) and (\ref{c4gq112}), as expected.  
Thus far, two primary fields, (\ref{c4p92}) and (\ref{c4o92'}),
were found, 
which will play the role of the undetermined spin-$\frac{9}{2}$ current
$O_{\frac{9}{2}''}(z)$.

The OPE between spin-$3$ current 
(\ref{c4w}) and spin-$4$ current (\ref{c4o4}) was considered.
Therefore,
\bea
&& W(z) \; O_4 (w) = \frac{1}{(z-w)^4} \, \frac{48\sqrt{6}}{185} \, W(w) 
\nonu \\
&& +\frac{1}{(z-w)^3} \, \left[ (\frac{1}{3}) \frac{48\sqrt{6}}{185} \, \pa W + \sqrt{\frac{2}{3}} O_{4''} 
+ \frac{12}{481} \left( G U -\frac{\sqrt{6}}{3} \pa W \right) \right](w) 
\nonu \\
&& 
+\frac{1}{(z-w)^2} \, 
\left[ (\frac{1}{14}) \frac{48\sqrt{6}}{185} \, \pa^2 W + (\frac{3}{8}) 
\sqrt{\frac{2}{3}} \pa O_{4''} 
+ (\frac{3}{8}) 
\frac{12}{481} \pa \left( G U -\frac{\sqrt{6}}{3} \pa W \right) \right. \nonu \\
&& \left. 
+5 \sqrt{\frac{2}{3}}
O_5 
+
Q_{5} \right](w) 
\nonu \\
 && 
+\frac{1}{(z-w)} \, 
\left[ (\frac{1}{84}) \frac{48\sqrt{6}}{185} \, \pa^3 W + 
(\frac{1}{12}) \sqrt{\frac{2}{3}} \pa^2 O_{4''} 
+ (\frac{1}{12}) 
\frac{12}{481} \pa^2 \left( G U -\frac{\sqrt{6}}{3} \pa W \right) \right. \nonu \\
&& \left. 
+
(\frac{2}{5}) 5 \sqrt{\frac{2}{3}}
\pa O_5 +
(\frac{2}{5}) \pa Q_{5} + Q_6 \right](w). 
\label{c4wo4}
\eea
By identifying the highest singular term from the explicit expression of the 
OPE,
the descendant fields can be expressed 
with correct coefficient functions for the spin-$3$ current, $W(w)$,
 on the right hand side.
The first nontrivial third-order term can be analyzed.
The OPE between $T(z)$ and 
 $ \left(\{W O_{4}\}_{-3} -(\frac{1}{3}) 
\frac{48\sqrt{6}}{185} \pa W\right)(w)$ can be calculated, 
and this OPE does not produce any higher order 
( $> 2$) singular terms. This leads to the appearance of a primary field. 
What of the OPE between $G(z)$ and  
 $ \left(\{W O_{4}\}_{-3} -(\frac{1}{3}) 
\frac{48\sqrt{6}}{185} \pa W\right)(w)$?
The third-order term of this OPE contains 
$\frac{4}{37} U(w)$. Therefore, the 
primary field containing 
$G U(w)$ can be extracted from the third-order pole. 
The spin-$4$ current (\ref{c4o4''}) 
remains as described previously.
Because  the third-order singular terms are determined completely, 
let us move on the second-order term.
According to (\ref{PhiPhi}), 
the first line of the second-order 
terms in (\ref{c4wo4}) can be extracted 
from the entire second-order pole.
The OPE can then be calculated 
with $T(z)$ and $ \left(\{W O_{4}\}_{-2} -\mbox{first line}\right)(w)$.
The nontrivial part of this OPE contains $\frac{1248\sqrt{6}}{1295} W(w)$ at 
the fourth-order pole.
For the OPE between $G(z)$ and  $ \left(\{W O_{4}\}_{-2} -\mbox{first line}\right)(w)$,
there are $\frac{573}{518} U(w)$ in the fourth-order term and 
$(-\frac{1}{5}) \frac{573}{518} \pa U(w)$ in the third-order term.
Therefore, the two quasi-primary fields 
corresponding to $T W(w)$(and derivative term) 
and $G \pa U(w)$(and other terms)  
respectively, were subtracted.
The spin-$5$ primary field (\ref{c4o5})
was obtained 
after subtracting these two 
quasi-primary fields properly in the second-order pole.
Moreover, the following spin-$5$ quasi-primary field 
can be obtained
\bea
Q_5(z) =\frac{458}{1665} \sqrt{\frac{2}{3}} 
\left( T W -\frac{3}{14} \pa^2 W\right)(z)
- \frac{19}{3330} 
\left( G \pa U -\frac{5}{3} \pa G U -\frac{\sqrt{6}}{7} \pa^2 W \right)(z)
\label{c45quasi}
\eea
and its OPEs with $T(z)$ and $G(z)$
are given by
\bea
T(z) \; Q_5(w) & = & \frac{1}{(z-w)^4} \, \frac{1248\sqrt{6}}{1295} W(w) +{\cal O}((z-w)^{-2}), 
\nonu \\
G(z) \; Q_5 (w) & = & -\frac{1}{(z-w)^4} \, \frac{573}{2590} U(w) -
\frac{1}{(z-w)^3} \, (-\frac{1}{5}) \frac{573}{2590} 
\pa U(w) \nonu \\
& + & {\cal O}((z-w)^{-2}).
\label{c45quasiOPEs}
\eea
The first equation of (\ref{c45quasiOPEs}) des not have 
a third-order pole.
The entire structure of the second-order pole is fixed, and 
the last first-order pole can be described. 
By subtracting the  three descendant fields (coming from spin-$3$ primary 
field and two spin-$4$ quasi-primary fields) 
and the remaining two descendant fields coming from 
the spin-$5$ quasi-primary fields appearing in 
the second-order pole, there is a spin-$6$ quasi-primary field that can be
expressed as follows: 
\bea
Q_6(z) & = & 
\frac{32}{12025} \left( G \pa^2 U - 4 \pa G \pa U +\frac{5}{2} \pa^2 G U -
\frac{1}{2\sqrt{6}} \pa^3 W  \right)(z) \nonu \\
& - &  \frac{192}{12025} \sqrt{6} \,
\left( T \pa W  -\frac{3}{2}  \pa T W -\frac{1}{8} \pa^3 W \right)(z) 
+\frac{1}{2} \sqrt{\frac{3}{2}} \, \left( T O_{4''} -\frac{1}{6} \pa^2 O_{4''} \right)(z)  
\nonu \\
& - & \frac{1}{4} \sqrt{\frac{3}{2}} 
\left( G O_{\frac{9}{2}'} -\frac{1}{9} \pa^2 O_{4''}\right)(z).
\label{c46quasi}
\eea
The following OPEs
can be calculated easily
\bea
T(z) \; Q_6 (w) & = &  \frac{1}{(z-w)^5} \, \frac{288}{925} \sqrt{6} \, W(w)
\nonu \\
& + & \frac{1}{(z-w)^4} \, \left[ (-\frac{1}{6}) \frac{288}{925} \sqrt{6} \, 
\pa W + 4 \sqrt{\frac{2}{3}} \,
O_{4''} + \frac{48}{481} \left(G U -\frac{\sqrt{6}}{3} 
\pa W \right)  \right] (w) \nonu \\
& + & {\cal O}((z-w)^{-2}),
\nonu \\
G(z) \; Q_6 (w)  & = &   \frac{1}{(z-w)^5} \, \frac{64}{185}  \, U(w)
 +  \frac{1}{(z-w)^4} \, (-\frac{2}{5}) \frac{64}{185} \pa U (w) \nonu \\
& + &   \frac{1}{(z-w)^3}  \left[  (\frac{1}{30}) \frac{64}{185} \pa^2 U 
+ \frac{32}{481} \sqrt{6} \left(G W -\frac{1}{6\sqrt{6}} 
\pa^2 U \right) \right. \nonu \\
&- &   \left. \frac{64}{481} \left( T U -\frac{1}{4} \pa^2 U \right)
-\frac{8}{3} \sqrt{\frac{2}{3}} \, O_{\frac{9}{2}'} 
\right] (w)  +  {\cal O}((z-w)^{-2}).
\label{c46quasiOPEs}
\eea
The first equation of 
(\ref{c46quasiOPEs}) does not have a third-order pole.

Therefore, 
the supercurrent, $\hat{O}_{4'}(Z)$ and $\hat{O}_{\frac{9}{2}}(Z)$,
 in the list of 
(\ref{higherspincurrents}) was constructed. 
Originally, the expression, $\hat{O}_{4'}(Z)$, was reported by \cite{ASS}.
Their component fields can be expressed as 
(\ref{c4o4''}), (\ref{c4o92'}) and (\ref{c4o5}).
Furthermore, the remaining component field can be expressed as
\bea
O_{\frac{9}{2}''}(z) = O_{\frac{9}{2}'}(z) +\frac{8}{5} P_{\frac{9}{2}}(z),
\label{c4o92''}
\eea
where the component fields are given in (\ref{c4o92'}) and (\ref{c4p92}).
The OPE between $G(z)$ and $O_{\frac{9}{2}''}(w)$
can be checked 
 to determine  if (\ref{c4o92''}) is the right superpartner of the 
current $O_5(z)$. 
The results showed 
that the first-order pole provides the spin-$5$ as with (\ref{c4go72}).  
Moreover, the OPE between $G(z)$ and $O_5(w)$ provides 
the correct singular terms 
where the second order pole has $9 O_{\frac{9}{2}''}(w)$ and the first-order pole 
has $\pa O_{\frac{9}{2}''}(w)$, as expressed in 
(\ref{c4go4}) \footnote{The 
OPE $G(z) \; P_{\frac{9}{2}}(w) =- \frac{1}{(z-w)^2} 5 O_{4''}(w) +
\frac{1}{(z-w)} \frac{5}{8} \left( O_5 -\pa O_{4''} \right)(w) +\cdots$
can be obtained conveniently. Therefore, one should 
consider equation (\ref{c4o92''}) to remove the unwanted terms $O_{4''}(w)$ and its descendant 
field.}. 
The following superfusion rule $
\left[\hat{W} \right] \left[\hat{O}_{\frac{7}{2}} \right]  =   \left[\hat{W} \right] + \left[\hat{O}_{4'}
\right] +\left[\hat{O}_{\frac{9}{2}} \right]$ can be considered.
The OPE between $\hat{W}(Z_1)$ and $\hat{O}_{4'}(Z_2)$ can be calculated 
to find the remaining higher spin currents.

$\bullet$ The construction of $\hat{O}_{\frac{11}{2}}(Z)$ and $\hat{O}_{6}(Z)$

This section describes 
the OPE between the spin-$\frac{5}{2}$ current (\ref{c4u}) and  spin-$4$
current (\ref{c4o4''}).
\bea
&& U(z) \; O_{4''} (w) = \frac{1}{(z-w)^3} \, \frac{888}{65} O_{\frac{7}{2}}(w) 
+\frac{1}{(z-w)^2} \, \left[ (\frac{2}{7}) \frac{888}{65} \pa O_{\frac{7}{2}}  -\frac{88}{65}
O_{\frac{9}{2}} \right](w)
\nonu \\
&& +\frac{1}{(z-w)} \, \left[ (\frac{3}{56}) \frac{888}{65} \pa^2 O_{\frac{7}{2}}  
-(\frac{1}{3}) \frac{88}{65} \pa
O_{\frac{9}{2}} + \frac{288}{61} \left( T O_{\frac{7}{2}} -\frac{3}{16} \pa^2 O_{\frac{7}{2}} \right) \right.
\nonu \\
&&  + \frac{58\sqrt{6}}{793} 
\left( G P_{4'}^{uu} -\frac{4\sqrt{6}}{9} \pa O_{\frac{9}{2}}-\frac{\sqrt{6}}{56} \pa^2 
O_{\frac{7}{2}} \right) \nonu \\
&& \left. -\frac{5493}{3965} \sqrt{\frac{3}{2}} 
\left( G P_{4'}^{ww} -\frac{2}{9} \sqrt{\frac{2}{3}} \pa O_{\frac{9}{2}}
-\frac{1}{14} \sqrt{\frac{2}{3}} \pa^2 O_{\frac{7}{2}} \right) 
+ O_{\frac{11}{2}} \right](w) +\cdots. 
\label{c4uo4''}
\eea
From the highest singular term in (\ref{c4uo4''}), 
the structure constant for the spin-$\frac{7}{2}$ current can be 
determined and the formula (\ref{PhiPhi}) with this numerical value
gives the explicit form for some part of 
the next singular terms.  
As stated before, 
this descendant field in the second-order singular term
can be extracted
and the remaining term can be expressed 
in terms of the spin-$\frac{9}{2}$ current (\ref{c4o92}).
The OPE between the $T(z)$ and $( \{ U O_{4''} \}_{-1} -(\frac{3}{56}) 
\frac{888}{65} \pa^2 O_{\frac{7}{2}}  
+ (\frac{1}{3}) \frac{88}{65}
\pa O_{\frac{9}{2}} )(w)$ can be calculated to determine the nontrivial 
first-order singular term completely.
This leads to the nontrivial fourth-order 
pole with $\frac{2997}{65} O_{\frac{7}{2}}(w)$,
the quasi-primary field containing $T O_{\frac{7}{2}}$
can be extracted  from the first-order pole.
Furthermore,
the OPE between $G(z)$ and 
 $( \{ U O_{4''} \}_{-1} -(\frac{3}{56}) 
\frac{888}{65} \pa^2 O_{\frac{7}{2}}  
+ (\frac{1}{3}) \frac{88}{65}
\pa O_{\frac{9}{2}} )(w)$ has a nontrivial third-order singular term with $\left(-\frac{439}{195} 
\sqrt{\frac{2}{3}} P_{4'}^{uu} + \frac{526}{65} \sqrt{\frac{2}{3}}  P_{4'}^{ww} \right)(w)$.
The two quasi-primary fields containing 
$G P_{4'}^{uu}(w)$ and $G P_{4'}^{ww}(w)$ can be considered.
Finally, the new spin-$\frac{11}{2}$ current can be 
derived by rearranging the first-order terms as done previously,
\bea
O_{\frac{11}{2}}(z) = \frac{3}{1586\sqrt{2}} \, f^{abc} \psi^a \psi^b \pa^4 \psi^c(z) +
\mbox{other lower order derivative terms},
\label{c4o112}
\eea
which is the $\theta$-independent term of $\hat{O}_{\frac{11}{2}}(Z)$.

Consider the spin-$\frac{5}{2}$ current (\ref{c4u}) 
and the spin-$\frac{9}{2}$ current (\ref{c4o92'}) to determine 
other unknown higher spin currents.
The result can be expressed as
\bea
&& U(z) \; O_{\frac{9}{2}'} (w) =
\frac{1}{(z-w)^3} \, \left[ \frac{116}{65}\sqrt{\frac{2}{3}} P_{4'}^{uu} -
\frac{584\sqrt{6}}{65} P_{4'}^{ww}  \right](w)
\nonu \\
&&+ \frac{1}{(z-w)^2} \, \left[ (\frac{1}{4}) 
\frac{116}{65}\sqrt{\frac{2}{3}} \pa P_{4'}^{uu} -
(\frac{1}{4}) \frac{584\sqrt{6}}{65} \pa P_{4'}^{ww} \right. \nonu \\
&& + \left. \frac{792}{65} 
\left( G O_{\frac{7}{2}} +\frac{1}{4\sqrt{6}} \pa P_{4'}^{uu}
-\frac{\sqrt{6}}{4} \pa P_{4'}^{ww} \right)  \right](w) 
\nonu \\
&& + \frac{1}{(z-w)} \, \left[ (\frac{1}{24}) \frac{116}{65}\sqrt{\frac{2}{3}} \pa^2 P_{4'}^{uu} -
(\frac{1}{24}) \frac{584\sqrt{6}}{65} \pa^2 P_{4'}^{ww} \right.
\nonu \\
&& +(\frac{3}{10}) \frac{792}{65} 
\pa \left( G O_{\frac{7}{2}} +\frac{1}{4\sqrt{6}} \pa P_{4'}^{uu}
-\frac{\sqrt{6}}{4} \pa P_{4'}^{ww} \right)   \nonu \\
&&  + \frac{41524\sqrt{6}}{91195} \left( T P_{4'}^{uu} -\frac{1}{6} \pa^2 P_{4'}^{uu} \right) - 
\frac{328993\sqrt{6}}{91195} 
\left( T P_{4'}^{ww} -\frac{1}{6} \pa^2 P_{4'}^{ww} \right) \nonu \\
&&  - \frac{43068}{19825} \left( G \pa O_{\frac{7}{2}} -\frac{7}{3} \pa G O_{\frac{7}{2}} +
\frac{1}{9\sqrt{6}} \pa^2 P_{4'}^{uu} -\frac{1}{3} \sqrt{\frac{2}{3}} \pa^2 P_{4'}^{ww} 
\right)  \nonu \\
&& \left. - \frac{231}{299} \left( G O_{\frac{9}{2}} -\frac{2}{63} \sqrt{\frac{2}{3}} \pa^2 P_{4'}^{uu} +
\frac{1}{21\sqrt{6}} \pa^2 P_{4'}^{ww} \right) + P_6\right] (w) +\cdots. 
\label{c4uo92'}
\eea
In this case, the first nontrivial primary fields in the right hand 
side are given by (\ref{c4'uu}) and (\ref{c4'ww}).
How can the quasi-primary field be extracted in the second-order pole?
Actually, there exists a primary field of spin-$5$ but its OPE with $G(z)$
exhibits unusual behavior. See Appendices $B$ and $C$. 
A third-order pole exists, which is expressed as
$\frac{4884}{65} O_{\frac{7}{2}}(w)$.
The primary field can be written 
as the one in the third term of second-order pole in 
(\ref{c4uo92'}).
By taking the derivative to these second-order terms with the 
appropriate coefficients  from (\ref{PhiPhi})
and subtracting them, the nontrivial terms remain 
in the first-order singular terms.
Therefore, it is important to determine if there are 
four quasi-primary fields and a single primary field 
right. 
To accomplish this, the OPE between $T(z)$ and 
$( \{ U O_{\frac{9}{2}'} \}_{-1} -\mbox{first three descendant fields})(w)$
should be calculated.
The fourth-order term of this OPE is given as 
$\frac{406}{65} \sqrt{\frac{2}{3}} P_{4'}^{uu}(w) -\frac{2044\sqrt{6}}{65} P_{4'}^{ww}(w)$.
The quasi-primary field containing $T P_{4'}^{uu}(w)$
and the quasi-primary field $T P_{4'}^{ww}(w)$ with possible derivative 
terms can be  considered.
Moreover, the OPE $G(z)$ with $( \{ U O_{\frac{9}{2}'} \}_{-1} -\mbox{first three descendant fields})(w)$
produces 
$-\frac{3108}{325} O_{\frac{7}{2}}(w)$ at the fourth-order term and 
$(-\frac{1}{7}) \frac{3108}{325} \pa O_{\frac{7}{2}}(w) -\frac{2464}{195} O_{\frac{9}{2}}(w)$
at the third-order pole.
This allows the two additional quasi-primary fields  to be taken, 
as shown in (\ref{c4uo92'}).   
Finally, subtracting the above four quasi-primary 
fields properly (with correct coefficients) leaves 
the following new spin-$6$ primary field:
\bea
P_6(z) = \frac{17}{8418\sqrt{6}} \, \psi^a \pa^5 \psi^a(z)
 +
\mbox{other lower order derivative terms}. 
\label{c4p6}
\eea
This will play the role of the final spin-$6$ current 
that this study is interested in.
Thus far, two primary fields (\ref{c4o112}) and (\ref{c4p6})
were found.

The OPE between 
the spin-$3$ current (\ref{c4w}) and spin-$4$ (\ref{c4o4''})
current are described as follows:
\bea
&& W(z) \; O_{4''} (w) =
\frac{1}{(z-w)^3} \, \left[ -\frac{328}{195} P_{4'}^{uu} +\frac{304}{65} P_{4'}^{ww}  \right](w) 
\nonu \\
&& + \frac{1}{(z-w)^2} \, 
\left[ -(\frac{3}{8}) \frac{328}{195} \pa P_{4'}^{uu} + (\frac{3}{8}) 
\frac{304}{65} \pa P_{4'}^{ww}  +\frac{132\sqrt{6}}{65} \left( G O_{\frac{7}{2}} +\frac{1}{4\sqrt{6}} \pa P_{4'}^{uu}
-\frac{\sqrt{6}}{4} \pa P_{4'}^{ww}  \right)  \right](w) 
\nonu \\
&&  + \frac{1}{(z-w)} \, 
\left[ -(\frac{1}{12}) \frac{328}{195} \pa^2 P_{4'}^{uu} + (\frac{1}{12}) 
\frac{304}{65} \pa^2 P_{4'}^{ww}  
+
(\frac{2}{5}) 
\frac{132\sqrt{6}}{65} \pa \left( G O_{\frac{7}{2}} +\frac{1}{4\sqrt{6}} \pa P_{4'}^{uu}
-\frac{\sqrt{6}}{4} \pa P_{4'}^{ww}  \right) \right.  \nonu \\ 
&& 
 - \frac{16896}{91195} \left( T P_{4'}^{uu} -\frac{1}{6} \pa^2 P_{4'}^{uu} \right) - 
\frac{24772}{91195} 
\left( T P_{4'}^{ww} -\frac{1}{6} \pa^2 P_{4'}^{ww} \right) \nonu \\
&&  - \frac{4736\sqrt{6}}{10675} \left( G \pa O_{\frac{7}{2}} -\frac{7}{3} \pa G O_{\frac{7}{2}} +
\frac{1}{9\sqrt{6}} \pa^2 P_{4'}^{uu} -\frac{1}{3} \sqrt{\frac{2}{3}} \pa^2 P_{4'}^{ww} 
\right)  \nonu \\
&& \left. - \frac{66\sqrt{6}}{1495} \left( G O_{\frac{9}{2}} -\frac{2}{63} \sqrt{\frac{2}{3}} \pa^2 P_{4'}^{uu} +
\frac{1}{21\sqrt{6}} \pa^2 P_{4'}^{ww} \right) + P_{6'}\right] (w) +\cdots. 
\label{c4wo4''}
\eea
The right hand side appears similar to the OPE (\ref{c4uo92'}).
For given structure constants on the two spin-$4$ fields, 
its descendant field terms are fixed completely  and 
the OPE between $T(z)$ with 
$( \{ W O_{4''} \}_{-2} -\mbox{two descendant fields})(w)$
are calculated, which 
means
there is no higher order singular term (order greater than $2$) 
suggesting that 
the extra terms  should correspond to the primary field.
On the other hand, 
the OPE between $G(z)$ and
$( \{ W O_{4''} \}_{-2} -\mbox{two descendant fields})(w)$
provides a 
nontrivial third-order pole with $\frac{814\sqrt{6}}{65} O_{\frac{7}{2}}(w)$.
The 
corresponding primary field  can be expressed 
in terms of $G O_{\frac{7}{2}}(w)$ plus other 
derivative terms in (\ref{c4wo4''}).
Based on these results for the second-order pole, three 
derivative terms can be obtained correctly
in the first-order singular terms (coming from the second-order pole).
The OPE between $T(z)$ and $( \{ W O_{4''} \}_{-1} -\mbox{three descendant fields})(w)$ can be calculated,
which will determine the possible quasi-primary fields 
that need to be considered. 
A fourth-order pole exists in this OPE, which is expressed as 
$-\frac{1312}{195} P_{4'}^{uu}(w) +\frac{1216}{65} P_{4'}^{ww}(w)$.
This suggests 
that the quasi-primary field should contain $T P_{4'}^{uu}(w)$
and $T P_{4'}^{ww}(w)$, respectively.
For the OPE $G(z) \; ( \{ W O_{4''} \}_{-1} -\mbox{three descendant fields})(w)$,
there is 
$\frac{3256\sqrt{6}}{325} O_{\frac{7}{2}}(w)$ at the fourth-order pole
and $(-\frac{2}{7}) \frac{3256\sqrt{6}}{325} \pa O_{\frac{7}{2}}(w)$
at the third-order pole.
Finally, after  extracting the new four quasi-primary fields from the first-order pole,
the following new spin-$6$ primary field
can be derived
\bea
P_{6'}(z) =
 \frac{121}{820755} \, \psi^a \pa^5 \psi^a(z)
 +
\mbox{other lower order derivative terms}. 
\label{c4p6'}
\eea
This is another candidate for the spin-$6$ current in the list 
(\ref{higherspincurrents}).

As the spin-$\frac{11}{2}$ current (\ref{c4o112})
has been found, its superpartner $O_6(z)$ current
should be determined.
By calculating 
the OPE $G(z)$ with $O_{\frac{11}{2}}(w)$, 
which should generate $O_6(w)$, 
it can be seen that it consists of a linear combination of
the previous spin-$6$ fields (\ref{c4p6}) and (\ref{c4p6'}) 
\bea
O_6(z) = -P_6(z) + \sqrt{6} P_{6'}(z). 
\label{c4o6}
\eea
Similarly, the OPE between $G(z)$ and the current $O_6(w)$
can be calculated. The results show 
that the second order pole has $11 O_{\frac{11}{2}}(w)$ 
and the first-order pole has $\pa O_{\frac{11}{2}}(w)$, as expected. 
Therefore, two unknown spin-$\frac{13}{2}$ currents
and its superpartner spin-$6$ current remain. 
On the other hand, the last spin-$6$ current 
can be obtained from the previous independent spin-$6$ currents.  
Effectively, one is left with the highest spin-$\frac{13}{2}$ current
in (\ref{higherspincurrents}).

Consider the last most complicated OPE between the spin-$3$ current 
(\ref{c4w}) and  spin-$\frac{9}{2}$ current (\ref{c4o92'})  to
determine the last unknown spin-$\frac{13}{2}$ current:
\bea
&& W(z) \; O_{\frac{9}{2}'} (w) =
\frac{1}{(z-w)^4} \, \frac{444\sqrt{6}}{65} O_{\frac{7}{2}}
+\frac{1}{(z-w)^3} \, \left[ (\frac{2}{7}) 
\frac{444\sqrt{6}}{65} \pa O_{\frac{7}{2}} -\frac{88}{13} \sqrt{\frac{2}{3}} O_{\frac{9}{2}} \right] (w)
\nonu \\
&& +\frac{1}{(z-w)^2} \, \left[ (\frac{3}{56}) 
\frac{444\sqrt{6}}{65} \pa^2 O_{\frac{7}{2}} -(\frac{1}{3}) 
\frac{88}{13} \sqrt{\frac{2}{3}} \pa O_{\frac{9}{2}} +\frac{1}{\sqrt{6}} O_{\frac{11}{2}}+
\frac{19224\sqrt{6}}{3965} 
\left( T O_{\frac{7}{2}} -\frac{3}{16} \pa^2 O_{\frac{7}{2}} \right) \right. 
\nonu \\
&& \left. 
 + \frac{8342}{3965} 
\left( G P_{4'}^{uu} -\frac{4\sqrt{6}}{9} \pa O_{\frac{9}{2}}-\frac{\sqrt{6}}{56} \pa^2 
O_{\frac{7}{2}} \right) 
-\frac{102117}{7930}  
\left( G P_{4'}^{ww} -\frac{2}{9} \sqrt{\frac{2}{3}} \pa O_{\frac{9}{2}}
-\frac{1}{14} \sqrt{\frac{2}{3}} \pa^2 O_{\frac{7}{2}} \right) \right] (w)
\nonu \\
&& +\frac{1}{(z-w)} \, \left[ (\frac{1}{126}) 
\frac{444\sqrt{6}}{65} \pa^3 O_{\frac{7}{2}} -(\frac{1}{15}) 
\frac{88}{13} \sqrt{\frac{2}{3}} \pa^2 O_{\frac{9}{2}} +(\frac{4}{11}) \frac{1}{\sqrt{6}} \pa O_{\frac{11}{2}}
\right.
\nonu \\ 
&& +
(\frac{4}{11}) \frac{19224\sqrt{6}}{3965} \pa
\left( T O_{\frac{7}{2}} -\frac{3}{16} \pa^2 O_{\frac{7}{2}} \right)  
 + (\frac{4}{11}) \frac{8342}{3965} \pa
\left( G P_{4'}^{uu} -\frac{4\sqrt{6}}{9} \pa O_{\frac{9}{2}}-\frac{\sqrt{6}}{56} \pa^2 
O_{\frac{7}{2}} \right) \nonu \\
&&  -(\frac{4}{11}) \frac{102117}{7930}  
\pa \left( G P_{4'}^{ww} -\frac{2}{9} \sqrt{\frac{2}{3}} \pa O_{\frac{9}{2}}
-(\frac{1}{14}) \sqrt{\frac{2}{3}} \pa^2 O_{\frac{7}{2}} \right)  \nonu \\
&&  - \frac{2464}{1495} \sqrt{\frac{2}{3}} 
\left( T O_{\frac{9}{2}} -\frac{3}{20} \pa^2 O_{\frac{9}{2}} \right)-\frac{2368}{715} \sqrt{\frac{2}{3}} 
\left( T \pa O_{\frac{7}{2}} -\frac{7}{4} \pa T O_{\frac{7}{2}}-\frac{1}{9} \pa^3 O_{\frac{7}{2}} 
\right) \nonu \\
&&   - \frac{6104}{16445} 
\left( G \pa P_{4'}^{uu} -\frac{8}{3} \pa G P_{4'}^{uu} -\frac{\sqrt{6}}{5} \pa^2 O_{\frac{9}{2}} -
\frac{2}{63}\sqrt{\frac{2}{3}} \pa^3 O_{\frac{7}{2}} \right) \nonu \\
&& \left. +\frac{8064}{3289} 
\left( G \pa P_{4'}^{ww} -\frac{8}{3} \pa G P_{4'}^{ww} -\frac{1}{5\sqrt{6}}
\pa^2 O_{\frac{9}{2}} -\frac{8}{189} \sqrt{\frac{2}{3}} \pa^3 O_{\frac{7}{2}} \right) +
\frac{2}{11} \sqrt{\frac{2}{3}} O_{\frac{13}{2}} \right] (w) \nonu \\
&& +\cdots.
\label{c4wo92'}
\eea
The independent fields in the right hand side up to the second-order singular terms 
are the same as that in the OPE (\ref{c4uo4''}).
First of all, the third-order pole in (\ref{c4wo92'}) 
should be checked to determine if it
contains 
a primary field of a spin-$\frac{9}{2}$ current (\ref{c4o92}) 
after subtracting the descendant 
field for $O_{\frac{7}{2}}(w)$.
For the second-order pole, 
new structures should be identified 
after subtracting the right quasi-primary fields with the
correct 
coefficients.
Calculate 
the OPE between $T(z)$ and $(\{W O_{\frac{9}{2}'} \}_{-2} - (\frac{3}{56}) 
\frac{444\sqrt{6}}{65} \pa^2 O_{\frac{7}{2}} +(\frac{1}{3}) 
\frac{88}{13} \sqrt{\frac{2}{3}} \pa O_{\frac{9}{2}})(w)$.  
This OPE 
contains $\frac{3441}{65}\sqrt{\frac{3}{2}} O_{\frac{7}{2}}(w)$ at the fourth-order pole.
The OPE between $G(z)$ and $(\{W O_{\frac{9}{2}'} \}_{-2} - (\frac{3}{56}) 
\frac{444\sqrt{6}}{65} \pa^2 O_{\frac{7}{2}} +(\frac{1}{3}) 
\frac{88}{13} \sqrt{\frac{2}{3}} \pa O_{\frac{9}{2}})(w)$ produces 
$\frac{6887}{585} P_{4'}^{uu}(w) -\frac{14126}{195} P_{4'}^{ww}(w)$ at the 
third-order pole.
The correct three quasi-primary fields with the correct coefficients 
can then be subtracted from the second-order pole, leaving
 the spin-$\frac{11}{2}$ current (\ref{c4o112}).
The complete structure 
of the first-order singular terms can now be determined.
The OPE between the $T(z)$ and $(\{W O_{\frac{9}{2}'} \}_{-1} - 
\mbox{derivative terms})(w)$ contains
$\frac{8288\sqrt{6}}{715} O_{\frac{7}{2}}(w)$ at the fifth-order pole and 
$(-\frac{1}{7})\frac{8288\sqrt{6}}{715} \pa O_{\frac{7}{2}}(w)-\frac{616\sqrt{6}}{65} 
O_{\frac{9}{2}}(w)$ at the fourth-order pole.
Similarly, 
from 
the OPE $G(z)$ with $(\{W O_{\frac{9}{2}'} \}_{-1} - 
\mbox{derivative terms})(w)$, 
$-\frac{896}{715} P_{4'}^{uu}(w) -\frac{24192}{715} P_{4'}^{ww}(w)$ 
and $-(-\frac{1}{8}) \frac{896}{715} \pa P_{4'}^{uu}(w) -(-\frac{1}{8}) 
\frac{24192}{715} \pa P_{4'}^{ww}(w)$ can be observed a the 
four-th order and  third-order pole, respectively.
From this analysis, the four 
possible  quasi-primary fields with fixed coefficients can be expressed.
The properties of  
the currents $T(z)$ or $G(z)$ and the quasi-primary fields are reported in 
Appendices $B$ and $C$.
The general structures of quasi-primary fields for given 
known quasi-primary fields, $\Phi_i(z)$ and $\Phi_j(z)$ found by 
\cite{BFKNRV} are explained.
Interestingly, once the OPE $\Phi_i(z) \; \Phi_j(w)$
is found, 
the quasi-primary fields containing the derivatives 
of these two quasi-primary fields are determined completely.
The following spin-$\frac{13}{2}$ primary current, which is 
the highest spin current in the list 
of (\ref{higherspincurrents}), can be derived 
by subtracting the above terms in the first-order singular terms,
\bea
O_{\frac{13}{2}}(z) =
 \frac{16\sqrt{2}}{4485} \, f^{abc} \psi^a \psi^b \pa^5 \psi^c(z) +
\mbox{other lower order derivative terms}. 
\label{c4o132}
\eea
The OPE $G(z)$ with 
the spin-$\frac{13}{2}$ current (\ref{c4o132}) should be calculated to 
determine its superpartner.
Finally, its correct superpartner is 
\bea
O_{6'}(z) =  P_6(z) + \frac{7}{2} \sqrt{\frac{3}{2}} P_{6'}(z). 
\label{c4o6'}
\eea
Furthermore, 
the OPE between $G(z)$ and $O_{6'}(w)$
leads to the first-order pole,
$O_{\frac{13}{2}}(w)$, as expected \footnote{Of course, the OPEs between 
$G(z)$ and $P_6(w)$ (or $P_{6'}(w)$) can be obtained from the standard results
for the OPEs between $G(z)$ and $O_6(w)$ (and $O_{6'}(w)$) with (\ref{c4o6}) and (\ref{c4o6'}).}. 
Therefore, 
the supercurrents, $\hat{O}_{\frac{11}{2}}(Z)$ and $\hat{O}_6(Z)$,
are constructed in the list of 
(\ref{higherspincurrents}).
The former consists of (\ref{c4o112}) and (\ref{c4o6}) and the latter is given by 
(\ref{c4o6'}) and (\ref{c4o132}).
The ${\cal N}=1$ superfusion rule is given by
$\left[\hat{W} \right] \left[\hat{O}_{4'} \right]  =  
\left[\hat{O}_{\frac{7}{2}} \right] + \left[\hat{O}_{4} \right] +
\left[\hat{O}_{\frac{11}{2}} \right] +\left[\hat{O}_6 \right]$.

In summary, the OPEs described thus far are given by (\ref{c4uu}), (\ref{c4wu}), (\ref{c4ww}),
(\ref{uo72}), (\ref{c4uo4}), (\ref{c4wo72}), (\ref{c4wo4}), (\ref{c4uo4''}), (\ref{c4uo92'}), 
(\ref{c4wo4''}), and (\ref{c4wo92'}). 
In principle, the other OPEs  not calculated in this paper
can be determined.
During these computations, 
the $12$ higher spin primary currents, where the spins are greater than
$3$,
can be found. Moreover, there are four quasi-primary fields 
and a primary field but these can be obtained from the known primary currents $T(z), G(z), U(z), W(z), O_{4''}(z)$ or
$O_{\frac{9}{2}'}(z)$. 
The OPEs found
in ${\cal N}=1$ superspace should be written down.
All these computations are based on the $c=4$ eight free fermion model.
In the next section, this model is generalized to the $c<4$ coset model.

\section{Higher spin currents in the 
${\cal N}=1$ supersymmetric coset minimal model (\ref{centralc1})}

Consider 
the perturbations of the $k \rightarrow \infty $ model described in the 
previous subsection.
Two spin-$1$ currents exist, 
$J^a(z)$ and $K^a(z)$ of level $3$ and $k$, which generate 
the algebra 
$(A_2^{(1)} \oplus A_2^{(1)})$. 
The OPE between the spin-$1$ currents, $J^a(z)$, is given in (\ref{JJ}) and the corresponding OPE for 
the spin-$1$ current $K^a(z)$ is 
\bea
K^a (z) \; K^b(w) = -\frac{1}{(z-w)^2} \frac{k}{2} \delta^{ab} + 
\frac{1}{(z-w)} f^{abc} K^c(w) +\cdots.
\label{KK}
\eea
The diagonal subalgebra $A_2^{(1)}$ generates the spin-$1$ current $J^a(z) +K^a(z)$ with level
$k+3$.

The coset Virasoro algebra is generated using 
the following Sugawara stress energy tensor
\bea
T(z) =-\frac{1}{6} J^a J^a(z) -\frac{1}{(k+3)} K^a K^a(z) +\frac{1}{(k+6)}
(J^a +K^a) (J^a +K^a)(z), 
\label{cosett}
\eea
which commutes with the above spin-$1$ current  $J^a(z) +K^a(z)$, as expected.
As the $k \rightarrow \infty$,  
the above (\ref{cosett}) becomes (\ref{c4t}).
The OPE of this spin-$2$ current and itself is given by (\ref{tt}), 
using the OPEs (\ref{JJ}) 
and (\ref{KK}), 
where the coset central charge is
characterized by the following function of $k$ 
\bea
c = 4 \left[ 1- \frac{18}{(k+3)(k+6)} \right],  \qquad k=1, 2, \cdots.
\label{centralc}
\eea
When $k=1$, this central charge reduces to the one in 
the minimal extension given in  subsection $2.2$.

By requiring that the spin-$\frac{3}{2}$ current should commute with the diagonal spin-$1$ current and 
should transform as a 
primary field under the stress energy tensor (\ref{cosett}), 
\bea
G(z) = -\frac{2k}{3\sqrt{3(k+3)(k+6)}} \psi^a \left( J^a -\frac{9}{k} K^a \right)(z),
\label{cosetg}
\eea
which satisfies (\ref{gg}) with the central charge (\ref{centralc}),
is derived, as reported in \cite{ASS}. 
In addition, 
this current reduces to (\ref{c4g}) as $k$ approaches $\infty$.

Similarly, the higher spin-$3$ current can be fixed using 
above regularity condition under the diagonal 
spin-$1$ current and primary condition using the stress energy tensor $T(z)$. 
In addition, the highest singular term should behave 
as $\frac{c}{3}$.
Therefore, it can be expressed  as \cite{ASS}
\bea
W(z) & = & \frac{2i}{9(k+3)(k+6)\sqrt{30(2k+3)(2k+15)}} d^{abc} \left[
  k(k+3)(2k+3)  J^a J^b J^c \right. \nonu \\
&& \left. -18 (k+3)(2k+3) J^a J^b K^c 
+ 162(k+3) J^a K^b K^c -162 K^a K^b K^c \right](z),
\label{Wk}
\eea
which reduces to the previous expression (\ref{c4w}) for $k \rightarrow 
\infty$. 
As performed previously, its fermionic superpartner can be obtained from 
the spin-$\frac{3}{2}$ current (\ref{cosetg}), which leads to 
the derivation reported in reference \cite{ASS}
\bea
U(z) & = & \frac{2\sqrt{6} i}{15\sqrt{10(k+3)(k+6)(2k+3)(2k+15)}}
d^{abc}
\left[ k (2k+3) \psi^a J^b J^c \right. \nonu \\
&-& \left. 15 (2k+3) \psi^a J^b K^c + 90 \psi^a K^b K^c  \right](z),
\label{Uk}
\eea
which also becomes (\ref{c4u}) for $k \rightarrow \infty$.

The $12$ higher spin currents 
in (\ref{higherspincurrents})
can be constructed for the $c < 4$ coset model.

$\bullet$ The construction of $\hat{O}_{\frac{7}{2}}(Z)$ and $\hat{O}_4(Z)$

The OPE between the spin-$\frac{5}{2}$ current (\ref{Uk}) and itself
can be calculated.
The only difference between the $c=4$ model and  $c<4$ model is 
the $k$-dependence in front of (\ref{Uk}) and there is 
an extra current $K(z)$ dependence. 
Therefore, the calculations are more involved. 
Nevertheless,
\bea
U(z) \; U(w) 
 & = & \frac{1}{(z-w)^5} \, \frac{8k(9+k)}{5(3+k)(6+k)}  
+ \frac{1}{(z-w)^3} \, 2T(w) +
\frac{1}{(z-w)^2} \, \pa T(w) \nonu \\
&+&\frac{1}{(z-w)} \, 
\left[ \frac{3}{10} \pa^2 T +\frac{9(3+k)(6+k)}{2(66+63k+7k^2)} \left( T^2 -\frac{3}{10} \pa^2 T 
\right) +  P_{4}^{uu} + P_{4'}^{uu} \right](w) \nonu \\
& + & \cdots.
\label{uulast}
\eea
The algebraic structure is the same as (\ref{c4uu}) except that 
the $k$-dependence occurs in many places.
The stress energy tensor is given by (\ref{cosett}).
The $k$-dependent   primary spin-$4$,  which is a generalization of (\ref{p4-uu}), 
is given  by
\footnote{The generalization of footnote \ref{gt} is obtained:
$G(z) \; \left( -\frac{7}{10} \pa^2 T + 
\frac{17(3+k)(6+k)}{6(66+63k+7k^2)} \left( T^2 -\frac{3}{10} \pa^2 T 
\right) + G \pa G \right)(w) = -\frac{1}{(z-w)^4} \left[
\frac{(-42+99 k+11 k^2) (378+333 k+37 k^2)}{8 (3+k) (6+k)(66+63 k+7 k^2)}
\right] G(w) -\frac{1}{(z-w)^3} \, (-\frac{1}{3}) \left[
\frac{(-42+99 k+11 k^2) (378+333 k+37 k^2)}{8 (3+k) (6+k)(66+63 k+7 k^2)} 
\right]
\pa G(w) +
{\cal O}((z-w)^{-2})$. The fourth-order and third-order 
singular terms can be derived. Of course, there is an overall 
factor in (\ref{puuexp}) that was not considered in this computation. }
\bea
 P_{4}^{uu}(z) & = &  
\frac{3(3+k)(6+k)(498+225k+25k^2)}{(-42+99k+11k^2)(378+333k+37k^2)} 
\nonu \\
& \times & \left[ -\frac{7}{10} \pa^2 T + 
\frac{17(3+k)(6+k)}{6(66+63k+7k^2)} \left( T^2 -\frac{3}{10} \pa^2 T 
\right) + G \pa G \right](z).
\label{puuexp}
\eea
In addition, the $k$-dependent generalization of (\ref{c4'uu})
can be obtained
as
\bea
P_{4'}^{uu}(z) & = & - 
\frac{2 (-1+k) k (9+k) 
(18+17 k+9 k^2)}{(6+k) (-42+99 k+11 k^2)(378+333 k+37 k^2)}
\psi^a 
\pa^3 \psi^a(z) \nonu \\
& + & \mbox{other} \,\, 1970 \,\, \mbox{terms}.
\label{uuexp}
\eea
Only the $\psi^a(z)$ dependent terms have a 
$(k-1)$ factor in (\ref{uuexp}).
The presence of this current was reported 
in reference \cite{ASS}, where there is no explicit form for this current. 
Note that the $k$-dependence in front of the quasi-primary field occurs,
whereas 
there is no $k$-dependence in front of the stress energy tensor and its descendant fields in (\ref{uulast}). 
Of course, the central term in the highest singular term of (\ref{uulast}) 
is the usual expression, $\frac{2}{5} c$, where $c$ is given by 
(\ref{centralc}). The results appear to be 
different from the original equation $(4.16)$ 
in reference 
\cite{ASS} but they are the 
same by manipulating the singular terms appropriately.
The quasi-primary field can be obtained 
from (\ref{uulast}) rather than from old one.
Moreover, (\ref{uulast}) reduces to  
(\ref{c4uu}) as $k \rightarrow \infty$.
The OPE between $T(z)$ and the quasi-primary field $(T^2 -\frac{3}{10} \pa^2 T)(w)$
has a nontrivial $k$-dependence and this OPE is presented in Appendix $D$.
Also see Appendix $E$.

Now move the following OPE between the spin-$3$ current (\ref{Wk}) and 
 spin-$\frac{5}{2}$ current (\ref{Uk}), 
\bea
W(z) \; U(w) 
 & = & \frac{1}{(z-w)^4} \, \frac{3}{\sqrt{6}} G(w)  +
\frac{1}{(z-w)^3} \, (\frac{2}{3}) \frac{3}{\sqrt{6}} \pa G(w) \nonu \\
& + & \frac{1}{(z-w)^2}
\left[ (\frac{1}{4}) \frac{3}{\sqrt{6}} \pa^2 G +
\frac{11\sqrt{6}(3+k)(6+k)}{(378+333k+37k^2)} 
\left( G T -\frac{1}{8} \pa^2 G \right) +O_{\frac{7}{2}}  \right](w)
\nonu \\
& + & \frac{1}{(z-w)} \left[ 
 (\frac{1}{15}) \frac{3}{\sqrt{6}} \pa^3 G + (\frac{4}{7})
\frac{11\sqrt{6}(3+k)(6+k)}{(378+333k+37k^2)} 
\pa \left( G T -\frac{1}{8} \pa^2 G \right) + 
(\frac{4}{7}) \pa O_{\frac{7}{2}} \right. \nonu \\
& + &  \left.  
\frac{4\sqrt{6}(3+k)(6+k)}{7(-42+99k+11k^2)} 
\left( \frac{4}{3} T \pa G - G \pa T -\frac{4}{15} \pa^3 G 
\right)  + O_{\frac{9}{2}} 
\right](w) +\cdots.
\label{wuexp}
\eea
The $k$-dependence occurs in front of the quasi-primary field. 
(\ref{wuexp}) leads to (\ref{c4wu}) as $k \rightarrow \infty$.
The $k$ dependent spin-$\frac{7}{2}$ (the corresponding $k\rightarrow \infty$ limit was given in (\ref{c4o72})) 
has the following form
\bea
O_{\frac{7}{2}}(z) & = & 
-\frac{\sqrt{2} (-1+k) k (9+k) (9+2 k)}{(15+2 k) 
\sqrt{18+9 k+k^2} (378+333 k+37 k^2)}
\, f^{abc} \, \psi^a \psi^b \pa^2 \psi^c(z) \nonu \\
& + & 
\mbox{other} \,\, 746  \,\, \mbox{terms},
\label{co72exp}
\eea
where only $\psi^a(z)$-dependent terms have the 
$(k-1)$ factor in (\ref{co72exp}).
$K^a$-dependent terms and mixed terms exist between $\psi^a(z)$ and $K^a(z)$.
Furthermore, the spin-$\frac{9}{2}$ current, which generalizes the previous expression (\ref{c4o92}), has 
\bea
O_{\frac{9}{2}}(z) & = &  
-\frac{8 \sqrt{2} (-1+k) k (1+k) (9+k)}{21 (15+2 k) \sqrt{18+9 k+k^2} 
(-42+99 k+11 k^2)}
\, f^{abc} \, 
\psi^a \psi^b \pa^3 \psi^c(z) 
 \nonu \\
& + & \mbox{other} \,\,  3624 \,\, \mbox{terms}.
\label{co92exp}
\eea
Only $\psi^a(z)$ dependent terms have a $(k-1)$ factor in (\ref{co92exp}).
The OPEs between the $T(z)$ and quasi-primary fields appearing in (\ref{wuexp})
contain the $k$-dependence, and these OPEs are given in Appendix $D$. 
Similarly, the OPEs between the $G(z)$ and those quasi-primary fields 
are given in Appendix $E$ where the $k$-dependence can be found explicitly.

The generalization of (\ref{c4ww}) can be obtained and the spin-$3$ current OPE is given by
\bea
W(z) \; W(w) =
 & = & \frac{1}{(z-w)^6} \, \frac{4k(9+k)}{3(3+k)(6+k)} +\frac{1}{(z-w)^4} \, 2 T(w)
+\frac{1}{(z-w)^3} \, \pa T(w) \nonu \\
& + & \frac{1}{(z-w)^2} \left[ (\frac{3}{10}) \pa^2 T +\frac{16(3+k)(6+k)}{3(66+63k+7k^2)} 
\left( T^2 -\frac{3}{10} \pa^2 T 
\right) +P_{4}^{ww} +P_{4'}^{ww} \right](w)
\nonu \\
& + & \frac{1}{(z-w)} \left[ (\frac{1}{15}) 
\pa^3 T +(\frac{1}{2})   \frac{16(3+k)(6+k)}{3(66+63k+7k^2)} 
\pa \left( T^2 -\frac{3}{10} \pa^2 T 
\right)\right. \nonu \\
& + & \left. (\frac{1}{2}) 
\pa P_{4}^{ww} + (\frac{1}{2}) \pa P_{4'}^{ww} \right](w)+  \cdots,
\label{wwexp}
\eea
where the $k$-dependent generalization of (\ref{p4-ww}),  
spin-$4$ primary field, has the following form
\bea
 P_4^{ww}(z) & = &  -\frac{48(-1+k)(3+k)(6+k)(10+k)}
{(-42+99k+11k^2)(378+333k+37k^2)} 
\nonu \\
& \times & \left[ -\frac{7}{10} \pa^2 T + \frac{17(3+k)(6+k)}{6(66+63k+7k^2)} 
\left( T^2 -\frac{3}{10} \pa^2 T 
\right) + G \pa G \right](z).
\label{expexp}
\eea
Owing to the $(k-1)$ factor in (\ref{expexp}), 
this current vanishes at the ``minimal'' extension described before. 
Another spin-$4$ primary current is obtained, 
which generalizes equation (\ref{c4'ww})
\bea
P_{4'}^{ww}(z) & = &
\frac{4 (-1+k)^2 k (9+k) (27+k)}{3 (15+2 k) (-42+99 k+11 k^2) (378+333 k+37 k^2)}
\psi^a \pa^3 \psi^a(z) \nonu \\ 
& + & \mbox{other} \,\, 1818  \,\, \mbox{terms},
\label{wwexp1}
\eea
where only $\psi^a(z)$ dependent terms have a $(k-1)$ factor in (\ref{wwexp1}).
The central term $\frac{c}{3}$ in (\ref{wwexp}) can be derived easily.

Therefore, as for the infinite $k$ case, 
the two primary currents (\ref{co72exp}) and (\ref{co92exp}) can be derived.
The new primary currents (\ref{c4o4}) and (\ref{c4o4'})
are constructed from the other two primary fields, (\ref{uuexp}) 
and (\ref{wwexp1}).
In other words, four independent currents in (\ref{higherspincurrents}) are  
found while calculating 
the OPEs (\ref{uulast}), (\ref{wuexp}) and (\ref{wwexp}).
The OPEs between the spin-$\frac{3}{2}$ current
and  the above four independent currents are the same as those in 
(\ref{c4go72}), (\ref{c4go4}), (\ref{OPEgo92}) and (\ref{OPEgo4}).

$\bullet$ The construction of $\hat{O}_{4'}(Z)$ and $\hat{O}_\frac{9}{2}(Z)$

Consider the OPE between the spin-$\frac{5}{2}$ current (\ref{Uk}) 
and spin-$\frac{7}{2}$ current (\ref{co72exp})
\bea
&& U(z) \; O_{\frac{7}{2}} (w)  =   \frac{1}{(z-w)^3}\, c_{uow} \, W(w) \nonu \\
&&  + 
\frac{1}{(z-w)^2} \left[ \frac{1}{3} \, c_{uow} \, \pa W - c_{uogu} \, \left( 
G U -\frac{\sqrt{6}}{3} \pa W \right) + O_{4''} \right](w) 
\nonu \\
&& +  \frac{1}{(z-w)} \left[ \frac{1}{14} \, c_{uow} \, \pa^2 W -\frac{3}{8} 
\, c_{uogu} \, \pa \left( 
G U -\frac{\sqrt{6}}{3} \pa W \right) + \frac{3}{8} \pa O_{4''} \right. \nonu \\
&& +  \left. c_{uotw} \, \left( T W -\frac{3}{14} \pa^2 W  \right) +
c_{uogu'} \, \left( G \pa U -\frac{5}{3} \pa G U -\frac{\sqrt{6}}{7} \pa^2 W 
\right) + O_5 \right](w) \nonu \\
&& +\cdots.
\label{5272}
\eea
Here the structure constants can be written as
\bea
c_{uow} & = & \frac{36 (-1+k) (10+k) (9+2 k)^2}{5 (3+2 k) (15+2 k)(378+333 k+37 k^2)},\nonu \\
c_{uogu} & = &   \frac{6 \sqrt{6} (-1+k) (3+k) (6+k) (10+k) (9+2 k)^2}{(3+2 k) (15+2 k) 
(90+117 k+13 k^2)(378+333 k+37 k^2)},\nonu \\
c_{uotw} & = & \frac{4 (3+k) (6+k) (9+2 k)^2 
(198+1719 k+191 k^2)}{45 
(3+2 k) (15+2 k)(74+45 k+5 k^2)(378+333 k+37 k^2)},
\nonu \\
c_{uogu'} & = & \frac{(3+k) (6+k) (9+2 k)^2 (5562+2583 
k+287 k^2)}{(30 \sqrt{6} 
(3+2 k) (15+2 k) (74+45 k+5 k^2)(378+333 k+37 k^2)}.
\label{constants}
\eea
Note that the OPE between the current $G(z)$ and spin-$4$ 
primary field in (\ref{5272}) appearing in  Appendix $E$ is used to 
determine the complete coefficient functions in the right hand 
side of (\ref{5272}). 
Moreover, the OPEs between the spin-$2$ current $T(z)$ or $G(z)$
and the quasi-primary fields of spin $5$ can be obtained from 
Appendices $D$ or $E$.
$(k-1)$ factors exist in the first two structure 
constants in (\ref{constants}).
Therefore,
how  can the coefficient $c_{uogu}$ be obtained explicitly?
The algebraic structure in (\ref{uo72}) could explain an infinite $k$. 
Because the coefficient, $c_{uow}$, is fixed from the third-order pole,
the first term in the second-order pole is determined.
By introducing the undetermined 
coefficient $c_{uogu}$ in front of the primary field
of spin-$4$, the OPE $G(z)$ with 
$ \left( \{ U O_{\frac{7}{2}}\}_{-2} -\frac{1}{3} \, c_{uow} \, \pa W + c_{uogu} \, ( 
G U -\frac{\sqrt{6}}{3} \pa W )\right)(w)$ can be calculated,
where $c_{uow}$ is given in (\ref{constants}).
The requirement that the third-order pole should vanish 
(i.e. primary condition) then determines the constant 
$c_{uogu}$ explicitly, leaving    
 the spin-$4$ primary field, which is given by 
\bea
O_{4''}(z) & = &  
\frac{384 i \sqrt{\frac{2}{5}} (-1+k) k (1+k) (9+2 k) \sqrt{45+36 k+4 k^2}}{
(3+k) (3+2 k) (15+2 k) (90+117 k+13 k^2)}
\psi^1 \psi^2 \psi^3 \psi^4 \psi^5 \psi^6 \psi^7 \psi^8(z) \nonu \\
& + &
\mbox{other}  \,\,  1376   \,\, \mbox{terms},
\label{aboveexp}
\eea
where only $\psi^a(z)$ dependent terms have a
$(k-1)$ factor in (\ref{aboveexp}).

What happens in the next-order pole?
Because the algebraic structure is known 
completely except for the $k$-dependent coefficient functions,
the OPEs between $T(z)$($G(z)$) using 
$ \left( \{ U O_{\frac{7}{2}}\}_{-1} - \mbox{three descendant terms}-\mbox{two quasi-primary terms} 
\right)(w)$ can be calculated using the 
undetermined coefficients $c_{uotw}$ and $c_{uogu'}$.
This should transform very specially.
In other words,  there should be
no  higher order terms where the order is greater than $2$ (once again 
the primary condition).
This enables the above two constants to be fixed, as in (\ref{constants}).  
Therefore, the following spin-$5$ primary current 
remains
\bea
O_{5}(z) & = & -
\frac{2 i \sqrt{\frac{2}{15}} (-1+k) k 
(9+k) (9+2 k)^2 (9+4 k)}{
45 (3+k) (6+k) (15+2 k) \sqrt{45+36 k+4 k^2} (74+45 k+5 k^2)}
f^{abc} d^{ade} \, \psi^b \psi^c \psi^d \pa^3 \psi^e (z) \nonu \\
& + & 
\mbox{other} \,\, 9603  \,\, \mbox{terms},
\label{5exp}
\eea
where
only $\psi^a(z)$ dependent terms have a $(k-1)$ factor in (\ref{5exp}).
The corresponding $k \rightarrow \infty$ limit expressions are 
(\ref{c4o4''}) and (\ref{c4o5}), respectively.

The OPE between the spin-$\frac{5}{2}$ current and 
spin-$4$ current can be expressed as follows:
\bea
&& U(z) \; O_4 (w) = \frac{1}{(z-w)^4} \, c_{uou} \, U(w) +
\frac{1}{(z-w)^3} \, \frac{1}{5} \, c_{uou} \, \pa U(w)  
\nonu \\
&& +\frac{1}{(z-w)^2} \, \left[ \frac{1}{30} \, c_{uou} \, \pa^2 U +
c_{uogw} \, \left( G W -\frac{1}{6\sqrt{6}} \pa^2 U \right) +
c_{uotu} \, \left( T U -\frac{1}{4} \pa^2 U
\right) + P_{\frac{9}{2}}\right](w) 
\nonu \\
&& + \frac{1}{(z-w)} \left[ \frac{1}{210} \, c_{uou} \, \pa^3 U +
\frac{1}{3}  \, c_{uogw} \, \pa \left( G W -\frac{1}{6\sqrt{6}} \pa^2 U \right) +
\frac{1}{3} \, c_{uotu} \, \pa \left( T U -\frac{1}{4} \pa^2 U
\right) \right. \nonu \\
&& \left. + \frac{1}{3} \pa P_{\frac{9}{2}} + Q_{\frac{11}{2}} \right],
\label{uo4exp}
\eea
where the correct spin-$4$ current is the sum of the 
previous spin-$4$ currents
\footnote{\label{spin4foot} 
Another spin-$4$ current exists $
O_{4'}(z) = \frac{1}{8} \left( \frac{16}{7} \sqrt{\frac{2}{3}} P_{4'}^{uu} 
-\frac{4}{7} \sqrt{6} P_{4'}^{ww}\right)(z)$, which is equal to the 
relationship of
(\ref{c4o4'})
with (\ref{uuexp}) and (\ref{wwexp1}).}
with (\ref{uuexp}) and (\ref{wwexp1}) 
\bea
O_4(z) = \left( -\frac{1}{\sqrt{6}} P_{4'}^{uu} +\sqrt{6} P_{4'}^{ww} \right)(z).
\label{spin4spin4}
\eea
This is identical to (\ref{c4o4}).
The structure constants in (\ref{uo4exp}) are given by 
\bea
c_{uou} & = & \frac{6 \sqrt{6} (-1+k) (10+k) (9+2 k)^2}{(3+2 k) (15+2 k)(378+333 k+37 k^2)},
\label{exp1} \\
c_{uogw} & = & \frac{12 (-1+k) (3+k) (6+k) (10+k) (9+2 k)^2 
(1290+1197 k+133 k^2)}{5 (3+2 k) (15+2 k) 
(74+45 k+5 k^2) (90+117 k+13 k^2) (378+333 k+37 k^2)},\nonu \\
c_{uotu} & = & \frac{8 \sqrt{\frac{2}{3}} (-1+k) (3+k) (6+k) (10+k) (9+2 k)^2 
(90+261 k+29 k^2)}{(3+2 k) (15+2 k) 
(74+45 k+5 k^2) (90+117 k+13 k^2) (378+333 k+37 k^2)},
\nonu
\eea
where all of these have $(k-1)$ factors in their expressions.
Of course, these constants (\ref{exp1}) reduce 
to the ones appearing in (\ref{c4uo4}) for 
an infinite $k$ limit.

Once again, the coefficients, 
$c_{uogw}$ and $c_{uotu}$, appearing in the second-order pole
are determined by evaluating the OPEs between $T(z)$(and similarly $G(z)$), 
and the 
whole second-order pole terms subtract the first three terms 
where the coefficient, $c_{uou}$ is known from the higher order terms.
The disappearance 
of the higher order singular terms, where the order is greater than 
$2$, fixes the above unknown two coefficients, which  
are given in (\ref{exp1}), leaving
the primary field in this singular terms.
The spin-$\frac{9}{2}$ primary current, which generalizes the previous expression (\ref{c4p92}),
can be obtained by the following:
\bea
P_{\frac{9}{2}}(z) & = &
\frac{n_1}{d_1} d^{abc} f^{bde} f^{cfg} \psi^a \psi^d \psi^e \psi^f \pa^2 \psi^g(z) + 
\mbox{other} \,\, 4671  \,\,  \mbox{terms}, 
\label{9halfexp}
\eea
where the intermediate $k$-dependent expressions are
\bea
n_1  & = & 
2 i \sqrt{\frac{2}{5}} (-1+k) k (9+2 k) (79110+149883 k+86489 k^2+19367 k^3+1741 
k^4+50 k^5),
\label{expp} \\
d_1 &= & 5 (3+k) (15+2 k) (74+45 k+5 k^2) (90+117 k+13 k^2) 
\sqrt{810+1053 k+441 k^2+72 k^3+4 k^4}.
\nonu
\eea
In this case, only $\psi^a(z)$ dependent terms 
have a $(k-1)$ factor in (\ref{9halfexp}). Of course, expression 
(\ref{expp}) reduces to
the numerical coefficient in (\ref{c4p92}).  

What about the first-order singular terms?
Because the second-order terms are 
determined, their descendant fields can be found 
with the known coefficient functions.
By introducing the arbitrary three coefficient functions in (\ref{q112}),
the equation can be solved in such a way that the whole first-order terms 
subtracted from above four known descendant field terms is equal to
the quasi-primary field $Q_{\frac{11}{2}}(w)$. 
This provides all the information for the three unknown coefficient 
functions that were introduced.
Therefore,
the general expression containing (\ref{q112}) can be obtained 
\bea
Q_{\frac{11}{2}}(z) & = & -\frac{8 \sqrt{\frac{2}{3}} 
(3+k) (6+k) (9+2 k)^2 (18+657 k+73 k^2)}{15 (3+2 k) 
(15+2 k) (90+117 k+13 k^2) (378+333 k+37 k^2)}
\nonu \\
&\times & \left( T \pa U -\frac{5}{4} \pa T U -\frac{1}{7} \pa^3 U \right)(z) \nonu \\
& + & 
\frac{2 (3+k) (6+k) (9+2 k)^2 (498+225 k+25 k^2)}{5 (3+2 k) 
(15+2 k) (90+117 k+13 k^2) (378+333 k+37 k^2)}
\nonu \\
& \times & \left( G \pa W -2 \pa G W -\frac{1}{21}\, \sqrt{\frac{2}{3}} 
\, \pa^3 U \right)(z) \nonu \\
& + & \frac{3 (3+k) (6+k)}{(3+2 k) (15+2 k)} 
\left( G O_{4''} -\frac{2}{9} \pa O_{\frac{9}{2}'} \right)(z).
\label{q112exp}
\eea
Compared to (\ref{q112}), $k$-dependent coefficient functions 
are in front of the 
three independent quasi-primary fields in (\ref{q112exp}). The 
OPEs can be calculated in a similar way as that used  
in (\ref{c4tq112}) and (\ref{c4gq112}), and they will show an explicit 
$k$-dependence in the right hand side of the OPEs. 
Compared to the previous OPE (\ref{5272}), there was no need to calculate 
the OPEs
between the $T(z)$( or $G(z)$) and some terms in the first-order pole.
This is because 
there are no other quasi-primary fields in the first-order terms.  
Here, the generalization of (\ref{c4o92'}) can be given by
\bea
O_{\frac{9}{2}'}(z) & = & 
-\frac{8 i \sqrt{\frac{2}{5}} (-1+k) k (1+k) (9+k) (9+2 k) 
\sqrt{45+36 k+4 k^2}}{
5 (3+k) (3+2 k) (15+2 k) \sqrt{18+9 k+k^2} (90+117 k+13 k^2)}
\nonu \\
&\times & d^{abc} f^{bde} f^{cfg} \psi^a \psi^d \psi^e \psi^f \pa^2 \psi^g(z)  + 
\mbox{other} \,\, 4430 \,\,  \mbox{terms}. 
\label{9halftwoexp}
\eea
Only the 
$\psi^a(z)$ dependent terms have $(k-1)$ factors in (\ref{9halftwoexp}).

Consider the OPE (\ref{c4wo72}) when $k$ is finite
\bea
&& W(z) \; O_{\frac{7}{2}} (w) = \frac{1}{(z-w)^4} \,c_{wou} \, U(w) +
\frac{1}{(z-w)^3} \, \frac{2}{5}\,  c_{wou} \, \pa U(w)
\nonu \\
&& + \frac{1}{(z-w)^2} \left[  \frac{1}{10} \, c_{wou} \, \pa^2 U  + c_{wotu} \, \left(
T U -\frac{1}{4} \pa^2 U \right) +
c_{wogw} \,  \left( G W -\frac{1}{6\sqrt{6}} \pa^2 U \right)+ c_{wop} \, P_{\frac{9}{2}}
\right. \nonu \\
&&  \left. + c_{woo'} \, O_{\frac{9}{2}'} \right](w)
+\frac{1}{(z-w)} \left[ 
 \frac{2}{105} \, c_{wou} \, \pa^3 U  +\frac{4}{9} c_{wotu} \pa \left(
T U -\frac{1}{4} \pa^2 U \right) \right. \nonu \\
&& \left. +
\frac{4}{9} \, 
c_{wogw} \, \pa \left( G W -\frac{1}{6\sqrt{6}} \pa^2 U \right)
+ \frac{4}{9} \, c_{wop} \, \pa P_{\frac{9}{2}} 
+  \frac{4}{9} \, c_{woo'} \, \pa O_{\frac{9}{2}'} +Q_{\frac{11}{2}'} \right](w) +\cdots.
\label{OPEOPE}
\eea
The three structure constants in (\ref{OPEOPE}) that depend on $k$ explicitly
ccan be written in terms of
\bea
c_{wou} & = & \frac{6 (-1+k) (10+k) (9+2 k)^2}{(3+2 k) (15+2 k)(378+333 k+37 k^2)} , \qquad
c_{wop} =\frac{1}{\sqrt{6}}, \qquad
c_{woo'} =\frac{1}{\sqrt{6}}, \label{Coeff} \\
c_{wotu} & = &  \frac{4 (-1+k) (3+k) (6+k) (10+k) (9+2 k)^2 (846+927 k+103 k^2)}
{(3 (3+2 k) (15+2 k)(74+45 k+5 k^2)(90+117 k+13 k^2) 
(378+333 k+37 k^2)},\nonu \\
c_{wogw} & = & \frac{4 \sqrt{6} (-1+k) (3+k) (6+k) (10+k) (9+2 k)^2 
(90+261 k+29 k^2)}{5 (3+2 k) (15+2 k) (74+45 k+5 k^2) (90+117 k+13 k^2) (378+333 k+37 k^2)}.
\nonu
\eea
In particular, they contain the $(k-1)$ factor.
These terms in (\ref{Coeff}) become those in (\ref{c4wo72}).
Because the algebraic structure is known for an infinite $k$,  
four undetermined coefficients, $c_{wotu}, c_{wogw}, c_{wop}$ and 
$c_{woo'}$ are taken 
in front of two quasi-primary fields and 
two primary fields (given by (\ref{9halfexp}) 
and (\ref{9halftwoexp})), 
respectively. Note that the constant $c_{wou}$ can be fixed from the higher order terms.
The second order terms can then be expressed in a similar manner to that 
in (\ref{OPEOPE}).
On the other hand, the explicit second-order pole from WZW currents
is known.
By equating these two, the unknown four coefficient functions
can be obtained  as given in (\ref{Coeff}).  

Furthermore, the spin-$\frac{11}{2}$ quasi-primary field can be obtained. 
As done in the OPE (\ref{uo4exp}), because there is no other 
quasi-primary field except this spin-$\frac{11}{2}$ current, 
the explicit form  for this field can be derived as follows:
\bea
Q_{\frac{11}{2}'}(z)  & = &  -\frac{2 
\sqrt{\frac{2}{3}} (3+k) (6+k) (9+2 k)^2 (18+657 k+73 k^2)}{
15 (3+2 k) (15+2 k) (90+117 k+13 k^2) (378+333 k+37 k^2) }
\nonu \\
& \times & \left( G \pa W -2 \pa G W -\frac{1}{21} \sqrt{\frac{2}{3}} \pa^3 U 
\right)(z) \nonu \\
& + & \frac{16 (3+k) (6+k) (9+2 k)^2 (738+9 k+k^2)}{
135 (3+2 k) (15+2 k) (90+117 k+13 k^2) (378+333 k+37 k^2)}
 \nonu \\
& \times & \left( T \pa U -\frac{5}{4} \pa T U -\frac{1}{7} 
\pa^3 U \right)(z)
\nonu \\
& + &  
\frac{\sqrt{\frac{3}{2}} (3+k) (6+k)}{(3+2 k) (15+2 k)}
\left( G O_{4''} -\frac{2}{9} \pa O_{\frac{9}{2}'}\right)(z).
\label{that1} 
\eea
As stated before, 
the OPEs can also be calculated 
as in (\ref{c4tq112'}) and (\ref{c4gq112'}).
Even for 
the $k$-dependent coefficients, this field (\ref{that1}) is a
quasi-primary field because 
the three terms are quasi-primary fields.

For the OPE between the spin-$3$ current and the spin-$4$ current (\ref{spin4spin4}), 
the following 
$k$-dependent expression can be obtained, which appeared in (\ref{c4wo4}) 
\bea
&& W(z) \; O_4 (w) = \frac{1}{(z-w)^4} \, c_{wow} \, W(w) 
\nonu \\
&& +\frac{1}{(z-w)^3} \, \left[ \frac{1}{3} \, c_{wow} \, \pa W +  c_{woo} \, O_{4''} 
+ c_{wogu} \, \left( G U -\frac{\sqrt{6}}{3} \pa W \right) \right](w) 
\nonu \\
&& 
+\frac{1}{(z-w)^2} \, 
\left[ \frac{1}{14} c_{wow} \, \pa^2 W + \frac{3}{8} \, c_{woo} \,  \pa O_{4''} 
+ \frac{3}{8} 
\, c_{wogu} \,  \pa \left( G U -\frac{\sqrt{6}}{3} \pa W \right) 
+c_{woo'} \, O_5 + Q_{5} \right](w) 
\nonu \\
 && 
+\frac{1}{(z-w)} \, 
\left[ \frac{1}{84} c_{wow} \, \pa^3 W + \frac{1}{12} \, c_{woo} \, \pa^2 O_{4''} 
+ \frac{1}{12} 
\, c_{wogu} \, \pa^2 \left( G U -\frac{\sqrt{6}}{3} \pa W \right) \right. \nonu \\
&& \left. 
+
\frac{2}{5} \, c_{woo'} \,
\pa O_5 + \frac{2}{5} \, \pa Q_{5} + Q_6 \right](w) +\cdots.
\label{wo4exp}
\eea
The structure constants can be expressed in terms of
\bea
c_{wow} & = & \frac{48 \sqrt{6} (-1+k) (10+k) (9+2 k)^2}{5 (3+2 k) (15+2 k) (378+333 k+37 k^2)},
\qquad c_{woo}  =  \sqrt{\frac{2}{3}} , 
\qquad 
c_{woo'}  =   5 \sqrt{\frac{2}{3}},
\nonu \\
c_{wogu} & = &  \frac{12 (-1+k) (3+k) (6+k) (10+k) (9+2 k)^2}{(3+2 k) (15+2 k) 
(90+117 k+13 k^2) (378+333 k+37 k^2)}.
\label{coeffcoeff}
\eea

As stated before, the coefficient, 
$c_{wow}$, can be fixed easily from the fourth-order pole.
Because the algebraic structure is determined from the 
infinite $k$ result,  two unknown 
coefficients, $c_{woo}$ and $c_{wogu}$,  are placed in the two primary fields,
respectively.
One of the primary fields 
was given in (\ref{aboveexp}).
The two coefficients can then be fixed without difficulty.
Now the focus shifts to the next order singular terms.
The easiest way to obtain the quasi-primary fields in 
(\ref{wo4exp}) can be seen in the following example. 
Once the second-order  pole 
in (\ref{wo4exp}) is found, 
the arbitrary  coefficient function $c_{woo'}$ and two additional 
coefficients can be placed in the quasi-primary field in (\ref{c45quasi}).    
Then all the coefficients can be fixed, as in 
(\ref{coeffcoeff}) and the spin-$5$ quasi-primary 
field with $k$-dependent coefficients
can be expressed as
\bea
Q_5(z) & = &
\frac{2 \sqrt{\frac{2}{3}} (9+2 k)^2 
(-168156+101412 k+104013 k^2+20610 k^3+1145 k^4)}{45 (3+2 k) (15+2 k) 
(74+45 k+5 k^2) (378+333 k+37 k^2)} \nonu \\
& \times &
\left( T W -\frac{3}{14} \pa^2 W\right)(z)
\nonu \\
& - & \frac{(3+k) (6+k) (9+2 k)^2 
(-5166+855 k+95 k^2)}{90 (3+2 k) (15+2 k) 
(74+45 k+5 k^2) (378+333 k+37 k^2)} \nonu \\
& \times &
\left( G \pa U -\frac{5}{3} \pa G U -\frac{\sqrt{6}}{7} \pa^2 W \right)(z).
\label{quasiquasi5}
\eea
Similarly, the spin-$6$ quasi-primary field 
can be analyzed in a similar way to that done in (\ref{quasiquasi5})
using the four unknown coefficients in (\ref{c46quasi}).
The OPEs can be constructed in a similar way to that 
in (\ref{c45quasiOPEs}). These can be obtained 
by equating the first-order pole to the above expressions with 
four unknown coefficients.
One has the following spin-$6$ quasi-primary field   
\bea
Q_6(z) & = & 
\frac{32 (-1+k) (3+k) (6+k) (10+k) 
(9+2 k)^2}{25 (3+2 k) (15+2 k) (90+117 k+13 k^2) (378+333 k+37 k^2)}
\nonu \\
&\times & \left( G \pa^2 U - 4 \pa G \pa U +\frac{5}{2} \pa^2 G U -
\frac{1}{2\sqrt{6}} \pa^3 W  \right)(z) \nonu \\
& - &  
\frac{192 \sqrt{6} (-1+k) (3+k) (6+k) (10+k) (9+2 k)^2}{
25 (3+2 k) (15+2 k) (90+117 k+13 k^2) (378+333 k+37 k^2)}
\nonu \\
& \times &
\left( T \pa W  -\frac{3}{2}  \pa T W -\frac{1}{8} \pa^3 W \right)(z) 
\nonu \\
& + & 
\frac{\sqrt{6} (3+k) (6+k)}{(3+2 k) (15+2 k)}
\left( T O_{4''} -\frac{1}{6} \pa^2 O_{4''} \right)(z)  
\nonu \\
& - & 
\frac{\sqrt{\frac{3}{2}} (3+k) (6+k)}{(3+2 k) (15+2 k)}
\left( G O_{\frac{9}{2}'} -\frac{1}{9} \pa^2 O_{4''}\right)(z).
\label{quasispin6}
\eea
The OPEs between the spin-$2$ current (or the spin-$\frac{3}{2}$ current) and the current (\ref{quasispin6})
can be calculated in a similar way to that in (\ref{c46quasiOPEs}).

As in the infinite $k$ case (\ref{c4o92''}), the following relation
with (\ref{9halfexp}) and (\ref{9halftwoexp})
exists
\bea
O_{\frac{9}{2}''}(z) = O_{\frac{9}{2}'}(z) +\frac{8}{5} P_{\frac{9}{2}}(z).
\label{todayexp}
\eea

Therefore,  two primary currents (\ref{aboveexp}) and (\ref{5exp}) 
can be found  for the infinite $k$ case. 
From the other two primary fields (\ref{9halfexp}) and (\ref{9halftwoexp}),
the primary current (\ref{todayexp}) can be constructed.
In other words, the four independent currents in (\ref{higherspincurrents}) are  
found when calculating  
the OPEs (\ref{5272}), (\ref{uo4exp}), (\ref{OPEOPE}) and (\ref{wo4exp}).
Four 
quasi-primary fields, which 
can be written in terms of known higher spin currents as well as the stress energy 
tensor and its superpartner can be found. 

$\bullet$ The construction of $\hat{O}_{\frac{11}{2}}(Z)$ and $\hat{O}_6(Z)$

The OPE between the spin-$\frac{5}{2}$ current 
(\ref{Uk}) 
and  spin-$4$ current (\ref{aboveexp}), 
corresponding to the infinite $k$ result (\ref{c4uo4''}),
can be calculated
\bea
&& U(z) \; O_{4''} (w) = \frac{1}{(z-w)^3} \, c_{uoo} \, O_{\frac{7}{2}}(w) 
+\frac{1}{(z-w)^2} \, \left[ \frac{2}{7} \, c_{uoo} \,  \pa O_{\frac{7}{2}}  + c_{uoo'} \,
O_{\frac{9}{2}} \right](w)
\nonu \\
&& +\frac{1}{(z-w)} \, \left[ \frac{3}{56} \, c_{uoo} \,  \pa^2 O_{\frac{7}{2}}  -\frac{1}{3} \, c_{uoo'} \,
\pa O_{\frac{9}{2}} + c_{uoto} \, \left( T O_{\frac{7}{2}} -\frac{3}{16} \pa^2 O_{\frac{7}{2}} \right) \right.
\nonu \\
&& \left. + c_{uogp} \,
\left( G P_{4'}^{uu} -\frac{4\sqrt{6}}{9} \pa O_{\frac{9}{2}}-\frac{\sqrt{6}}{56} \pa^2 
O_{\frac{7}{2}} \right) -c_{uogp'} \,
\left( G P_{4'}^{ww} -\frac{2}{9} \sqrt{\frac{2}{3}} \pa O_{\frac{9}{2}}
-\frac{1}{14} \sqrt{\frac{2}{3}} \pa^2 O_{\frac{7}{2}} \right) \right.
\nonu \\
&&  \left.+ O_{\frac{11}{2}} \right](w) +\cdots. 
\label{this}
\eea
The structure constants in (\ref{this}) can be obtained
\bea
c_{uoo}  & = &  \frac{24 (1+k) (8+k) (378+333 k+37 k^2)}{5 (3+k) (6+k) (90+117 k+13 k^2)} ,\nonu \\
c_{uogw} &  = &  -\frac{2 (9+2 k)^2 (-42+99 k+11 k^2)}{5 (3+k) (6+k) (90+117 k+13 k^2)},\nonu \\
c_{uoto}  & = &  \frac{288 (1+k) (8+k) (846+585 k+65 k^2)}{5 (90+117 k+13 k^2) (954+549 k+61 k^2)},
\nonu \\
c_{uogp} & = & \frac{n_1}{d}, \qquad
c_{uogp'}   =  \frac{n_2}{d}, 
\label{conscons}
\eea
where the numerators and denominator in the last two coefficients 
are functions of $k$
\bea
n_1 & \equiv & 2 \sqrt{6} 
(2135484+3378672 k+2386233 k^2+869670 k^3+165765 k^4+15660 k^5+580 k^6), \nonu \\
n_2 & \equiv &  -3 \sqrt{\frac{3}{2}} (13838364+31369356 k+26544483 k^2+10463418 
k^3+2064411 k^4 \nonu \\
& + & 197748 k^5+7324 k^6), \nonu \\
d & \equiv & 5 (3+2 k) (15+2 k) (90+117 k+13 k^2) (954+549 k+61 k^2).
\label{conscons1}
\eea
The OPEs between the spin-$2$ and spin-$\frac{3}{2}$ currents, $T(z)$ and 
$G(z)$,
with the quasi-primary fields in (\ref{this}) 
can be found in Appendices $D$ and $E$, as described before.
The structure of the third- and second-order poles can be deteremined
in a straightforward manner.
In the first-order term, the first two descendant terms
are known and three unknown coefficients $c_{uoto}, c_{uogp}$
and $c_{uogp'}$ are introduced. 
The next step is to determine  how to 
obtain these $k$-dependent coefficients.
Because 
the remaining term can be  characterized by the primary field
of spin-$\frac{11}{2}$, 
there should only be the second-order and first-order terms
when the OPE between the currents 
$T(z)$ or $G(z)$ and the first-order pole
is calculated after
subtracting the above five terms including two derivative terms.
This condition fixes the above unknown three coefficients,
which  are given 
in (\ref{conscons}) and (\ref{conscons1}), leaving 
the spin-$\frac{11}{2}$ primary current,  
which is given by
\bea
O_{\frac{11}{2}}(z)  & = &  
\frac{n_1}{d_1} f^{abc} \psi^a \psi^b \pa^4 \psi^c(z)  + 
\mbox{other} \,\, 22096 \,\, \mbox{terms}, \nonu \\
n_1 & \equiv &  3 \sqrt{2} (-1+k) k (1+k) (9+k) (9+2 k) \nonu \\
& \times & (-23328-31338 k-12843 k^2-1370 
k^3+53 k^4+10 k^5),  \nonu \\
d_1  & \equiv & 
10 (3+k) (6+k) (3+2 k) 
(15+2 k)^2 \sqrt{18+9 k+k^2} (90+117 k+13 k^2) \nonu \\
& \times & (954+549 k+61 k^2),
\label{112112}
\eea 
which generalizes the previous expression (\ref{c4o112}) for an infinite $k$.

The OPE between the spin-$\frac{5}{2}$ current (\ref{Uk}) 
and  spin-$\frac{9}{2}$ current (\ref{9halftwoexp}), 
corresponding to the previous result 
(\ref{c4uo92'}), for the finite $k$
can be  summarized by
\bea
&& U(z) \; O_{\frac{9}{2}'} (w) =
\frac{1}{(z-w)^3} \, \left[ c_{uop} \, P_{4'}^{uu} -
c_{uop'} P_{4'}^{ww}  \right](w)
\nonu \\
&&+ \frac{1}{(z-w)^2} \, \left[ \frac{1}{4} \, c_{uop} \, \pa P_{4'}^{uu} -
\frac{1}{6} \, c_{uop'} \,  \pa P_{4'}^{ww} +c_{uogo} \,
\left( G O_{\frac{7}{2}} +\frac{1}{4\sqrt{6}} \pa P_{4'}^{uu}
-\frac{\sqrt{6}}{4} \pa P_{4'}^{ww} \right)  \right](w) 
\nonu \\
&& + \frac{1}{(z-w)} \, \left[ \frac{1}{24} \, c_{uop} \, \pa^2 P_{4'}^{uu} -
\frac{1}{24} \, c_{uop'} \, \pa^2 P_{4'}^{ww} +\frac{3}{10} \, c_{uogo} \,
\pa \left( G O_{\frac{7}{2}} +\frac{1}{4\sqrt{6}} \pa P_{4'}^{uu}
-\frac{\sqrt{6}}{4} \pa P_{4'}^{ww} \right)  \right. \nonu \\
&&  + c_{uotp} \, \left( T P_{4'}^{uu} -\frac{1}{6} \pa^2 P_{4'}^{uu} \right) - 
c_{uotp'} \,
\left( T P_{4'}^{ww} -\frac{1}{6} \pa^2 P_{4'}^{ww} \right) \nonu \\
&&  - c_{uogo'} \, \left( G \pa O_{\frac{7}{2}} -\frac{7}{3} \pa G O_{\frac{7}{2}} +
\frac{1}{9\sqrt{6}} \pa^2 P_{4'}^{uu} -\frac{1}{3} \sqrt{\frac{2}{3}} \pa^2 P_{4'}^{ww} 
\right)  \nonu \\
&& \left. - c_{uogo''} \, \left( G O_{\frac{9}{2}} -\frac{2}{63} \sqrt{\frac{2}{3}} \pa^2 P_{4'}^{uu} +
\frac{1}{21\sqrt{6}} \pa^2 P_{4'}^{ww} \right) + P_6\right] (w) +\cdots. 
\label{expexp1}
\eea
The structure constants in (\ref{expexp1}) can be expressed as  
\bea
c_{uop} & = &  \frac{4 \sqrt{\frac{2}{3}} (10368+5562 k+2967 k^2+522 k^3+29 k^4)}{5 (3+k) (6+k) 
(90+117 k+13 k^2)}, \nonu \\
c_{uop'} & = & \frac{8 \sqrt{6} (6966+12069 k+7254 k^2+1314 k^3+73 k^4)}{5 
(3+k) (6+k) (90+117 k+13 k^2)}, \nonu \\
c_{uogo} & = &  \frac{792 (1+k) (8+k)}{5 (90+117 k+13 k^2)}, \qquad
c_{uotp}  =  \frac{n_1}{d}, 
\qquad
c_{uotp'}  =   \frac{n_2}{d},  \nonu \\
c_{uogo'} & = & -\frac{12 (1+k) (8+k) 
(378+333 k+37 k^2) (1818+873 k+97 k^2)}{25 
(-6+9 k+k^2) (90+117 k+13 k^2) (954+549 k+61 k^2)}, \nonu \\
c_{uogo''} & = & -\frac{21 (9+2 k)^2 (66+45 k+5 k^2) 
(-42+99 k+11 k^2)}{5 (3+2 k) (15+2 k) (90+117 k+13 k^2) 
(366+207 k+23 k^2)},
\label{complex}
\eea
where the numerators and denominator in (\ref{complex}) of the 
fourth and fifth
constants
are given by
\bea
n_1 & \equiv & 28 
\sqrt{\frac{2}{3}} 
(830045232+8372710800 k+18353902752 k^2+19040042412 k^3+
11562105645 k^4 \nonu \\
& + & 4408598988 k^5+1070601270 k^6+164251116 k^7+15373601 k^8+
800820 k^9+17796 k^{10}),
\nonu \\
n_2 & \equiv & 
-7 \sqrt{6} (9794995632+61010651520 k+137265126552 k^2+158181370992 k^3 \nonu \\
& + & 105657645375 k^4
 +  42990128568 k^5+10861840470 k^6+1705114836 k^7 \nonu \\
& + & 161571871 k^8+8459820 k^9+187996 k^{10}),
\nonu \\
d & \equiv & 5 (3+2 k) (15+2 k) 
(-6+9 k+k^2) (90+117 k+13 k^2) (366+207 k+23 k^2) \nonu \\
& \times & (954+549 k+61 k^2).
\label{someexp}
\eea
The third-order pole  in the right hand 
side can be determined easily.
The coefficient for the quasi-primary field in the second-order pole can be fixed 
without difficulty.
The next step is to determine
how the four quasi-primary fields and  single primary field 
can exist in the last first-order term. 
Four unknown coefficients can be introduced because the derivative terms
are fixed completely.
The procedure 
done for the infinite $k$ case can be used. The four coefficients 
can be  determined.
Finally, after subtracting these four quasi-primary fields correctly,
the following new spin-$6$ primary field remains
\bea
P_6(z) & = & \frac{n_1}{d}
 \, \psi^a \pa^5 \psi^a(z)
 +
\mbox{other lower order derivative terms} , \nonu \\
n_1  & \equiv &   (-1+k) k (1+k) (9+k) (9+2 k) 
(-1365527808-3366282888 k-3098773908 k^2 \nonu \\
& - & 1210990014 k^3-58275207 k^4+110774898 k^5+38348106 k^6+5639106 k^7
+398009 k^8 \nonu \\
& + & 11050 k^9), \nonu \\
d & \equiv & 75 \sqrt{6} (3+k) (6+k)^2 (3+2 k) (15+2 k)^2 
(-6+9 k+k^2) (90+117 k+13 k^2) \nonu \\
& \times & 
(366+207 k+23 k^2) (954+549 k+61 k^2).
\label{spin6spin6}
\eea
As in the previous case, the factor $(k-1)$ is contained 
in this $\psi^a(z)$-dependent spin-$6$ current.

Similarly, 
following OPE between the spin-$3$ current (\ref{Wk}) and the spin-$4$ current 
(\ref{aboveexp}),  
corresponding to the previous result (\ref{c4wo4''}),
can be derived:
\bea
&& W(z) \; O_{4''} (w) =
\frac{1}{(z-w)^3} \, \left[ -c_{wop}\, P_{4'}^{uu} + c_{wop'} \, P_{4'}^{ww}  \right](w) 
\nonu \\
&& + \frac{1}{(z-w)^2} \, 
\left[ -\frac{3}{8} \, c_{wop} \, \pa P_{4'}^{uu} + \frac{3}{8} 
\, c_{wop'} \, \pa P_{4'}^{ww}  + c_{wogo} \, \left( G O_{\frac{7}{2}} +\frac{1}{4\sqrt{6}} \pa P_{4'}^{uu}
-\frac{\sqrt{6}}{4} \pa P_{4'}^{ww}  \right)  \right](w) 
\nonu \\
&&  + \frac{1}{(z-w)} \, 
\left[ -\frac{1}{12} \, c_{wop} \, \pa^2 P_{4'}^{uu} + \frac{1}{12} 
\, c_{wop'} \, \pa^2 P_{4'}^{ww}  
+
\frac{2}{5} \, c_{wogo} \, \pa \left( G O_{\frac{7}{2}} +\frac{1}{4\sqrt{6}} \pa P_{4'}^{uu}
-\frac{\sqrt{6}}{4} \pa P_{4'}^{ww}  \right) \right.   \nonu \\ 
&& 
 - c_{wotp} \, \left( T P_{4'}^{uu} -\frac{1}{6} \pa^2 P_{4'}^{uu} \right) - 
c_{wotp'} \,
\left( T P_{4'}^{ww} -\frac{1}{6} \pa^2 P_{4'}^{ww} \right) \nonu \\
&&  - c_{wogo'} \, \left( G \pa O_{\frac{7}{2}} -\frac{7}{3} \pa G O_{\frac{7}{2}} +
\frac{1}{9\sqrt{6}} \pa^2 P_{4'}^{uu} -\frac{1}{3} \sqrt{\frac{2}{3}} \pa^2 P_{4'}^{ww} 
\right)  \nonu \\
&& \left. - c_{wogo''} \, \left( G O_{\frac{9}{2}} -\frac{2}{63} \sqrt{\frac{2}{3}} \pa^2 P_{4'}^{uu} +
\frac{1}{21\sqrt{6}} \pa^2 P_{4'}^{ww} \right) + P_{6'}\right] (w) +\cdots. 
\label{beforefinal}
\eea
The unknown coefficients in (\ref{beforefinal}) are fixed by
\bea
c_{wop} & = & -\frac{8 (-36+369 k+41 k^2)}{15 (90+117 k+13 k^2)}, \nonu \\
c_{wop'} & = & \frac{8 (2106+6129 k+3759 k^2+684 k^3+38 k^4)}{5 (3+k) (6+k) 
(90+117 k+13 k^2)}, \nonu \\
c_{wogo} & = & \frac{132 \sqrt{6} (1+k) (8+k)}{5 (90+117 k+13 k^2)}, \qquad
c_{wotp}  =  \frac{n_1}{d}, \qquad
c_{wotp'}  =  \frac{n_2}{d}, \nonu \\
c_{wogo'} & = & -\frac{64 \sqrt{6} (1+k) (8+k) (369+234 k+26 k^2) (378+333 k+37 k^2)}{
175 (-6+9 k+k^2) (90+117 k+13 k^2) (954+549 k+61 k^2)}, 
\nonu \\
c_{wogo''} & = & -\frac{6 \sqrt{6} (1+k) (8+k) (9+2 k)^2 (-42+99 k+11 k^2)}{
5 (3+2 k) (15+2 k) (90+117 k+13 k^2) (366+207 k+23 k^2)},
\nonu \\
n_1 & \equiv &
-\frac{32}{3} (3+k) (6+k) (-109854468-570043980 k-583026525 k^2-175058712 k^3\nonu \\
& + & 15297117 k^4
 +  16269390 k^5+2997578 k^6+228096 k^7+6336 k^8), \nonu \\
n_2 & \equiv & -4 (42504047568+157399614720 k+224326680552 k^2+165246862320 k^3 \nonu \\
& + & 70324852617 k^4 
+  18329683752 k^5+3051110298 k^6+334279692 k^7+24334537 k^8\nonu \\
& + & 1114740 k^9+24772 k^{10}),
\label{coefficientcoefficient}
\eea
where $d$ in (\ref{coefficientcoefficient}) is the same that in (\ref{someexp}).
To arrive at the final result (\ref{beforefinal}), 
the OPEs between the spin-$2$ and spin-$\frac{3}{2}$ currents
with the quasi-primary fields appearing in (\ref{beforefinal})
should be obtained. 
These can be found in  Appendices $D$ and $E$.
The structure constants on the two spin-$4$ fields and 
its descendant fields are fixed completely.
The nontrivial second-order pole can be determined in a similar manner.
Based on the results for the second-order pole, 
the correct three derivatives can be obtained 
terms in the first-order singular terms and  
the coefficients appearing in the quasi-primary fields can be determined.
After  extracting these four quasi-primary fields from the first-order pole,
the following new spin-$6$ primary field
can be derived
\bea
P_{6'}(z) & = & \frac{n_1}{d} \, \psi^a \pa^5 \psi^a(z)
 +
\mbox{other lower order derivative terms}, \nonu \\ 
 n_1 & \equiv &  2 (-1+k) k (1+k) (9+k) (9+2 k) (-114528816-144741492 k-9500652 k^2
\nonu \\
& + & 61972749 k^3+35947494 k^4+8541978 k^5+996804 k^6+56069 k^7
+1210 k^8), \nonu \\
 d & \equiv &  225 (3+k) (6+k) (3+2 k) (15+2 k)^2 
(-6+9 k+k^2) (90+117 k+13 k^2) \nonu \\
& \times & (366+207 k+23 k^2) (954+549 k+61 k^2).
\label{otherspin6}
\eea
The following spin-$6$ current, which is a superpartner of 
the spin-$\frac{11}{2}$ current (\ref{112112}), 
together with (\ref{spin6spin6}) and (\ref{otherspin6})
can be constructed
\bea
O_6(z) = -P_6(z) + \sqrt{6} P_{6'}(z), 
\label{spin6expexp}
\eea
which is the same as (\ref{c4o6}).

The final OPE, which is more involved, can be calculated
\bea
&& W(z) \; O_{\frac{9}{2}'} (w) =
\frac{1}{(z-w)^4} \, c_{woo} \, O_{\frac{7}{2}}
+\frac{1}{(z-w)^3} \, \left[ \frac{2}{7} 
\, c_{woo} \,  \pa O_{\frac{7}{2}} - c_{woo'} \, O_{\frac{9}{2}} \right] (w)
\nonu \\
&& +\frac{1}{(z-w)^2} \, \left[ \frac{3}{56} 
\, c_{woo} \, \pa^2 O_{\frac{7}{2}} -\frac{1}{3} 
\, c_{woo'} \, \pa O_{\frac{9}{2}} +c_{woo''} \, O_{\frac{11}{2}} \right. \nonu \\
&& +
c_{woto} \,
\left( T O_{\frac{7}{2}} -\frac{3}{16} \pa^2 O_{\frac{7}{2}} \right)
 + c_{wogp} \,
\left( G P_{4'}^{uu} -\frac{4\sqrt{6}}{9} \pa O_{\frac{9}{2}}-\frac{\sqrt{6}}{56} \pa^2 
O_{\frac{7}{2}} \right) \nonu \\
&& \left. -c_{wogp'} \,
\left( G P_{4'}^{ww} -\frac{2}{9} \sqrt{\frac{2}{3}} \pa O_{\frac{9}{2}}
-\frac{1}{14} \sqrt{\frac{2}{3}} \pa^2 O_{\frac{7}{2}} \right) \right] (w)
\nonu \\
&& +\frac{1}{(z-w)} \, \left[ \frac{1}{126} 
\, c_{woo} \, \pa^3 O_{\frac{7}{2}} -\frac{1}{15} 
c_{woo'} \, \pa^2 O_{\frac{9}{2}} +\frac{4}{11} \, c_{woo''} \, \pa O_{\frac{11}{2}} \right. \nonu \\
&&  +
\frac{4}{11} \, c_{woto} \,  \pa
\left( T O_{\frac{7}{2}} -\frac{3}{16} \pa^2 O_{\frac{7}{2}} \right)  
 + \frac{4}{11} \,  c_{wogp} \,  \pa
\left( G P_{4'}^{uu} -\frac{4\sqrt{6}}{9} \pa O_{\frac{9}{2}}-\frac{\sqrt{6}}{56} \pa^2 
O_{\frac{7}{2}} \right) \nonu \\
&& -\frac{4}{11} \, c_{wogp'} \,  
\pa \left( G P_{4'}^{ww} -\frac{2}{9} \sqrt{\frac{2}{3}} \pa O_{\frac{9}{2}}
-\frac{1}{14} \sqrt{\frac{2}{3}} \pa^2 O_{\frac{7}{2}} \right)  \nonu \\
&&  - c_{woto'} 
\left( T O_{\frac{9}{2}} -\frac{3}{20} \pa^2 O_{\frac{9}{2}} \right)- c_{woto''}
\left( T \pa O_{\frac{7}{2}} -\frac{7}{4} \pa T O_{\frac{7}{2}}-\frac{1}{9} \pa^3 O_{\frac{7}{2}} 
\right) \nonu \\
&&   - c_{wogp''} \,
\left( G \pa P_{4'}^{uu} -\frac{8}{3} \pa G P_{4'}^{uu} -\frac{\sqrt{6}}{5} \pa^2 O_{\frac{9}{2}} -
\frac{2}{63}\sqrt{\frac{2}{3}} \pa^3 O_{\frac{7}{2}} \right) \nonu \\
&& \left. + c_{wogp'''} \,
\left( G \pa P_{4'}^{ww} -\frac{8}{3} \pa G P_{4'}^{ww} -\frac{1}{5\sqrt{6}}
\pa^2 O_{\frac{9}{2}} -\frac{8}{189} \sqrt{\frac{2}{3}} \pa^3 O_{\frac{7}{2}} \right) +
\frac{2}{11} \sqrt{\frac{2}{3}} O_{\frac{13}{2}} \right] (w) \nonu \\
&& +\cdots.
\label{final}
\eea
The structure constants in (\ref{final}) are given by
\bea
c_{woo} & = & \frac{12 
\sqrt{6} (1+k) (8+k) (378+333 k+37 k^2)}{
5 (3+k) (6+k) (90+117 k+13 k^2)}, \nonu \\
c_{woo'} & = & -\frac{2 \sqrt{\frac{2}{3}} 
(9+2 k)^2 (-42+99 k+11 k^2)}{(3+k) (6+k) (90+117 k+13 k^2)}, \qquad
c_{woo''}  =  \frac{1}{\sqrt{6}}, \nonu \\
c_{woto} & = & \frac{216 \sqrt{6} (1+k) (8+k) 
(1354+801 k+89 k^2)}{5 
(90+117 k+13 k^2) (954+549 k+61 k^2)}, \qquad
c_{wogp}  =  \frac{n_1}{d},
\qquad
c_{wogp'}  =  \frac{n_2}{d},
\nonu \\
c_{woto'} & = &  
-\frac{56 \sqrt{\frac{2}{3}} (9+2 k)^2 (-42+99 k+11 k^2)}{
5 (90+117 k+13 k^2) (366+207 k+23 k^2)}, \nonu \\
c_{woto''} & = & -\frac{64 \sqrt{\frac{2}{3}} (1+k) (8+k) 
(378+333 k+37 k^2)}{55 (-6+9 k+k^2) 
(90+117 k+13 k^2)}, 
\nonu \\
c_{wogp''} & = & -\frac{56 
(152280+426492 k+372846 k^2+151785 k^3
+30505 k^4+2943 k^5+109 k^6)}{55 
(-6+9 k+k^2) (90+117 k+13 k^2) 
(366+207 k+23 k^2)}, \nonu \\
c_{wogp'''} & = & \frac{
84 (629532+1636308 k+1520823 k^2+647478 k^3
+133171 k^4+12960 k^5+480 k^6)}{
55 (-6+9 k+k^2 ) (90+117 k+13 k^2) (366+207 k+23 k^2)},
\nonu \\
n_1 & \equiv & 4 (24802524+60056964 k+54194319 k^2+22722930 k^3+
4640895 k^4 \nonu \\
& + & 450468 k^5+16684 k^6),
\nonu \\
n_2 & \equiv & -3 
(195174684+484795692 k+441009171 k^2+185289498 k^3+37865451 k^4 \nonu \\
&& + 3676212 k^5  +  136156 k^6), \nonu \\
d & \equiv & 10 (3+2 k) (15+2 k) 
(90+117 k+13 k^2) (954+549 k+61 k^2).
\label{compcoeff}
\eea
How does one obtain a complete set of structure constants?
In principle, the procedure used  for the infinite $k$ case
can be followed.
On the other hand, the first-order pole is rather complicated.
First of all, 
the first-order singular terms should be calculated completely but
this is  time consuming. 
After that,
the derivative terms appearing in the 
first-order terms
are known from the higher-order singular terms.
This leaves four quasi-primary fields and one 
additional primary field of spin-$\frac{13}{2}$,
which is the last element of the higher spin current in the list 
(\ref{higherspincurrents}).

One way to determine 
the unknown $k$-dependent coefficient functions and 
explicit form the spin-$\frac{13}{2}$ is 
to write down the spin-$\frac{13}{2}$ current 
using its superpartner $O_{6'}(z)$.
The current $O_{6'}(z)$ can be determined, which is 
similar to (\ref{spin6expexp})
because $P_6(z)$ and $P_{6'}(z)$ can be obtained from (\ref{expexp1}) 
and (\ref{beforefinal})
\bea
O_{6'}(z) =  P_6(z) + \frac{7}{2} \sqrt{\frac{3}{2}} P_{6'}(z),
\label{aboveabove}
\eea
which is the same as (\ref{c4o6'}).
By equating the first-order pole from the WZW currents to 
the sum of the derivative terms, quasi-primary fields with four unknown 
coefficients and spin-$\frac{13}{2}$ current, the four unknown
coefficients can be fixed, as expressed in (\ref{compcoeff}).
The last spin-$\frac{13}{2}$ current, which is a superpartner 
of (\ref{aboveabove}), can be given by 
\bea
O_{\frac{13}{2}}(z) & = & \frac{n}{d}
\, f^{abc} \psi^a \psi^b \pa^5 \psi^c(z) +
\mbox{other lower order derivative terms}, \nonu \\
n &\equiv& 16 \sqrt{2} (-1+k) k (1+k) (9+k) (9+2 k)^2 
(-1998-123 k+573 k^2+103 k^3+5 k^4), \nonu \\
d & \equiv &
75 (3+k) (6+k) (15+2 k) 
(-6+9 k+k^2) \sqrt{18+9 k+k^2} (90+117 k+13 k^2)\nonu \\
& \times & (366+207 k+23 k^2).  
\label{finalhigherspin}
\eea
The OPEs between the spin-$2$ and spin-$\frac{3}{2}$ currents
with the seven quasi-primary fields appearing in (\ref{final}) 
can be found in Appendices $D$ and $E$.

Therefore, the higher spin currents are given by 
(\ref{Uk}), (\ref{Wk}), (\ref{co72exp}), (\ref{co92exp}),
 (\ref{aboveexp}), (\ref{5exp}), (\ref{spin4spin4}), footnote \ref{spin4foot}, (\ref{9halftwoexp}),
(\ref{todayexp}), (\ref{112112}), (\ref{spin6expexp}), (\ref{aboveabove}), and 
(\ref{finalhigherspin}), 
and some of the OPEs between them are computed.
In these calculations, the infinite $k$ case in previous section 
is  crucial because the algebraic structure in the OPEs 
is common to each other. 
Although all the singular terms in the OPEs can be obtained 
by defining 
OPE between the current $\psi^a(z)$ and  current $K^a(z)$, it
is difficult to 
express those singular items in terms of the quasi-primary fields and 
higher spin currents.
Note that
the quasi-primary fields can be 
expressed  as those higher spin currents and the stress energy tensor with its superpartner. 
The right hand side of the remaining OPEs  not considered 
in this study should contain only the known primary currents
(and their composite fields that can be either quasi-primary fields or primary fields) 
in the list (\ref{higherspincurrents}).  

\section{Conclusions and outlook }

In this paper we have found 
concrete expressions for 
the higher spin currents in the list of (\ref{higherspincurrents}) of
the $c \leq 4$ model in terms of the WZW currents $\psi^a(z)$ and $K^a(z)$.
They satisfy the following superfusion rules
\bea
\left[\hat{W} \right] \left[\hat{W} \right] & = &  \left[\hat{I} \right] + \left[\hat{O}_{\frac{7}{2}}
\right] +\left[\hat{O}_4 \right],
\nonu \\
\left[\hat{W} \right] \left[\hat{O}_{\frac{7}{2}} \right] & = &  \left[\hat{W} \right] + \left[\hat{O}_{4'}
\right] +\left[\hat{O}_{\frac{9}{2}} \right],
\nonu \\
\left[\hat{W} \right] \left[\hat{O}_{4'} \right] & = & 
\left[\hat{O}_{\frac{7}{2}} \right] + \left[\hat{O}_{4} \right] +
\left[\hat{O}_{\frac{11}{2}} \right] +\left[\hat{O}_6 \right].
\label{fusionrule}
\eea
All the coefficients in the OPEs (\ref{fusionrule}) are fixed.
In the third superfusion rule, some of the algebraic structures of the first 
one occur.
The remaining OPEs were not calculated. 
According to the observation of \cite{ASS},   
they will, in general, take the form $\left[ \hat{I}
\right]+ \left[\hat{W} \right] + \left[ \hat{O}_{\frac{7}{2}} \right] +\left[\hat{O}_{4} \right] + 
 \left[\hat{O}_{4'} \right] +\left[\hat{O}_{\frac{9}{2}} \right] 
+\left[\hat{O}_{\frac{11}{2}} \right] +\left[\hat{O}_6 \right]$, 
in the right hand side.
Finding a concrete expression for a 
quasi-primary field of 
some given spin in terms of $16$ currents is a nontrivial task.
For the most complicated 
OPE between the spin-$\frac{13}{2}$ current and itself,
the singular terms have a $13$-th order pole, $\cdots$, 
second-order pole, and 
first-order pole. The highest quasi-primary field 
of spin-$12$ can then appear in the 
first-order singular term and should be written in terms of 
the known higher spin currents.
Possibly formula (\ref{Nahmformula}) will be helpful in
finding the structure of this quasi-primary field.

An interesting direction is to generalize the coset 
discussed in this paper to the case of a general $N$,
\bea
\frac{\widehat{SU}(N)_k \oplus \widehat{SU}(N)_N}
{\widehat{SU}(N)_{k+N}}, \qquad
c = \frac{(N^2-1)}{2} \left[ 1- \frac{2N^2}{(k+N)(k+2N)} \right] < 
 \frac{(N^2-1)}{2}.  
\label{othercoset}
\eea
This will be  a supersymmetric extension of $W_N$ algebra.
As pointed out in reference \cite{SS}, 
the lowest model in the series of coset models ($k=1$ or $c=\frac{(-1+N)(1+3N)}{2(1+2N)}$)
has ``minimal'' super $W_N$ algebra where there are  
supercurrents of spins
$\frac{3}{2}, \frac{5}{2}, \cdots, (N-\frac{1}{2})$. 
For the general $k > 1$, extra additional currents  should appear. 
As in the present paper, 
it is better to look at the $k \rightarrow 
\infty$ model (or $c=\frac{(N^2-1)}{2}$ fermion model in the adjoint representation of 
$SU(N)$) first because it contains all the algebraic structures and is simpler than 
the more general $c < \frac{(N^2-1)}{2}$ model.
All the 
OPEs in sections $2$ and $3$ should be generalized to the OPEs with 
$N$-dependence. 
In the context of minimal model holography \cite{GG,GG1}, 
the correct normalizations should be calculated for the higher spin currents 
with spins greater than $3$. 
In the original paper \cite{ASS}, the character technique was used to 
generate the complete currents in $c=4$ eight fermion model. 
Generalizing this for an arbitrary $N$ would be interesting study.
As a first step, it will also be useful to consider the $N=4$ case.

A study oF the most general coset models
by 
$\frac{\widehat{SU}(N)_k \oplus \widehat{SU}(N)_l}
{\widehat{SU}(N)_{k+l}}$ with a central charge would be interesting.
For this particular case ($l=N$), this model reduces to the above
model (\ref{othercoset}).
For the higher spin $3, 4$  currents, 
the explicit construction was calculated as mentioned in the introduction. 
Obtaining other higher spin currents explicitly, e.g.
 the spin-$5$ Casimir operator, is an open problem.

In reference \cite{Ahn1202}, the coset model was based on the group 
$SO(N)$ with a given central charge. Therefore, 
more study will be needed to determine if 
the present result can be applied to the different minimal model. 

The following  
describes some partial results in CFT that hold for the general $N$.
The spin-$2$ and spin-$\frac{3}{2}$ currents can be generalized 
without difficulty from (\ref{cosett}) and (\ref{cosetg}), respectively.
To match with the convention of reference \cite{Ahn2011},  
the spin-$1$ currents are rescaled 
together using rescaled structure constants,
the second order pole  for the OPE $J^a (z) \, 
J^b(w)$ given in (\ref{JJ})  has $-N \,\delta^{ab}$,
and  the one for the OPE $K^a(z) \, K^b(w)$ given in (\ref{KK}) has 
$-k \,\delta^{ab}$. 
The relative $N$-dependent  coefficients for $T(z)$ and $G(z)$ 
are determined by calculating the OPEs $T(z) \, T(w)$ and $G(z) \, G(w)$  
completely to satisfy (\ref{tt}) and (\ref{gg}), respectively, 
with the central 
charge (\ref{othercoset}).
For the spin-$3$ current, the formula can be obtained from 
reference \cite{BBSS2} (or 
the OPE $W(z) \, W(w)$, with four unknown coefficients, 
can be used 
explicitly as in \cite{BBSS2} or the regularity of the spin-$3$ 
current with the 
diagonal spin-$1$ current and the primary condition of spin-$3$ current 
under the stress energy tensor can be applied \cite{Ahn2011}). 
For its superpartner, $U(z)$, 
there are 
several ways to determine the complete relative $N$-dependent coefficients.
The OPE between $G(z)$ and $W(w)$ can be used and  
$U(w)$ can be read off from (\ref{gw}) (or the OPE
$U(z) \, U(w)$ calculated  to fix the coefficients).

The two ${\cal N}=1$ supermultiplets with complete 
$N$-dependent coefficients are listed as follows.
\bea
T(z) & = & -\frac{1}{4N} \, J^a J^a(z) -\frac{1}{2(k+N)} \, K^a K^a(z) +
\frac{1}{2(k+2N)} \,
(J^a +K^a) (J^a +K^a)(z), 
\nonu \\
G(z) & = &  -\frac{\sqrt{2} k}{3\sqrt{N(k+N)(k+2N)}} 
\, \psi^a \, \left( J^a -\frac{3N}{ k} K^a \right)(z),
\nonu \\
W(z) & = &  \frac{i \, d^{abc}}{
6N(N+k)(2N+k) \sqrt{6(N+2k)(5N+2k)(N^2-4)}} \left[
  k(k+N)(2k+N)  J^a J^b J^c \right. \nonu \\
&- & \left. 6N(k+N)(2k+N) J^a J^b K^c 
+ 18N^2(k+N) J^a K^b K^c -6N^3 K^a K^b K^c \right](z),
\nonu \\
U(z) & = &
\frac{i \, d^{abc}}{\sqrt{
50N(k+N)(2k+N) (k+2 N) (2 k+5 N) (-4+N^2)}} 
\left[  k (2 k+N)  
\psi^a J^b J^c \right. \nonu \\
&-&  \left.  5N (2 k+N) \psi^a J^b K^c  
 +   10 N^2
\psi^a K^b K^c  \right](z).
\label{newUk}
\eea
These reduce to the previous 
results when $N=3$ up to the overall constants (due to the different 
normalizations).
Note that there are half-integer (higher) spin currents.
How can these $N$-dependent expressions be interpreted?
The zero modes for these currents (in particular, the higher
spin currents) can be analyzed in a similar  to that in
references \cite{GH,Ahn2011}, which  will 
lead to three-point functions (the generalization of \cite{GGHR}) 
with the scalars for all values 
of the 't Hooft coupling in the large $(N,k)$ 't Hooft limit.    

Regarding the symmetry behind the currents in (\ref{newUk}), 
it is natural to ask how the OPEs between these currents arise.
From expressions (\ref{newUk}), the following nontrivial
OPEs between the two lowest higher spin currents, $W(z)$ and $U(z)$, 
are expected in the more general coset model (\ref{othercoset}): 
\bea
W(z) \; W(w) 
& = & \frac{1}{(z-w)^6} \, \frac{c}{3} +\frac{1}{(z-w)^4} \, 2 T(w)
+\frac{1}{(z-w)^3} \, \pa T(w) \nonu \\
& + & \frac{1}{(z-w)^2} \left[ (\frac{3}{10}) \pa^2 T +\frac{32}{22+5c} 
\left( T^2 -\frac{3}{10} \pa^2 T 
\right) +P_{4}^{ww} +P_{4'}^{ww} \right](w)
\nonu \\
& + & \frac{1}{(z-w)} \left[ (\frac{1}{15}) 
\pa^3 T +(\frac{1}{2})   \frac{32}{22+5c} 
\pa \left( T^2 -\frac{3}{10} \pa^2 T 
\right)\right. \nonu \\
& + & \left. (\frac{1}{2}) 
\pa P_{4}^{ww} + (\frac{1}{2}) \pa P_{4'}^{ww} \right](w)+  \cdots,
\nonu \\
W(z) \; U(w) 
 & = & \frac{1}{(z-w)^4} \, \frac{3}{\sqrt{6}} G(w)  +
\frac{1}{(z-w)^3} \, (\frac{2}{3}) \frac{3}{\sqrt{6}} \pa G(w) \nonu \\
& + & \frac{1}{(z-w)^2}
\left[ (\frac{1}{4}) \frac{3}{\sqrt{6}} \pa^2 G +
\frac{11\sqrt{6}}{(4c+21)} 
\left( G T -\frac{1}{8} \pa^2 G \right) +O_{\frac{7}{2}}  \right](w)
\nonu \\
& + & \frac{1}{(z-w)} \left[ 
 (\frac{1}{15}) \frac{3}{\sqrt{6}} \pa^3 G + (\frac{4}{7})
\frac{11\sqrt{6}}{(4c+21)} 
\pa \left( G T -\frac{1}{8} \pa^2 G \right) + 
(\frac{4}{7}) \pa O_{\frac{7}{2}} \right. \nonu \\
& + &  \left.  
\frac{12\sqrt{6}}{7(10c-7)} 
\left( \frac{4}{3} T \pa G - G \pa T -\frac{4}{15} \pa^3 G 
\right)  + O_{\frac{9}{2}} 
\right](w) +\cdots,
\nonu \\
U(z) \; U(w) 
 & = & \frac{1}{(z-w)^5} \, \frac{2c}{5}  
+ \frac{1}{(z-w)^3} \, 2T(w) +
\frac{1}{(z-w)^2} \, \pa T(w) \nonu \\
&+&\frac{1}{(z-w)} \, 
\left[ (\frac{3}{10}) \pa^2 T 
+\frac{27}{22+5c} \left( T^2 -\frac{3}{10} \pa^2 T 
\right) +  P_{4}^{uu} + P_{4'}^{uu} \right](w) \nonu \\
& + & \cdots,
\label{newuulast}
\eea
where the two primary fields \cite{HR,ASS} corresponding to
(\ref{expexp}) and  (\ref{puuexp})  can be expressed as 
\bea
 P_4^{ww}(z) & = &  \frac{8(10-7c)}
{(4c+21)(10c-7)} 
 \left[ -\frac{7}{10} \pa^2 T + \frac{17}{22+5c} 
\left( T^2 -\frac{3}{10} \pa^2 T 
\right) + G \pa G \right](z),
\nonu \\
 P_{4}^{uu}(z) & = &  
-\frac{3(2c-83)}{(4c+21)(10c-7)} 
 \left[ -\frac{7}{10} \pa^2 T + 
\frac{17}{22+5c} \left( T^2 -\frac{3}{10} \pa^2 T 
\right) + G \pa G \right](z).
\label{newpuuexp}
\eea
The $k$-dependent coefficient functions 
appearing in (\ref{wwexp}), (\ref{wuexp}) and (\ref{uulast}) 
are simply replaced with the central charge (\ref{centralc}) and    
generalized to the $N$-dependent  central charge (\ref{othercoset}).
 (\ref{newuulast}) has  no self couping constant.
In other words, there are no $W(w)$-terms in the OPE $W(z) \, W(w)$ and 
$U(w)$-terms in the OPE $U(z) \, U(w)$.
This allows only $c$-dependent coefficient functions to be observed.
The classical $c \rightarrow \infty $ limit \cite{BW} of (\ref{newuulast})
suggests that the $\pa^2 T(w), \pa^3 T(w), \pa^2 G(w), \pa^3 G(w)$-terms 
with $c$-dependent 
coefficients on the right hand side of (\ref{newuulast}) vanish and the 
only $G \pa G(z)$ term in (\ref{newpuuexp}) survives in this classical limit. 

Note that the OPE $W(z) \, W(w)$ contains a composite current 
$G \pa G(w)$ (even at the classical level) 
due to the ${\cal N}=1$ supersymmetry. 
Furthermore, it has another spin-$4$ current.
This is a new feature compared to the bosonic theory \cite{BS,BBSS1}. 
In other words, compared to the standard Zamolodchikov's $W_3$ algebra,
there are two additional primary fields, $P_4^{ww}(w)$ and $P_{4'}^{ww}(w)$.
Compared to the bosonic extended algebra 
\cite{BBSS1}, there is an  additional primary field $P_4^{ww}(w)$.
The OPEs (\ref{newuulast}) hold for any 
$N$, and are exact and complete expressions except  
that the Casimir operators
of spin $s$ with $s=\frac{7}{2}, 4, 4, \frac{9}{2}$ for general 
$N$ are unknown. On the other hand, 
they can be generated by $W(z)$ and $U(z)$
in (\ref{newUk}) in addition to $T(z)$ and $G(z)$. 

The unknown four currents (so far the explicit forms in terms of
$\psi^a(z)$ and $K^a(z)$ for general $N$ are not known), 
$P_{4'}^{ww}(w), O_{\frac{7}{2}}(w), O_{\frac{9}{2}}(w)$, 
and 
$P_{4'}^{uu}(w)$ on the right hand side of (\ref{newuulast})
can be found, in principle, by calculating the OPEs between the explicit 
expressions (\ref{newUk}).
The spin-$\frac{7}{2}$ current, $O_{\frac{7}{2}}(z)$, 
is expected to contain the following nonderivative
composite operators by realizing the possible combinations from the 
currents $W(z)$ and $U(w)$, given in (\ref{newUk}):
\bea
&& d^{abc} d^{cde} \psi^a J^b J^d J^e, \qquad 
 d^{abc} d^{cde} \psi^a J^b J^d K^e, \qquad
 d^{abc} d^{cde} \psi^a J^d J^e K^b, \qquad
d^{abc} d^{cde} \psi^a J^b K^d K^e, \nonu \\
&& 
d^{abc} d^{cde} \psi^a J^d K^b K^e, \qquad
d^{abc} d^{cde} \psi^a K^b K^d K^e.
\label{spin72current}
\eea
Of course, there are different types of derivative terms due to  
normal ordering in the composite fields.  
Note that the quantity, $f^{abc} \psi^b J^c(z)$, 
is zero  due to (\ref{Jdef}).
The complete relative $N$-dependent coefficients can be determined 
by calculating the corresponding OPE $W(z) \, U(w)$ and extracting the 
structure of the second-order pole carefully.
Each $d$ symbol in (\ref{spin72current})
comes from the currents, $W(z)$ and $U(w)$, respectively.
Similarly, the spin $\frac{9}{2}$ current can be obtained from 
the first-order pole of this OPE.
For the spin-$4$ current, $P_{4'}^{ww}(z)$, the field contents 
can be obtained 
from the previous result \cite{Ahn2011} and the other terms in the 
second order pole in the OPE $W(z) \, W(w)$.
Finally, another spin-$4$ current, $P_{4'}^{uu}(z)$, 
can be determined from the OPE of $U(z) \, U(w)$ explicitly.
A further study should  determine these currents.

What happens with the self-coupling term?
The OPE of 
$O_4(z) \, O_4(w)$  
has a self coupling term $O_4(w)$ 
and one direct way to see this $N$-dependent self 
coupling constant is to calculate this OPE explicitly (eventhough this will be 
very complicated). A further study should determine 
other OPEs for general $N$ not considered in this  study. 

One application of these results is the ability to analyze 
a large $k$ limit with $N$ fixed ($c=4$ model) 
similar to that in reference 
\cite{GS}. The OPEs are summarized by (\ref{fusionrule}).
The asymptotic symmetry of the higher 
spin $AdS_3$ gravity (at the quantum level) 
can be summarized from the two dimensional CFT results
obtained thus far.
The OPEs (\ref{newuulast}) with (\ref{newpuuexp}) 
should provide  asymptotic 
symmetry algebra in the $AdS_3$ bulk theory   at
both the classical and quantum levels.
The three-point functions from the CFT computations  
should correspond to the three-point functions in the 
$AdS_3$ bulk theory. A further study should examine the corresponding 
bulk theory computations.  
See also recent papers \cite{CHR2,MZ}.
The coset model is 
a ${\cal N}=1$ version of the coset model studied in reference 
\cite{Schoutensetal}.
Therefore, the bulk theory would have  higher spin gauge symmetry
in $AdS_3$ string theory. This is because  the central charge 
(\ref{othercoset}) in 
this coset model is proportional to $N^2$ rather than $N$. 
The algebra described is 
larger than the conventional $W_N$ algebra: the existence of half-integer 
spin currents. 
See also relevant studies \cite{Ahn1206,Ahn1208}, where the 
${\cal N}=2$ minimal model holography can be determined 
using asymptotic symmetry algebra.

\vspace{.7cm}

\centerline{\bf Acknowledgments}

CA would like to thank his previous collaborators K. Schoutens and 
A. Sevrin for the work of \cite{ASS}. 
This study was supported by the Mid-career Researcher Program through
the National Research Foundation of Korea (NRF) grant 
funded by the Korean government (MEST) (No. 2012-045385).
CA acknowledges warm hospitality from 
the School of  Liberal Arts (and Institute of Convergence Fundamental
Studies), Seoul National University of Science and Technology.

\newpage

\appendix

\renewcommand{\thesection}{\large \bf \mbox{Appendix~}\Alph{section}}
\renewcommand{\theequation}{\Alph{section}\mbox{.}\arabic{equation}}

\section{The coefficients appearing in the 
descendant fields of quasi-primary or primary fields (\ref{PhiPhi})}

The introduction presented the OPE between the two quasi-primary fields (including the primary fields), 
where the coefficient functions that 
depend on the spins, $h_i, h_j, h_k$ and the number of derivatives $n$,
are given by
\bea
A_{i,j,k,n} \equiv \frac{1}{n!} \frac{\Gamma(h_i-h_j+h_k+n)}{\Gamma(h_i-h_j+h_k)} 
\frac{\Gamma(2h_k)}{\Gamma(2h_k+n)} =
\frac{1}{n!} \prod_{x=0}^{n-1} \frac{(h_i-h_j+h_k+x)}{ (2h_k+x)}.
\label{Coeff1}
\eea
Here, they are written in terms of the Pochhammer symbol because sometimes 
the denominator in  the original expression can have a zero values.
To avoid this feature, the ratio of two Gamma functions should be expressed
in terms of 
the Pochhammer symbol as shown in (\ref{Coeff1}).

Using the definition of the coefficients (\ref{Coeff1}),
those vanishing and nonvanishing 
coefficients appearing in the Sections $2$, and $3$,
should be checked as follows:
\bea
A_{2,2,2,1} &=& \frac{1}{2}, \qquad
A_{2,\frac{3}{2},\frac{3}{2},1} =\frac{2}{3}, 
\qquad
A_{\frac{3}{2},3,\frac{5}{2},1} =\frac{1}{5},
\qquad
A_{\frac{3}{2},4,\frac{7}{2},1} =\frac{1}{7},
\nonu \\
A_{\frac{3}{2},\frac{9}{2},4,1} & = & \frac{1}{8},
\qquad
A_{3,3,2,1}  =   \frac{1}{2}, \qquad 
A_{3,3,2,2} = \frac{3}{20},
\qquad
A_{3,3,2,3} = \frac{1}{30}, 
\nonu \\
A_{3,3,4,1} & = & \frac{1}{2}, 
\qquad
A_{3,\frac{5}{2},\frac{3}{2},1}  =   \frac{2}{3}, \qquad 
A_{3,\frac{5}{2},\frac{3}{2},2} = \frac{1}{4},
\qquad
A_{3,\frac{5}{2},\frac{3}{2},3} = \frac{1}{15}, 
\nonu \\ 
A_{3,\frac{5}{2},\frac{7}{2},1} & = & \frac{4}{7},
\qquad
A_{\frac{5}{2},\frac{5}{2},2,1}  =   \frac{1}{2}, \qquad 
A_{\frac{5}{2}, \frac{5}{2},2,2} = \frac{3}{20},
\qquad
A_{\frac{5}{2},\frac{7}{2},3,1} = \frac{1}{3}, 
\nonu \\
A_{\frac{5}{2},\frac{7}{2},3,2} & = & \frac{1}{14},
\qquad
A_{\frac{5}{2},\frac{7}{2},4,1}  =   \frac{3}{8}, \qquad 
A_{\frac{5}{2},4, \frac{5}{2},1} = \frac{1}{5},
\qquad
A_{\frac{5}{2},4,\frac{5}{2},2} = \frac{1}{30}, \nonu \\
A_{\frac{5}{2},4,\frac{5}{2},3} & = &  \frac{1}{210},
\qquad
A_{\frac{5}{2},4,\frac{9}{2},1}  =  \frac{1}{3}, \qquad 
A_{2,\frac{11}{2},\frac{5}{2},1} = -\frac{1}{5},
\qquad
A_{\frac{3}{2},\frac{11}{2},3,1} = -\frac{1}{6}, \nonu \\
A_{3,\frac{7}{2},\frac{5}{2},1} & = & \frac{2}{5},
\qquad
A_{3,\frac{7}{2},\frac{5}{2},2}  =   \frac{1}{10}, \qquad 
A_{3,\frac{7}{2},\frac{5}{2},3} = \frac{2}{105},
\qquad
A_{3,\frac{7}{2},\frac{9}{2},1} = \frac{4}{9}, \nonu \\ 
A_{3,4,3,1} & = & \frac{1}{3},
\qquad
A_{3,4,3,2}  =  \frac{1}{14}, \qquad 
A_{3,4,3,3} = \frac{1}{84},
\qquad
A_{3,4,4,1} = \frac{3}{8}, \nonu \\ 
A_{3,4,4,2} & = & \frac{1}{12},
\qquad
A_{3,4,5,1} = \frac{2}{5},
\qquad
A_{\frac{3}{2},5,\frac{5}{2},1}  =  -\frac{1}{5}, \qquad 
A_{2,6,3,1} = -\frac{1}{6},
\nonu \\
A_{\frac{3}{2},6,\frac{5}{2},1} & = &  -\frac{2}{5}, \qquad 
A_{\frac{3}{2},6,\frac{5}{2},2} = \frac{1}{30},
\qquad
A_{\frac{5}{2},4,\frac{7}{2},1}  =  \frac{2}{7}, \qquad 
A_{\frac{5}{2},4,\frac{7}{2},2} = \frac{3}{56},
\nonu \\
A_{\frac{5}{2},4,\frac{9}{2},1} & = & \frac{1}{3}, \qquad 
A_{\frac{5}{2},\frac{9}{2},4,1} = \frac{1}{4},
\qquad
A_{\frac{5}{2},\frac{9}{2},4,2}  =   \frac{1}{24}, \qquad 
A_{\frac{5}{2},\frac{9}{2},5,1} = \frac{3}{10},
\nonu \\
A_{3,4,4,1} & = &  \frac{3}{8}, \qquad 
A_{3,4,4,2} = \frac{1}{12},
\qquad
A_{3,4,5,1}  =   \frac{2}{5}, \qquad 
A_{3,\frac{9}{2},\frac{7}{2},1} = \frac{2}{7},
\nonu \\
A_{3,\frac{9}{2},\frac{7}{2},2} & = &  \frac{3}{56}, \qquad 
A_{3,\frac{9}{2},\frac{7}{2},3} = \frac{1}{126},
\qquad
A_{3,\frac{9}{2},\frac{9}{2},1}  =  \frac{1}{3}, \qquad 
A_{3,\frac{9}{2},\frac{9}{2},2} = \frac{1}{15},
\nonu \\
A_{3,\frac{9}{2},\frac{11}{2},1} & = &  \frac{4}{11}. 
\label{Acoefficient}
\eea

\section{The OPE between the stress energy tensor and the quasiprimary
or primary  fields in $c=4$ model}

To determine if a conformal field is quasi-primary field,
the OPE between the stress energy tensor $T(z)$ and a field $\Phi(w)$
should be calculated and 
the vanishing of third-order pole in the OPE $T(z) \; \Phi(w)$
should be chceked.
All the quasi-primary fields (where there are three primary fields) 
in sections $2$ and $3$ are listed as follows
\bea
&& T(z) \; \left(T T -\frac{3}{10} \pa^2 T \right)(w) = \frac{1}{(z-w)^4} \, \frac{42}{5} \, T(w) 
+{\cal O}((z-w)^{-2}), 
\nonu \\
&& T(z) \; \left(G T -\frac{1}{8} \pa^2 G \right)(w) = \frac{1}{(z-w)^4} \, \frac{37}{8} \,
G(w) + {\cal O}((z-w)^{-2}),
\nonu \\
&& T(z) \; \left(G \pa T -\frac{4}{3} T \pa G + \frac{4}{15} \pa^3 G \right)(w) =
\frac{1}{(z-w)^5} \, \frac{33}{5} \, G(w) 
\nonu \\
&& +\frac{1}{(z-w)^4} \, (-\frac{1}{3}) \frac{33}{5} \, \pa G(w) +{\cal O}((z-w)^{-2}),
\nonu \\
&& T(z) \; \left(G U -\frac{\sqrt{6}}{3} \pa W \right)(w)  = {\cal O}((z-w)^{-2}),
\nonu \\
&& T(z) \; \left( T W -\frac{3}{14} \pa^2 W \right)(w)   =  \frac{1}{(z-w)^4} \, \frac{71}{7} W(w) 
+{\cal O}((z-w)^{-2}), 
\nonu \\
&& T(z) \; \left( G \pa U -\frac{5}{3} \pa G U -\frac{\sqrt{6}}{7} \pa^2 W \right)(w)  = 
-\frac{1}{(z-w)^4} \, \frac{124}{7} \, \sqrt{\frac{2}{3}}  \, W(w) +{\cal O}((z-w)^{-2}), 
\nonu \\
&& T(z) \;  \left(G W -\frac{1}{6\sqrt{6}} \pa^2 U \right)(w)  =  \frac{1}{(z-w)^4} \, 
5 \sqrt{\frac{2}{3}}  \, U (w) +{\cal O}((z-w)^{-2}),
\nonu \\
&& T(z) \; \left(T U -\frac{1}{4} \pa^2 U \right)(w)  = 
 \frac{1}{(z-w)^4} \, 
\frac{33}{4}  \, U(w) +{\cal O}((z-w)^{-2}),
\nonu \\
&& T(z) \; \left( T O_{\frac{7}{2}} -\frac{3}{16} \pa^2 O_{\frac{7}{2}} \right)(w)  
= \frac{1}{(z-w)^4} \, 
\frac{193}{16}  \, O_{\frac{7}{2}}(w) +{\cal O}((z-w)^{-2}),
\nonu \\
&& T(z) \; \left( G P_{4'}^{uu} -\frac{4\sqrt{6}}{9} \pa O_{\frac{9}{2}}-\frac{\sqrt{6}}{56} \pa^2 
O_{\frac{7}{2}} \right)(w)  = \frac{1}{(z-w)^4} \, 
\frac{17}{4}\, \sqrt{\frac{3}{2}}  \, O_{\frac{7}{2}}(w) +{\cal O}((z-w)^{-2}),
\nonu \\
&& T(z) \; \left( G P_{4'}^{ww} -\frac{2}{9} \sqrt{\frac{2}{3}} \pa O_{\frac{9}{2}}
-\frac{1}{14} \sqrt{\frac{2}{3}} \pa^2 O_{\frac{7}{2}} \right) (w)
 =  \frac{1}{(z-w)^4} \, 
\frac{17}{\sqrt{6}}  \, O_{\frac{7}{2}}(w) +{\cal O}((z-w)^{-2}),
\nonu \\
&& T(z) \; \left( G O_{\frac{7}{2}} +\frac{1}{4\sqrt{6}} \pa P_{4'}^{uu}
-\frac{\sqrt{6}}{4} \pa P_{4'}^{ww} \right)(w)  = +{\cal O}((z-w)^{-2}),
\nonu \\
&& T(z) \; \left( T P_{4'}^{uu} -\frac{1}{6} \pa^2 P_{4'}^{uu} \right)(w)  = 
 \frac{1}{(z-w)^4} \, 14 \, P_{4'}^{uu}(w) +{\cal O}((z-w)^{-2}),
\nonu \\
&& T(z) \; \left( T P_{4'}^{ww} -\frac{1}{6} \pa^2 P_{4'}^{ww} \right)(w)  = 
 \frac{1}{(z-w)^4} \, 14 \, P_{4'}^{ww}(w) +{\cal O}((z-w)^{-2}),
\nonu \\
&& T(z) \; \left( G \pa O_{\frac{7}{2}} -\frac{7}{3} \pa G O_{\frac{7}{2}} +
\frac{1}{9\sqrt{6}} \pa^2 P_{4'}^{uu} -\frac{1}{3} \sqrt{\frac{2}{3}} \pa^2 P_{4'}^{ww} \right)(w)
 =  \nonu \\
&& \frac{1}{(z-w)^4} \, \left[ \frac{25}{3\sqrt{6}} \, P_{4'}^{uu} -25\sqrt{\frac{2}{3}} 
P_{4'}^{ww}\right] (w) +{\cal O}((z-w)^{-2}),
\nonu \\
&& T(z) \; \left( G O_{\frac{9}{2}} -\frac{2}{63} \sqrt{\frac{2}{3}} \pa^2 P_{4'}^{uu} +
\frac{1}{21\sqrt{6}} \pa^2 P_{4'}^{ww} \right)(w)  = 
 \nonu \\
&& \frac{1}{(z-w)^4} \, \left[ \frac{104}{21} \sqrt{\frac{2}{3}} \, 
P_{4'}^{uu} -\frac{26}{7}\sqrt{\frac{2}{3}} 
P_{4'}^{ww}\right] (w) +{\cal O}((z-w)^{-2}),
\nonu \\
&& T(z) \; \left( T O_{\frac{9}{2}} -\frac{3}{20} \pa^2 O_{\frac{9}{2}} \right)(w)  = 
 \frac{1}{(z-w)^4} \, \frac{319}{20} \, O_{\frac{9}{2}} (w) +{\cal O}((z-w)^{-2}),
\nonu \\
&& T(z) \; \left( 
T \pa O_{\frac{7}{2}} -\frac{7}{4} \pa T O_{\frac{7}{2}}-\frac{1}{9} \pa^3 O_{\frac{7}{2}} \right)(w)  = 
-\frac{1}{(z-w)^5} \, \frac{301}{12} O_{\frac{7}{2}} (w) \nonu \\
&& -
 \frac{1}{(z-w)^4} \, (-\frac{1}{7}) \frac{301}{12} \, \pa O_{\frac{7}{2}}(w) +{\cal O}((z-w)^{-2}),  
\nonu \\
&& T(z) \; \left( G \pa P_{4'}^{uu} -\frac{8}{3} \pa G P_{4'}^{uu} -\frac{\sqrt{6}}{5} \pa^2 O_{\frac{9}{2}} -
\frac{2}{63}\sqrt{\frac{2}{3}} \pa^3 O_{\frac{7}{2}} \right)(w)  = 
\nonu \\
&& -\frac{1}{(z-w)^5} \, \frac{50}{3} \, \sqrt{\frac{2}{3}} \, O_{\frac{7}{2}} (w)  +\frac{1}{(z-w)^4} \left[ -(-\frac{1}{7}) \frac{50}{3} 
\sqrt{\frac{2}{3}} \, \pa O_{\frac{7}{2}} -
\frac{286}{5} \sqrt{\frac{2}{3}} \, O_{\frac{9}{2}} \right] (w) +{\cal O}((z-w)^{-2}),
\nonu \\
&& T(z) \; \left( G \pa P_{4'}^{ww} -\frac{8}{3} \pa G P_{4'}^{ww} -\frac{1}{5\sqrt{6}}
\pa^2 O_{\frac{9}{2}} -\frac{8}{189} \sqrt{\frac{2}{3}} \pa^3 O_{\frac{7}{2}} \right)(w)
 = \nonu \\
&& -\frac{1}{(z-w)^5} \, \frac{200}{9} \, \sqrt{\frac{2}{3}} \, O_{\frac{7}{2}} (w) +\frac{1}{(z-w)^4} \left[ -(-\frac{1}{7}) \frac{200}{9} \sqrt{\frac{2}{3}} \, \pa O_{\frac{7}{2}} -
\frac{143}{15} \sqrt{\frac{2}{3}} \, O_{\frac{9}{2}} \right] (w) \nonu \\
&& +{\cal O}((z-w)^{-2}),
\nonu \\
&& T(z) \; \left( T \pa U -\frac{5}{4} \pa T U -\frac{1}{7} \pa^3 U \right)(w) =
-\frac{1}{(z-w)^5} \, \frac{345}{28} \, U(w) -\frac{1}{(z-w)^4} \, (-\frac{1}{5}) \,
\frac{345}{28} \, \pa U(w) \nonu \\
&& +  {\cal O}((z-w)^{-2}),
\nonu \\  
&& T(z) \; \left( G \pa W - 2 \pa G W -\frac{1}{21} \sqrt{\frac{2}{3}} \pa^3 U \right)(w)
= -\frac{1}{(z-w)^5} \, \frac{55}{7} \sqrt{\frac{2}{3}} \, U(w) \nonu \\
&& -\frac{1}{(z-w)^4} \, (-\frac{1}{5}) 
 \frac{55}{7} \sqrt{\frac{2}{3}} 
\,
\pa U(w) +  {\cal O}((z-w)^{-2}),    
\nonu \\
&& T(z) \; \left( G O_{4''} -\frac{2}{9} \pa O_{\frac{9}{2}'} \right)(w)
=+  {\cal O}((z-w)^{-2}),
\nonu \\
&& T(z) \; \left( G \pa^2 U - 4 \pa G \pa U +\frac{5}{2} \pa^2 G U -\frac{1}{2\sqrt{6}} \pa^3 W
 \right)(w) =\frac{1}{(z-w)^5} \, 9 \sqrt{6} \, W(w)
\nonu \\
&& +\frac{1}{(z-w)^4} \left[ (-\frac{1}{6}) 9\sqrt{6} \pa W +\frac{75}{2} 
\left( G U -\frac{\sqrt{6}}{3} \pa W \right) \right](w) 
+  {\cal O}((z-w)^{-2}),
\nonu \\
&& T(z) \; \left( T \pa W -\frac{3}{2} \pa T W -\frac{1}{8} \pa^3 W \right)(w)
=-\frac{1}{(z-w)^5} 18 W(w) -\frac{1}{(z-w)^4} (-\frac{1}{6}) 18 \pa W(w) 
\nonu \\
&& +  {\cal O}((z-w)^{-2}),
\nonu \\
&& T(z) \; \left( T O_{4''} -\frac{1}{6} \pa^2 O_{4''} \right)(w)
= \frac{1}{(z-w)^4} \, 14 O_{4''}(w) + {\cal O}((z-w)^{-2}),
\nonu \\
&& T(z) \; \left( G O_{\frac{9}{2}'} -\frac{1}{9} \pa^2 O_{4''} \right)(w)
= \frac{1}{(z-w)^4} \, \frac{52}{3} O_{4''}(w) + {\cal O}((z-w)^{-2}).
\label{stresswithquasi}
\eea
(\ref{stresswithquasi}) has no third-order pole.
The vanishing and nonvanishing 
coefficients appearing in (\ref{stresswithquasi})
can be checked 
using the definition of the coefficients (\ref{Coeff1})
as follows:
\bea
A_{2,4,2,1} & = & 0, \qquad 
A_{2,\frac{7}{2},\frac{3}{2},1} =0,
\qquad
A_{2,\frac{9}{2},\frac{3}{2},1}  =  -\frac{1}{3}, \qquad 
A_{2,\frac{9}{2},\frac{3}{2},2} =0,
\nonu \\
A_{2,\frac{9}{2},\frac{5}{2},1} & = & 0, \qquad 
A_{2,5,3,1} =0,
\qquad
A_{2,\frac{11}{2},\frac{5}{2},1} = -\frac{1}{5},
\qquad
A_{2,\frac{11}{2},\frac{5}{2},2} =0,
\nonu \\
A_{2,\frac{11}{2},\frac{7}{2},1}  & = &  0, \qquad 
A_{2,6,3,1} = -\frac{1}{6}, \qquad
A_{2,6,3,2} = 0,
\qquad
A_{2,6,4,1} =0,
\nonu \\
A_{2,\frac{13}{2},\frac{7}{2},1}  & = &  -\frac{1}{7}, \qquad 
A_{2,\frac{13}{2},\frac{7}{2},2} =0,
\qquad
A_{2, \frac{13}{2},\frac{9}{2},1}  =  0. 
\label{CoeffA}
\eea
For example, the first OPE in (\ref{stresswithquasi}) has no third-order 
singular term. This can be 
realized by the disappearance of $A_{2,4,2,1}$ in (\ref{CoeffA}).
In other words,  the disappearance 
of the third-order singular term cn be understood from the explicit 
WZW currents and this can be confirmed from (\ref{PhiPhi}).   

Note that the following quasi-primary fields 
\cite{Thielemans1} can be derived from (\ref{stresswithquasi})  
\bea
\left(T \Phi_i -\frac{3}{2(2h_i+1)} \pa^2 \Phi_i\right)(z).
\label{TPhi}
\eea
The relative coefficient in (\ref{TPhi}) is fixed from the definition of quasi-primary condition.
Therefore the other quasi-primary fields that contain the derivatives in the quadratic normal ordered product should be considered.

In reference \cite{BFKNRV}, 
any quasi-primary field can be written in terms of 
quadratic part and linear part 
\bea
\sum_{r=0}^n \, B_{i,j,n,r} \, \pa^r \, N (\Phi_j, \pa^{n-r} \Phi_i) 
+ \sum_{k: \, h_i+h_j-h_k \geq 1}  \, C_{ijk} \, C_{i,j,k,n} \, \pa^{h_i+h_j-h_k+n} \Phi_k.
\label{Nahmformula}
\eea
When $n=0$, the first term in (\ref{Nahmformula}) does not contain any derivatives.
Each coefficient functions are introduced as follows:
\bea
&& B_{i,j,n,r} \equiv (-1)^r \frac{\Gamma(n+1)}{\Gamma(r+1) \Gamma(n-r+1)}
\frac{\Gamma(2h_i+n)}{\Gamma(r+1) \Gamma(2h_i+n-r)}
\frac{\Gamma(r+1) \Gamma(2h_i+2h_j+2n-r-1)}{\Gamma(2h_i+2h_j+2n-1)},
\nonu \\
&& C_{i,j,k,n} \equiv -(-1)^n \frac{\Gamma(h_i+h_j-h_k+n)}{\Gamma(h_i+h_j-h_k) \Gamma(n+1)} 
\frac{\Gamma(n+1) \Gamma(2h_i+2h_j+n-1)}{\Gamma(2h_i+2h_j+2n-1)} \nonu \\
&&
\times \frac{\Gamma(2h_i+n)}{\Gamma(h_i+h_j-h_k+n+1) \Gamma(h_i-h_j+h_k)}
\frac{\Gamma(h_i+h_j-h_k) \Gamma(2h_k)}{\Gamma(h_i+h_j+h_k-1)}
\frac{1}{(h_i+h_j+h_k+n-1)}
\nonu \\
&& \times \frac{1}{\Gamma(h_i+h_j-h_k)}.
\label{BC}
\eea
Their binomial symbols can be rewritten simply as Gamma functions.
Although $B_{i,j,n,r} $ and $ C_{i,j,k,n}$ can be simplified further, 
they maintain their  present form.  
The structure constant, $C_{ijk}$, is the same as that in (\ref{PhiPhi})
where fields $\Phi_i(z)$ and $\Phi_j(w)$ 
appear in the first quadratic part in 
(\ref{Nahmformula}). This suggests that 
the complete structure of quasi-primary field is determined only 
if the OPE 
$\Phi_i(z) \; \Phi_{j}(w)$ is known  
because the second term contains the above structure constant $C_{ijk}$. Otherwise, this structure constant is not known and 
this unknown structure constant is fixed by the Jacobi identities
when the Jacobi identities mentioned before are used.

Note that their normal ordered product used in this paper 
is different from 
the one in reference \cite{BS}.
In other words \cite{Blumenhagenetal1},  
\bea
N (\Phi_j, \pa^{n-r} \Phi_i)(z)  = (\pa^{n-r} \Phi_i \, \Phi_j)(z). 
\label{trans}
\eea
Once  formula (\ref{Nahmformula}) is used,   
the convention (\ref{trans}) should be applied 
to the expression of the quasi-primary field. 
The quasi-primary field 
$\left(G \pa T -\frac{4}{3} T \pa G + \frac{4}{15} \pa^3 G \right)(z)$
can be obtained from (\ref{stresswithquasi}).
How can  this be observed from (\ref{Nahmformula})?
Consider $\Phi_i(z) =T(z), \Phi_j(z)=G(z)$.  
$T(z)\; G(w) 
= \frac{1}{(z-w)^2} \frac{3}{2} G(w) +\frac{1}{(z-w)} \pa G(w) +\cdots$.
From this, the field, $\Phi_k(z)=G(z)$, with $C_{ijk}=\frac{3}{2}$.
From (\ref{BC}), $B_{2,\frac{3}{2},1,0}=1, B_{2,\frac{3}{2},1,1}=-\frac{4}{7}$
and $C_{2,\frac{3}{2},\frac{3}{2},1}=\frac{8}{105}$.
$\frac{3}{7} N (G \pa T)(z) -\frac{4}{7} N (\pa G T)(z) 
+\frac{4}{35} \pa^3 G(z)$ can be 
obtained by substitiuting these numerical values in (\ref{Nahmformula}),
On the other hand, from (\ref{trans}),
$\frac{3}{7} \left( \pa T G -\frac{4}{3} T \pa G  +\frac{4}{15} \pa^3 
G \right)(z)$.
Furthermore, 
$\frac{3}{7} \left( G \pa T  -\frac{4}{3} T \pa G  
+\frac{4}{15} \pa^3 G \right)(z)$, which is proportional to 
the quasi-primary field  can be obtained
using the relation $G \pa T(z) = \pa T G(z)$.
Similar analysis and checks can be performed to determine 
if all the 
quasi-primary fields appearing 
in this paper can be read off from the general formula. 

All the quasi-primary fields can be checked 
using (\ref{Nahmformula}).
The relevant coefficient functions that appear in the 
quasi-primary fields in 
(\ref{stresswithquasi}) are listed as follows:
\bea
B_{2,\frac{3}{2},1,0} & = & 1, \qquad
B_{2,\frac{3}{2},1,1} = -\frac{4}{7}, \qquad
B_{\frac{5}{2},\frac{3}{2},1,0} =1, \qquad
B_{\frac{5}{2},\frac{3}{2},1,1} =-\frac{5}{8},
\nonu \\
B_{\frac{7}{2},\frac{3}{2},1,0}  & = & 1, \qquad
B_{\frac{7}{2},\frac{3}{2},1,1} = -\frac{7}{10}, \qquad
B_{\frac{7}{2},2,1,0} =1,
\qquad
B_{\frac{7}{2},2,1,1} =-\frac{7}{11},
\nonu \\
B_{4,\frac{3}{2},1,0} & = & 1,
\qquad
B_{4,\frac{3}{2},1,1} =-\frac{8}{11},\qquad
B_{\frac{5}{2},2,1,0} =1,
\qquad
B_{\frac{5}{2},2,1,1} =-\frac{5}{9},
\nonu \\
B_{3,\frac{3}{2},1,0} & = & 1,
\qquad
B_{3,\frac{3}{2},1,1} = -\frac{2}{3},
\qquad
B_{\frac{5}{2},\frac{3}{2},2,0} =1,\qquad
B_{\frac{5}{2},\frac{3}{2},2,1} =-\frac{6}{5},
\nonu \\
B_{\frac{5}{2},\frac{3}{2},2,2} & = & \frac{1}{3},
\qquad
B_{3,2,1,0} =1,
\qquad
B_{3,2,1,1} =-\frac{3}{5},\qquad
B_{\frac{3}{2},\frac{5}{2},0,0} =1,
\nonu \\
B_{\frac{3}{2},3,0,0} & = & 1,
\qquad
B_{\frac{3}{2},4,0,0} =1,
\qquad
B_{\frac{3}{2},\frac{7}{2},0,0} =1,
\qquad
B_{\frac{3}{2},\frac{9}{2},0,0} =1,
\label{Bcoeff}
\eea
and
\bea
C_{2,\frac{3}{2},\frac{3}{2},1} & = & \frac{8}{105},
\qquad
C_{\frac{5}{2},\frac{3}{2},3,1} =\frac{5}{28},
\qquad
C_{\frac{7}{2},\frac{3}{2},4,1} =\frac{7}{30},
\qquad
C_{\frac{7}{2},2,\frac{7}{2},1} =\frac{10}{99},
\nonu \\
C_{4,\frac{3}{2},\frac{7}{2},1} & = & \frac{16}{99},
\qquad 
C_{4,\frac{3}{2},\frac{9}{2},1} =\frac{14}{55},
\qquad
C_{\frac{5}{2},2,\frac{5}{2},1} =\frac{4}{63},
\qquad
C_{3,\frac{3}{2},\frac{5}{2},1} =\frac{8}{63},
\nonu \\
C_{\frac{5}{2},\frac{3}{2},3,2} &  = & -\frac{1}{18},
\qquad
C_{3,2,3,1} =\frac{1}{12},
\qquad
C_{\frac{3}{2},\frac{5}{2},3,0} =-\frac{1}{3},
\qquad
C_{\frac{3}{2},3,\frac{5}{2},0} =-\frac{1}{30},
\nonu \\
C_{\frac{3}{2},4,\frac{7}{2},0} & = & -\frac{1}{56},
\qquad
C_{\frac{3}{2},4,\frac{9}{2},0} =-\frac{2}{9},
\qquad
C_{\frac{3}{2},\frac{7}{2},4,0} =-\frac{1}{4},
\qquad
C_{\frac{3}{2},\frac{9}{2},4,0} =-\frac{1}{72}.
\label{Ccoeff}
\eea
The previous relation (\ref{TPhi}) can be obtained 
because $B_{2,i,0,0}=1$, $ C_{2,i,i,0}=-\frac{3}{2h_i(2h_i+1)}$ and $C_{2ii}=h_i$(
the second-order pole of OPE $\Phi_i(z) \; T(w)$ is given by
$\frac{1}{(z-w)^2} h_i \Phi_i(w)$ and there is also first-order singular term).

What happens when the spin-$\frac{3}{2}$ current $G(z)$ is combined 
with any primary field $\Phi_i(z)$ of spin-$h_i$?
As done in (\ref{Bcoeff}) and (\ref{Ccoeff}), 
 $B_{\frac{3}{2},i,0,0}=1$, $ C_{\frac{3}{2},i,i-\frac{1}{2},0}=-\frac{1}{4h_i(h_i-\frac{1}{2})}$ and $C_{2,i,i-\frac{1}{2}}=
2(h_i-\frac{1}{2})$ (when 
the second-order pole of OPE $\Phi_i(z) \; G(w)$ is given by
$\frac{1}{(z-w)^2} 2 (h_i-\frac{1}{2}) \Phi_{i-\frac{1}{2}}(w)$ and 
there exists a
first-order singular term),
which is similar  to (\ref{TPhi}).
Furthermore, when 
 the first-order pole of OPE $\Phi_i(z) \; G(w)$ is given by
$\frac{1}{(z-w)}  \Phi_{i+\frac{1}{2}}(w)$, 
 $C_{2,i,i+\frac{1}{2}}=1$ with 
 $B_{\frac{3}{2},i,0,0}=1$ and $ C_{\frac{3}{2},i,i+\frac{1}{2},0}=-\frac{1}{(h_i+\frac{1}{2})}$.
From these two cases, the following can be derived
\bea
\left( G \Phi_i -\frac{1}{2h_i} \pa^2 \Phi_{i-\frac{1}{2}} \right)(z), 
\qquad \mbox{or} \qquad
\left(G \Phi_i -\frac{1}{(h_i +\frac{1}{2})} \pa \Phi_{i+\frac{1}{2}} 
\right)(z).
\label{gope}
\eea
Therefore, all the quasi-primary fields containing 
$T(z)$  or $G(z)$ in this paper can be classified by (\ref{TPhi}) 
and (\ref{gope}). 

The mixed form of (\ref{gope}) should be used for
the case, $\Phi_i(z) =P_{4'}^{uu}(z)$ or $\Phi_i(z) = 
P_{4'}^{ww}(z)$. Note that the explicit OPEs are given in  
footnote \ref{unusual}.
The original expression (\ref{BC}) should be used
when the field, $\Phi_i(z)$ or $\Phi_j(w)$, 
does not contain $T(z)$ or $G(z)$. This will occur when 
the OPEs between the 
higher spin currents not considered in this paper are calculated.

\section{The OPE between the spin-$\frac{3}{2}$ current and the quasiprimary
or primary field  fields in $c=4$ model}

For the correct superpartner in the given superfield, 
it is important to know the OPE between the spin-$\frac{3}{2}$ current 
and arbitrary quasi-primary fields
\bea
&& G(z) \; \left(T T -\frac{3}{10} \pa^2 T \right)(w) = \frac{1}{(z-w)^4} \, \frac{51}{20} \, G(w) 
+\frac{1}{(z-w)^3} \, (-\frac{1}{3}) \frac{51}{20} \pa G(w)  +{\cal O}((z-w)^{-2}), 
\nonu \\
&& G(z) \; \left(G T -\frac{1}{8} \pa^2 G \right)(w) = \frac{1}{(z-w)^3} \, \frac{37}{6} \,
T(w) + {\cal O}((z-w)^{-2}),
\nonu \\
&& G(z) \; \left(G \pa T -\frac{4}{3} T \pa G + \frac{4}{15} \pa^3 G \right)(w) =
-\frac{1}{(z-w)^4} \, \frac{44}{5} \, T(w) 
\nonu \\
&& -\frac{1}{(z-w)^3} \, (-\frac{1}{4}) \frac{44}{5} \, \pa T(w) + {\cal O}((z-w)^{-2}),
\nonu \\
&& G(z) \; \left(G U -\frac{\sqrt{6}}{3} \pa W \right)(w)  = \frac{1}{(z-w)^3} \, \frac{13}{3} \,
U(w) +{\cal O}((z-w)^{-2}),
\nonu \\
&& G(z) \; \left( T W -\frac{3}{14} \pa^2 W \right)(w)   
=  \frac{1}{(z-w)^4} \, \frac{155}{14\sqrt{6}} \, U(w)
\nonu \\
&& +\frac{1}{(z-w)^3} \, (-\frac{1}{5}) \frac{155}{14\sqrt{6}} \pa U(w) +{\cal O}((z-w)^{-2}),
\nonu \\
&& G(z) \; \left( G \pa U -\frac{5}{3} \pa G U -\frac{\sqrt{6}}{7} \pa^2 W \right)(w)  = 
-\frac{1}{(z-w)^4} \, \frac{335}{21} \, U(w)
\nonu \\
&& -\frac{1}{(z-w)^3} \, (-\frac{1}{5}) \frac{335}{21} \pa U(w) +{\cal O}((z-w)^{-2}),
\nonu \\
&& G(z) \;  \left(G W -\frac{1}{6\sqrt{6}} \pa^2 U \right)(w)  = 
\frac{1}{(z-w)^3} \, \frac{25}{3}  W(w) +{\cal O}((z-w)^{-2}),
\nonu \\
&& G(z) \; \left(T U -\frac{1}{4} \pa^2 U \right)(w)  = 
\frac{1}{(z-w)^3} \, 2\sqrt{6} W(w)
+{\cal O}((z-w)^{-2}),
\nonu \\
&& G(z) \; \left( T O_{\frac{7}{2}} -\frac{3}{16} \pa^2 O_{\frac{7}{2}} \right)(w)  
=\frac{1}{(z-w)^3} \, \left[ -\frac{17}{8\sqrt{6}} P_{4'}^{uu} +\frac{17}{4} \sqrt{\frac{3}{2}} 
P_{4'}^{ww}\right](w) +{\cal O}((z-w)^{-2}),
\nonu \\
&& G(z) \; \left( G P_{4'}^{uu} -\frac{4\sqrt{6}}{9} \pa O_{\frac{9}{2}}-\frac{\sqrt{6}}{56} \pa^2 
O_{\frac{7}{2}} \right)(w)  = \nonu \\
&& 
\frac{1}{(z-w)^3} \, \left[ \frac{239}{36} P_{4'}^{uu} +\frac{17}{6}  
P_{4'}^{ww} \right](w) +{\cal O}((z-w)^{-2}),
\nonu \\
&& G(z) \; \left( G P_{4'}^{ww} -\frac{2}{9} \sqrt{\frac{2}{3}} \pa O_{\frac{9}{2}}
-\frac{1}{14} \sqrt{\frac{2}{3}} \pa^2 O_{\frac{7}{2}} \right) (w)
 = \nonu \\
&& \frac{1}{(z-w)^3} \, \left[ -\frac{17}{27} P_{4'}^{uu} +\frac{98}{9} 
P_{4'}^{ww} \right](w) +{\cal O}((z-w)^{-2}),
\nonu \\
&& G(z) \; \left( G O_{\frac{7}{2}} +\frac{1}{4\sqrt{6}} \pa P_{4'}^{uu}
-\frac{\sqrt{6}}{4} \pa P_{4'}^{ww} \right)(w)  =  
 \frac{1}{(z-w)^3} \, \frac{37}{6} \, O_{\frac{7}{2}}(w) +{\cal O}((z-w)^{-2}),
\nonu \\
&& G(z) \; \left( T P_{4'}^{uu} -\frac{1}{6} \pa^2 P_{4'}^{uu} \right)(w)  = 
\frac{1}{(z-w)^4} \, 5 \sqrt{\frac{3}{2}} \, O_{\frac{7}{2}} (w) \nonu \\
&& + 
 \frac{1}{(z-w)^3} \, \left[ (-\frac{1}{7}) 5\sqrt{\frac{3}{2}}\, 
\pa O_{\frac{7}{2}} + 13 \sqrt{\frac{2}{3}} O_{\frac{9}{2}} \right] (w) +{\cal O}((z-w)^{-2}),
\nonu \\
&& G(z) \; \left( T P_{4'}^{ww} -\frac{1}{6} \pa^2 P_{4'}^{ww} \right)(w)  =
\frac{1}{(z-w)^4} \, 10 \sqrt{\frac{3}{2}} \, O_{\frac{7}{2}} (w) \nonu \\
&& + 
 \frac{1}{(z-w)^3} \, \left[ (-\frac{1}{7}) 10 \sqrt{\frac{3}{2}}\, 
\pa O_{\frac{7}{2}} + \frac{13}{3\sqrt{6}} O_{\frac{9}{2}} \right] (w) +{\cal O}((z-w)^{-2}),
\nonu \\
&& G(z) \; \left( G \pa O_{\frac{7}{2}} -\frac{7}{3} \pa G O_{\frac{7}{2}} +
\frac{1}{9\sqrt{6}} \pa^2 P_{4'}^{uu} -\frac{1}{3} \sqrt{\frac{2}{3}} \pa^2 
P_{4'}^{ww} \right)(w)
= \nonu \\
&& -\frac{1}{(z-w)^4} \, \frac{77}{3} \, O_{\frac{7}{2}} (w) 
-
 \frac{1}{(z-w)^3} \, (-\frac{1}{7}) \frac{77}{3} \, 
\pa O_{\frac{7}{2}} (w) +{\cal O}((z-w)^{-2}),
\nonu \\
&& G(z) \; \left( G O_{\frac{9}{2}} -\frac{2}{63} \sqrt{\frac{2}{3}} \pa^2 P_{4'}^{uu} +
\frac{1}{21\sqrt{6}} \pa^2 P_{4'}^{ww} \right)(w)  = 
 \frac{1}{(z-w)^3} \, \frac{103}{9} \, 
O_{\frac{9}{2}} (w) +{\cal O}((z-w)^{-2}),
\nonu \\
&& G(z) \; \left( T O_{\frac{9}{2}} -\frac{3}{20} \pa^2 O_{\frac{9}{2}} \right)(w)  =  
 \frac{1}{(z-w)^4} \, \left[ \frac{208}{35} \sqrt{\frac{2}{3}} P_{4'}^{uu} -\frac{52}{35} 
\sqrt{6} \, P_{4'}^{ww} \right](w) \nonu \\
&& + \frac{1}{(z-w)^3} \, (-\frac{1}{8}) \pa  \left[ \frac{208}{35} \sqrt{\frac{2}{3}} P_{4'}^{uu} 
-\frac{52}{35} 
\sqrt{6} \, P_{4'}^{ww} \right](w) + {\cal O}((z-w)^{-2}),
\nonu \\
&& G(z) \; \left( T \pa O_{\frac{7}{2}} -\frac{7}{4} \pa T O_{\frac{7}{2}}-\frac{1}{9} \pa^3 O_{\frac{7}{2}} 
\right)(w)  =  
 \frac{1}{(z-w)^4} \, \left[ \frac{25}{6\sqrt{6}} \, P_{4'}^{uu} -\frac{25}{\sqrt{6}} 
 \, P_{4'}^{ww} \right](w) \nonu \\
&& +\frac{1}{(z-w)^3} \, 
 \left[(-\frac{1}{8}) \frac{25}{6\sqrt{6}} \, \pa P_{4'}^{uu} 
-(-\frac{1}{8}) \frac{25}{\sqrt{6}} 
 \, \pa P_{4'}^{ww} \right. \nonu \\
&& \left. -\frac{21}{4} \left( G O_{\frac{7}{2}}
+\frac{1}{4\sqrt{6}} \, \pa P_{4'}^{uu} - \frac{\sqrt{6}}{4} \, 
 \, \pa P_{4'}^{ww} \right) \right](w) +{\cal O}((z-w)^{-2}),
\nonu \\
&& G(z) \; \left( G \pa P_{4'}^{uu} -\frac{8}{3} \pa G P_{4'}^{uu} -\frac{\sqrt{6}}{5} \pa^2 O_{\frac{9}{2}} -
\frac{2}{63}\sqrt{\frac{2}{3}} \pa^3 O_{\frac{7}{2}} \right)(w)  = 
 \nonu \\
&& \frac{1}{(z-w)^4} \, \left[ -\frac{1444}{45} \, P_{4'}^{uu} +\frac{56}{15} 
 \, P_{4'}^{ww} \right](w)  +\frac{1}{(z-w)^3} \, 
 \left[ -(-\frac{1}{8}) \frac{1444}{45} \, 
\pa P_{4'}^{uu} + (-\frac{1}{8}) \frac{56}{15} \, 
 \, \pa P_{4'}^{ww} \right. \nonu \\
&& \left. -2 \sqrt{6} \, \left( G O_{\frac{7}{2}}  
+\frac{1}{4\sqrt{6}} \, \pa P_{4'}^{uu} - \frac{\sqrt{6}}{4} \, 
 \, \pa P_{4'}^{ww} \right) \right](w) +{\cal O}((z-w)^{-2}),
\nonu \\
&& G(z) \; \left( G \pa P_{4'}^{ww} -\frac{8}{3} \pa G P_{4'}^{ww} -\frac{1}{5\sqrt{6}}
\pa^2 O_{\frac{9}{2}} -\frac{8}{189} \sqrt{\frac{2}{3}} \pa^3 O_{\frac{7}{2}} \right)(w)
 =  \nonu \\
&& \frac{1}{(z-w)^4} \, \left[ -\frac{112}{135} \, P_{4'}^{uu} -\frac{1192}{45} 
 \, P_{4'}^{ww} \right](w) 
 +\frac{1}{(z-w)^3} \, 
 \left[ -(-\frac{1}{8}) \frac{112}{135} \, \pa P_{4'}^{uu} -(-\frac{1}{8}) \frac{1192}{45} \, 
 \, \pa P_{4'}^{ww} \right. \nonu \\
&& \left. -8 \sqrt{\frac{2}{3}} \, \left( G O_{\frac{7}{2}}  
+\frac{1}{4\sqrt{6}} \, \pa P_{4'}^{uu} - \frac{\sqrt{6}}{4} \, 
 \, \pa P_{4'}^{ww} \right) \right](w) +{\cal O}((z-w)^{-2}),
\nonu \\
&& G(z) \; \left( T \pa U -\frac{5}{4} \pa T U -\frac{1}{7} \pa^3 U \right)(w) =
-\frac{1}{(z-w)^4} \, \frac{33}{7} \sqrt{\frac{3}{2}} \, W(w) \nonu \\
&& +\frac{1}{(z-w)^3} \left[ -(-\frac{1}{6}) \frac{33}{7} \sqrt{\frac{3}{2}} \pa W -\frac{15}{4}
\left( G U -\frac{\sqrt{6}}{3} \pa W \right) \right] (w) +  {\cal O}((z-w)^{-2}),
\nonu \\  
&& G(z) \; \left( G \pa W - 2 \pa G W -\frac{1}{21} \sqrt{\frac{2}{3}} \pa^3 U \right)(w)
= -\frac{1}{(z-w)^4} \, \frac{116}{7} \,
 W(w) \nonu \\
&& + \frac{1}{(z-w)^3} \left[- (-\frac{1}{6}) 
\frac{116}{7} \pa W -5\sqrt{\frac{2}{3}} \left( G U -\frac{\sqrt{6}}{3} \pa W 
\right)\right](w) + {\cal O}((z-w)^{-2}),    
\nonu \\
&& G(z) \; \left( G O_{4''} -\frac{2}{9} \pa O_{\frac{9}{2}'} \right)(w)
=\frac{1}{(z-w)^3} \frac{64}{9} O_{4''}(w) +  {\cal O}((z-w)^{-2}),
\nonu \\
&& G(z) \; \left( G \pa^2 U - 4 \pa G \pa U +\frac{5}{2} \pa^2 G U -\frac{1}{2\sqrt{6}} \pa^3 W
 \right)(w) =\frac{1}{(z-w)^5} \, 85 \, U(w)
\nonu \\
&& +\frac{1}{(z-w)^4} (-\frac{2}{5}) 85 \pa U(w) \nonu \\
&& +\frac{1}{(z-w)^3} 
\left[  (\frac{1}{30}) 85 \pa^2 U - 2\sqrt{6} 
\left( G W -\frac{1}{6\sqrt{6}} \pa^2 U \right) + 10 \left( T U -\frac{1}{4} \pa^2 U  
\right)  \right](w) 
+  {\cal O}((z-w)^{-2}),
\nonu \\
&& G(z) \; \left( T \pa W -\frac{3}{2} \pa T W -\frac{1}{8} \pa^3 W \right)(w)
=-\frac{1}{(z-w)^5} \, \frac{5}{2} \sqrt{\frac{3}{2}} \, U(w) 
-\frac{1}{(z-w)^4} (-\frac{2}{5}) \frac{5}{2} \sqrt{\frac{3}{2}} \pa U(w) 
\nonu \\
&& +  \frac{1}{(z-w)^3} \left[- (\frac{1}{30}) 
\frac{5}{2} \sqrt{\frac{3}{2}} \pa^2 U  -\frac{9}{2} \left( G W -\frac{1}{6\sqrt{6}} \pa^2 U \right)
+5\sqrt{\frac{2}{3}} \left( T U -\frac{1}{4} \pa^2 U  \right)
\right] (w) 
\nonu \\
&& + {\cal O}((z-w)^{-2}),
\nonu \\
&& G(z) \; \left( T O_{4''} -\frac{1}{6} \pa^2 O_{4''} \right)(w)
= \frac{1}{(z-w)^3} \, \frac{13}{6} O_{\frac{9}{2}'}(w) + {\cal O}((z-w)^{-2}),
\nonu \\
&& G(z) \; \left( G O_{\frac{9}{2}'} -\frac{1}{9} \pa^2 O_{4''} \right)(w)
= \frac{1}{(z-w)^3} \, \frac{103}{9} O_{\frac{9}{2}'}(w) + {\cal O}((z-w)^{-2}).
\label{stresswithquasi1}
\eea

Using the definition of the coefficients (\ref{Coeff1}),
those vanishing and nonvanishing 
coefficients appearing in (\ref{stresswithquasi1}) can be checked
as follows:
\bea
A_{\frac{3}{2},4,\frac{3}{2},1}  & = &  -\frac{1}{3}, 
\qquad
A_{\frac{3}{2},4,\frac{3}{2},2}  =  0, 
\qquad
A_{\frac{3}{2},\frac{9}{2},2,1}   =    -\frac{1}{4}, 
\qquad
A_{\frac{3}{2},5,\frac{5}{2},1}   =   -\frac{1}{5}, 
\nonu \\
A_{\frac{3}{2},\frac{11}{2},3,1} & = & -\frac{1}{6},
\qquad
A_{\frac{3}{2},6, \frac{5}{2}, 1} =-\frac{2}{5},
\qquad
A_{\frac{3}{2},6, \frac{5}{2}, 2}  =  \frac{1}{30},
\qquad
A_{\frac{3}{2},6,\frac{7}{2},1}   =   -\frac{1}{7}, 
\nonu \\
A_{\frac{3}{2},\frac{13}{2},4,1}   & = &   -\frac{1}{8}.
\label{CoeffA1}
\eea
For example, the first OPE in (\ref{stresswithquasi1}) has a 
third-order singular term with a coefficient 
$-\frac{1}{3}$, which coincides with the value, 
$A_{\frac{3}{2},4,\frac{3}{2},1}   =   -\frac{1}{3}$.

\section{The OPE between the stress energy tensor and the quasi-primary
or primary   fields in the $c<4$ coset model}

In Appendices $B$ and $C$, the central charge $c$ was fixed to $c=4$.
Now this section considers 
the OPEs for general central charge $c<4$. The OPEs are given as follows:
\bea
&& T(z) \; \left(T T -\frac{3}{10} \pa^2 T \right)(w) = \frac{1}{(z-w)^4} \left[ \frac{6 (66+63 k+7 k^2
)}{5 (3+k) (6+k)}  \right] T(w) 
+{\cal O}((z-w)^{-2}), 
\nonu \\
&& T(z) \; \left(G T -\frac{1}{8} \pa^2 G \right)(w) = \frac{1}{(z-w)^4} \left[ 
\frac{378+333 k+37 k^2}{8 (3+k) (6+k)}  \right]
G(w) + {\cal O}((z-w)^{-2}),
\nonu \\
&& T(z) \; \left(G \pa T -\frac{4}{3} T \pa G + \frac{4}{15} \pa^3 G \right)(w) =
\frac{1}{(z-w)^5} \left[ \frac{3 (-42+99 k+11 k^2)}{5 (3+k) (6+k)}
 \right] G(w) 
\nonu \\
&& +\frac{1}{(z-w)^4} \, (-\frac{1}{3})  
\left[\frac{3 (-42+99 k+11 k^2)}{5 (3+k) (6+k)} \right] \pa G(w) +{\cal O}((z-w)^{-2}),
\nonu \\
&& T(z) \; \left(G U -\frac{\sqrt{6}}{3} \pa W \right)(w)  = {\cal O}((z-w)^{-2}),
\nonu \\
&& T(z) \; \left( T W -\frac{3}{14} \pa^2 W \right)(w)   =  \frac{1}{(z-w)^4} \left[ 
\frac{1026+639 k+71 k^2}{7 (3+k) (6+k)} \right] W(w) 
+{\cal O}((z-w)^{-2}), 
\nonu \\
&& T(z) \; \left( G \pa U -\frac{5}{3} \pa G U -\frac{\sqrt{6}}{7} \pa^2 W \right)(w)  = 
-\frac{1}{(z-w)^4} \, \frac{124}{7} \, \sqrt{\frac{2}{3}}  \, W(w) +{\cal O}((z-w)^{-2}), 
\nonu \\
&& T(z) \;  \left(G W -\frac{1}{6\sqrt{6}} \pa^2 U \right)(w)  =  \frac{1}{(z-w)^4} \, 
5 \sqrt{\frac{2}{3}}  \, U(w) +{\cal O}((z-w)^{-2}),
\nonu \\
&& T(z) \; \left(T U -\frac{1}{4} \pa^2 U \right)(w)  = 
 \frac{1}{(z-w)^4} \left[ \frac{3(150+99 k+11 k^2)}{4 (3+k) (6+k)} 
  \right] U(w) +{\cal O}((z-w)^{-2}),
\nonu \\
&& T(z) \; \left( T O_{\frac{7}{2}} -\frac{3}{16} \pa^2 O_{\frac{7}{2}} \right)(w)  
= \frac{1}{(z-w)^4} \left[ \frac{2898+1737 k+193 k^2}{16 (3+k) (6+k)}
\right] O_{\frac{7}{2}}(w) +{\cal O}((z-w)^{-2}),
\nonu \\
&& T(z) \; \left( G P_{4'}^{uu} -\frac{4\sqrt{6}}{9} \pa O_{\frac{9}{2}}-\frac{\sqrt{6}}{56} \pa^2 
O_{\frac{7}{2}} \right)(w)  = \frac{1}{(z-w)^4} \, 
\frac{17}{4}\, \sqrt{\frac{3}{2}}  \, O_{\frac{7}{2}}(w) +{\cal O}((z-w)^{-2}),
\nonu \\
&& T(z) \; \left( G P_{4'}^{ww} -\frac{2}{9} \sqrt{\frac{2}{3}} \pa O_{\frac{9}{2}}
-\frac{1}{14} \sqrt{\frac{2}{3}} \pa^2 O_{\frac{7}{2}} \right) (w)
 =  \frac{1}{(z-w)^4} \, 
\frac{17}{\sqrt{6}}  \, O_{\frac{7}{2}}(w) +{\cal O}((z-w)^{-2}),
\nonu \\
&& T(z) \; \left( G O_{\frac{7}{2}} +\frac{1}{4\sqrt{6}} \pa P_{4'}^{uu}
-\frac{\sqrt{6}}{4} \pa P_{4'}^{ww} \right)(w)  = +{\cal O}((z-w)^{-2}),
\nonu \\
&& T(z) \; \left( T P_{4'}^{uu} -\frac{1}{6} \pa^2 P_{4'}^{uu} \right)(w)  = 
 \frac{1}{(z-w)^4} \left[ \frac{2(108+63k+7k^2)}{(3+k)(6+k)} \right] P_{4'}^{uu}(w) +{\cal O}((z-w)^{-2}),
\nonu \\
&& T(z) \; \left( T P_{4'}^{ww} -\frac{1}{6} \pa^2 P_{4'}^{ww} \right)(w)  = 
 \frac{1}{(z-w)^4}   \left[ \frac{2(108+63k+7k^2)}{(3+k)(6+k)} \right] P_{4'}^{ww}(w) +{\cal O}((z-w)^{-2}),
\nonu \\
&& T(z) \; \left( G \pa O_{\frac{7}{2}} -\frac{7}{3} \pa G O_{\frac{7}{2}} +
\frac{1}{9\sqrt{6}} \pa^2 P_{4'}^{uu} -\frac{1}{3} \sqrt{\frac{2}{3}} \pa^2 P_{4'}^{ww} \right)(w)
 =  \nonu \\
&& \frac{1}{(z-w)^4} \, \left[ \frac{25}{3\sqrt{6}} \, P_{4'}^{uu} -25\sqrt{\frac{2}{3}} 
P_{4'}^{ww}\right] (w) +{\cal O}((z-w)^{-2}),
\nonu \\
&& T(z) \; \left( G O_{\frac{9}{2}} -\frac{2}{63} \sqrt{\frac{2}{3}} \pa^2 P_{4'}^{uu} +
\frac{1}{21\sqrt{6}} \pa^2 P_{4'}^{ww} \right)(w)  = 
 \nonu \\
&& \frac{1}{(z-w)^4} \, \left[ \frac{104}{21} \sqrt{\frac{2}{3}} \, 
P_{4'}^{uu} -\frac{26}{7}\sqrt{\frac{2}{3}} 
P_{4'}^{ww}\right] (w) +{\cal O}((z-w)^{-2}),
\nonu \\
&& T(z) \; \left( T O_{\frac{9}{2}} -\frac{3}{20} \pa^2 O_{\frac{9}{2}} \right)(w)  = 
 \frac{1}{(z-w)^4} \left[ \frac{(5022+2871k+319k^2)}{20(3+k)(6+k)} 
\right] O_{\frac{9}{2}} (w) +{\cal O}((z-w)^{-2}),
\nonu \\
&& T(z) \; \left( 
T \pa O_{\frac{7}{2}} -\frac{7}{4} \pa T O_{\frac{7}{2}}-\frac{1}{9} \pa^3 O_{\frac{7}{2}} \right)(w)  = 
-\frac{1}{(z-w)^5} \left[ \frac{7(342+387k+43k^2)}{12(3+k)(6+k)} 
\right] O_{\frac{7}{2}} (w) \nonu \\
&& -
 \frac{1}{(z-w)^4} \, (-\frac{1}{7}) \left[ 
 \frac{7(342+387k+43k^2)}{12(3+k)(6+k)} 
\right] \pa O_{\frac{7}{2}}(w) +{\cal O}((z-w)^{-2}),  
\nonu \\
&& T(z) \; \left( G \pa P_{4'}^{uu} -\frac{8}{3} \pa G P_{4'}^{uu} -\frac{\sqrt{6}}{5} \pa^2 O_{\frac{9}{2}} -
\frac{2}{63}\sqrt{\frac{2}{3}} \pa^3 O_{\frac{7}{2}} \right)(w)  = 
\nonu \\
&& -\frac{1}{(z-w)^5} \, \frac{50}{3} \, \sqrt{\frac{2}{3}} \, O_{\frac{7}{2}} (w)  
+\frac{1}{(z-w)^4} \left[ -(-\frac{1}{7}) \frac{50}{3} 
\sqrt{\frac{2}{3}} \, \pa O_{\frac{7}{2}} -
\frac{286}{5} \sqrt{\frac{2}{3}} \, O_{\frac{9}{2}} \right] (w) +{\cal O}((z-w)^{-2}),
\nonu \\
&& T(z) \; \left( G \pa P_{4'}^{ww} -\frac{8}{3} \pa G P_{4'}^{ww} -\frac{1}{5\sqrt{6}}
\pa^2 O_{\frac{9}{2}} -\frac{8}{189} \sqrt{\frac{2}{3}} \pa^3 O_{\frac{7}{2}} \right)(w)
 = \nonu \\
&& -\frac{1}{(z-w)^5} \, \frac{200}{9} \, \sqrt{\frac{2}{3}} \, O_{\frac{7}{2}} (w) 
+\frac{1}{(z-w)^4} \left[ -(-\frac{1}{7}) \frac{200}{9} \sqrt{\frac{2}{3}} \, \pa O_{\frac{7}{2}} -
\frac{143}{15} \sqrt{\frac{2}{3}} \, O_{\frac{9}{2}} \right] (w) +{\cal O}((z-w)^{-2}),
\nonu \\
&& T(z) \; \left( T \pa U -\frac{5}{4} \pa T U -\frac{1}{7} \pa^3 U \right)(w) =
-\frac{1}{(z-w)^5} \left[ \frac{15(78+207k+23k^2)}{28(3+k)(6+k)} \right] U(w) \nonu \\
&& -\frac{1}{(z-w)^4} \, (-\frac{1}{5}) \left[
 \frac{15(78+207k+23k^2)}{28(3+k)(6+k)} \right] \pa U(w) +  {\cal O}((z-w)^{-2}),
\nonu \\  
&& T(z) \; \left( G \pa W - 2 \pa G W -\frac{1}{21} \sqrt{\frac{2}{3}} \pa^3 U \right)(w)
= -\frac{1}{(z-w)^5} \, \frac{55}{7} \sqrt{\frac{2}{3}} \, U(w) \nonu \\
&& -\frac{1}{(z-w)^4} \, (-\frac{1}{5}) \,
\pa U(w) +  {\cal O}((z-w)^{-2}),    
\nonu \\
&& T(z) \; \left( G O_{4''} -\frac{2}{9} \pa O_{\frac{9}{2}'} \right)(w)
=+  {\cal O}((z-w)^{-2}),
\nonu \\
&& T(z) \; \left( G \pa^2 U - 4 \pa G \pa U +\frac{5}{2} \pa^2 G U -\frac{1}{2\sqrt{6}} \pa^3 W
 \right)(w) =\frac{1}{(z-w)^5} \, 9 \sqrt{6} \, W(w)
\nonu \\
&& +\frac{1}{(z-w)^4} \left[ -3 \sqrt{\frac{3}{2}} \pa W +\frac{75}{2} 
\left( G U -\frac{\sqrt{6}}{3} \pa W \right) \right](w) 
+  {\cal O}((z-w)^{-2}),
\nonu \\
&& T(z) \; \left( T \pa W -\frac{3}{2} \pa T W -\frac{1}{8} \pa^3 W \right)(w)
=-\frac{1}{(z-w)^5} \left[ \frac{18(6+9k+k^2)}{(3+k)(6+k)} \right] W(w) \nonu \\
&& -\frac{1}{(z-w)^4} (-\frac{1}{6}) 
\left[\frac{18(6+9k+k^2)}{(3+k)(6+k)} \right]
 \pa W(w)  +  {\cal O}((z-w)^{-2}),
\nonu \\
&& T(z) \; \left( T O_{4''} -\frac{1}{6} \pa^2 O_{4''} \right)(w)
= \frac{1}{(z-w)^4} \left[ \frac{2(108+63k+7k^2)}{(3+k)(6+k)} \right] 
O_{4''}(w) + {\cal O}((z-w)^{-2}), \nonu \\
&& T(z) \; \left( G O_{\frac{9}{2}'} -\frac{1}{9} \pa^2 O_{4''} \right)(w)
= \frac{1}{(z-w)^4} \, \frac{52}{3} O_{4''}(w) + {\cal O}((z-w)^{-2}).
\label{opetwithothers}
\eea
Of course, after the $k \rightarrow \infty$ limit, 
the above OPEs (\ref{opetwithothers}) become OPEs (\ref{stresswithquasi}).  
The $k$-dependence in the quasi-primary fields that contain the $T(w)$
can be derived 
because the OPE between $T(z)$ and $T(w)$ contains the central charge.
On the other hand, the 
OPEs between the $T(z)$ and quasi-primary fields which do not have 
$T(w)$ in their expression are the same as those for the $c=4$ model. 
In this case, the previous coefficients (\ref{CoeffA}) hold.

\section{The OPE between the spin-$\frac{3}{2}$ current and the quasi-primary
or primary   fields in the $c < 4$ coset model}

Similarly, the general central charge and OPEs are given by 
\bea
&& G(z) \; \left(T T -\frac{3}{10} \pa^2 T \right)(w) = \frac{1}{(z-w)^4} \, \frac{51}{20} \, G(w) 
+\frac{1}{(z-w)^3} \, (-\frac{1}{3}) \frac{51}{20} \pa G(w)  +{\cal O}((z-w)^{-2}), 
\nonu \\
&& G(z) \; \left(G T -\frac{1}{8} \pa^2 G \right)(w) = \frac{1}{(z-w)^3} \left[ 
\frac{378+333 k+37 k^2}{6 (3+k) (6+k)}
 \right]
T(w) + {\cal O}((z-w)^{-2}),
\nonu \\
&& G(z) \; \left(G \pa T -\frac{4}{3} T \pa G + \frac{4}{15} \pa^3 G \right)(w) =
-\frac{1}{(z-w)^4} \left[ 
\frac{4 (-42+99 k+11 k^2)}{5 (3+k) (6+k)}
 \right] T(w) 
\nonu \\
&& -\frac{1}{(z-w)^3} \, (-\frac{1}{4}) 
\left[\frac{4 (-42+99 k+11 k^2)}{5 (3+k) (6+k)}
 \right] \pa T(w) + {\cal O}((z-w)^{-2}),
\nonu \\
&& G(z) \; \left(G U -\frac{\sqrt{6}}{3} \pa W \right)(w)  = \frac{1}{(z-w)^3} \left[ 
\frac{90+117k+13k^2}{3(3+k)(6+k)} \right]
U(w) +{\cal O}((z-w)^{-2}),
\nonu \\
&& G(z) \; \left( T W -\frac{3}{14} \pa^2 W \right)(w)   
=  \frac{1}{(z-w)^4} \, \frac{155}{14\sqrt{6}} \, U(w)
\nonu \\
&& +\frac{1}{(z-w)^3} \, (-\frac{1}{5}) \frac{155}{14\sqrt{6}} \pa U(w) +{\cal O}((z-w)^{-2}),
\nonu \\
&& G(z) \; \left( G \pa U -\frac{5}{3} \pa G U -\frac{\sqrt{6}}{7} \pa^2 W \right)(w)  = 
-\frac{1}{(z-w)^4} \left[ \frac{5 (198+603 k+67 k^2)}{21 (3+k) (6+k)} 
\right] U(w)
\nonu \\
&& -\frac{1}{(z-w)^3} \, (-\frac{1}{5}) 
\left[\frac{5 (198+603 k+67 k^2)}{21 (3+k) (6+k)}  \right]
\pa U(w) +{\cal O}((z-w)^{-2}),
\nonu \\
&& G(z) \;  \left(G W -\frac{1}{6\sqrt{6}} \pa^2 U \right)(w)  = 
\frac{1}{(z-w)^3} \left[ \frac{306+225 k+25 k^2}{3 (3+k) (6+k)}  
\right] W(w) +{\cal O}((z-w)^{-2}),
\nonu \\
&& G(z) \; \left(T U -\frac{1}{4} \pa^2 U \right)(w)  = 
\frac{1}{(z-w)^3} \,  2\sqrt{6} W(w) +{\cal O}((z-w)^{-2}),
\nonu \\
&& G(z) \; \left( T O_{\frac{7}{2}} -\frac{3}{16} \pa^2 O_{\frac{7}{2}} \right)(w)  
=\frac{1}{(z-w)^3} \, 
 \left[ -\frac{17}{8\sqrt{6}} P_{4'}^{uu} +\frac{17}{4} \sqrt{\frac{3}{2}} 
P_{4'}^{ww}\right](w)
+{\cal O}((z-w)^{-2}),
\nonu \\
&& G(z) \; \left( G P_{4'}^{uu} -\frac{4\sqrt{6}}{9} \pa O_{\frac{9}{2}}-\frac{\sqrt{6}}{56} \pa^2 
O_{\frac{7}{2}} \right)(w)  = \nonu \\
&& 
\frac{1}{(z-w)^3} \, \left[ \frac{(2574+2151k+239k^2)}{36(3+k)(6+k)}  P_{4'}^{uu} +\frac{17}{6}  
P_{4'}^{ww} \right](w) +{\cal O}((z-w)^{-2}),
\nonu \\
&& G(z) \; \left( G P_{4'}^{ww} -\frac{2}{9} \sqrt{\frac{2}{3}} \pa O_{\frac{9}{2}}
-\frac{1}{14} \sqrt{\frac{2}{3}} \pa^2 O_{\frac{7}{2}} \right) (w)
 = \nonu \\
&& \frac{1}{(z-w)^3} \, \left[ -\frac{17}{27} P_{4'}^{uu} +\frac{2(666+441k+49k^2)}{9(3+k)(6+k)} 
P_{4'}^{ww} \right](w) +{\cal O}((z-w)^{-2}),
\nonu \\
&& G(z) \; \left( G O_{\frac{7}{2}} +\frac{1}{4\sqrt{6}} \pa P_{4'}^{uu}
-\frac{\sqrt{6}}{4} \pa P_{4'}^{ww} \right)(w)  =  
 \frac{1}{(z-w)^3} \left[ \frac{378+333k+37k^2}{6(3+k)(6+k)} 
\right] O_{\frac{7}{2}}(w) \nonu \\
&& +{\cal O}((z-w)^{-2}),
\nonu \\
&& G(z) \; \left( T P_{4'}^{uu} -\frac{1}{6} \pa^2 P_{4'}^{uu} \right)(w)  = 
\frac{1}{(z-w)^4} \, 5 \sqrt{\frac{3}{2}} \, O_{\frac{7}{2}} (w) \nonu \\
&& + 
 \frac{1}{(z-w)^3} \, \left[ (-\frac{1}{7}) 5\sqrt{\frac{3}{2}}\, 
\pa O_{\frac{7}{2}} + 13 \sqrt{\frac{2}{3}} O_{\frac{9}{2}} \right] (w) +{\cal O}((z-w)^{-2}),
\nonu \\
&& G(z) \; \left( T P_{4'}^{ww} -\frac{1}{6} \pa^2 P_{4'}^{ww} \right)(w)  =
\frac{1}{(z-w)^4} \, 10 \sqrt{\frac{3}{2}} \, O_{\frac{7}{2}} (w) \nonu \\
&& + 
 \frac{1}{(z-w)^3} \, \left[ (-\frac{1}{7}) 10 \sqrt{\frac{3}{2}}\, 
\pa O_{\frac{7}{2}} + \frac{13}{3\sqrt{6}} O_{\frac{9}{2}} \right] (w) +{\cal O}((z-w)^{-2}),
\nonu \\
&& G(z) \; \left( G \pa O_{\frac{7}{2}} -\frac{7}{3} \pa G O_{\frac{7}{2}} +
\frac{1}{9\sqrt{6}} \pa^2 P_{4'}^{uu} -\frac{1}{3} \sqrt{\frac{2}{3}} \pa^2 
P_{4'}^{ww} \right)(w)
= \nonu \\
&& -\frac{1}{(z-w)^4} \left[ \frac{7(54+99k+11k^2)}{3(3+k)(6+k)} 
\right] O_{\frac{7}{2}} (w) 
-
 \frac{1}{(z-w)^3} \, (-\frac{1}{7})   \left[ \frac{7(54+99k+11k^2)}{3(3+k)(6+k)} 
\right]
\pa O_{\frac{7}{2}} (w) \nonu \\
&& +{\cal O}((z-w)^{-2}),
\nonu \\
&& G(z) \; \left( G O_{\frac{9}{2}} -\frac{2}{63} \sqrt{\frac{2}{3}} \pa^2 P_{4'}^{uu} +
\frac{1}{21\sqrt{6}} \pa^2 P_{4'}^{ww} \right)(w)  = 
 \frac{1}{(z-w)^3} \left[ \frac{(1422+927k+103k^2)}{9(3+k)(6+k)} \right] 
O_{\frac{9}{2}} (w) \nonu \\
&& +{\cal O}((z-w)^{-2}),
\nonu \\
&& G(z) \; \left( T O_{\frac{9}{2}} -\frac{3}{20} \pa^2 O_{\frac{9}{2}} \right)(w)  =  
 \frac{1}{(z-w)^4} \, \left[ \frac{208}{35} \sqrt{\frac{2}{3}} P_{4'}^{uu} -\frac{52}{35} 
\sqrt{6} \, P_{4'}^{ww} \right](w) \nonu \\
&& + \frac{1}{(z-w)^3} \, (-\frac{1}{8}) \pa  \left[ \frac{208}{35} \sqrt{\frac{2}{3}} P_{4'}^{uu} 
-\frac{52}{35} 
\sqrt{6} \, P_{4'}^{ww} \right](w)  + {\cal O}((z-w)^{-2}),
\nonu \\
&& G(z) \; \left( T \pa O_{\frac{7}{2}} -\frac{7}{4} \pa T O_{\frac{7}{2}}-\frac{1}{9} \pa^3 O_{\frac{7}{2}} 
\right)(w)  =  
 \frac{1}{(z-w)^4} \, \left[ \frac{25}{6\sqrt{6}} \, P_{4'}^{uu} -\frac{25}{\sqrt{6}} 
 \, P_{4'}^{ww} \right](w) \nonu \\
&& +\frac{1}{(z-w)^3} \, 
 \left[(-\frac{1}{8}) \frac{25}{6\sqrt{6}} \, \pa P_{4'}^{uu} 
-(-\frac{1}{8}) \frac{25}{\sqrt{6}} 
 \, \pa P_{4'}^{ww} \right. \nonu \\
&& \left. -\frac{21}{4} \left( G O_{\frac{7}{2}}
+\frac{1}{4\sqrt{6}} \, \pa P_{4'}^{uu} - \frac{\sqrt{6}}{4} \, 
 \, \pa P_{4'}^{ww} \right) \right](w) +{\cal O}((z-w)^{-2}),
\nonu \\
&& G(z) \; \left( G \pa P_{4'}^{uu} -\frac{8}{3} \pa G P_{4'}^{uu} -\frac{\sqrt{6}}{5} \pa^2 O_{\frac{9}{2}} -
\frac{2}{63}\sqrt{\frac{2}{3}} \pa^3 O_{\frac{7}{2}} \right)(w)  = 
 \nonu \\
&& \frac{1}{(z-w)^4} \, \left[ -\frac{4(2178+3249k+361k^2)}{45(3+k)(6+k)} \, P_{4'}^{uu} +\frac{56}{15} 
 \, P_{4'}^{ww} \right](w)  \nonu \\
&& +\frac{1}{(z-w)^3} \, 
 \left[ -(-\frac{1}{8})  \, \frac{4(2178+3249k+361k^2)}{45(3+k)(6+k)}
\pa P_{4'}^{uu} + (-\frac{1}{8}) \frac{56}{15} \, 
 \, \pa P_{4'}^{ww} \right. \nonu \\
&& \left. -2 \sqrt{6} \, \left( G O_{\frac{7}{2}}  
+\frac{1}{4\sqrt{6}} \, \pa P_{4'}^{uu} - \frac{\sqrt{6}}{4} \, 
 \, \pa P_{4'}^{ww} \right) \right](w) +{\cal O}((z-w)^{-2}),
\nonu \\
&& G(z) \; \left( G \pa P_{4'}^{ww} -\frac{8}{3} \pa G P_{4'}^{ww} -\frac{1}{5\sqrt{6}}
\pa^2 O_{\frac{9}{2}} -\frac{8}{189} \sqrt{\frac{2}{3}} \pa^3 O_{\frac{7}{2}} \right)(w)
 =  \nonu \\
&& \frac{1}{(z-w)^4} \, \left[ -\frac{112}{135} \, P_{4'}^{uu} -\frac{8(522+1341k+149k^2)}{45(3+k)(6+k)} 
 \, P_{4'}^{ww} \right](w) \nonu \\
&& +\frac{1}{(z-w)^3} \, 
 \left[ -(-\frac{1}{8}) \frac{112}{135} \, \pa P_{4'}^{uu} -(-\frac{1}{8}) \frac{8(522+1341k+149k^2)}{45(3+k)(6+k)}  
 \, \pa P_{4'}^{ww} \right. \nonu \\
&& \left. -8 \sqrt{\frac{2}{3}} \, \left( G O_{\frac{7}{2}}  
+\frac{1}{4\sqrt{6}} \, \pa P_{4'}^{uu} - \frac{\sqrt{6}}{4} \, 
 \, \pa P_{4'}^{ww} \right) \right](w)  +{\cal O}((z-w)^{-2}),
\nonu \\
&& G(z) \; \left( T \pa U -\frac{5}{4} \pa T U -\frac{1}{7} \pa^3 U \right)(w) =
-\frac{1}{(z-w)^4} \, \frac{33}{7} \sqrt{\frac{3}{2}} \, W(w) \nonu \\
&& +\frac{1}{(z-w)^3} \left[ -(-\frac{1}{6}) \frac{33}{7} \sqrt{\frac{3}{2}} \pa W -\frac{15}{4}
\left( G U -\frac{\sqrt{6}}{3} \pa W \right) \right] (w) +  {\cal O}((z-w)^{-2}),
\nonu \\  
&& G(z) \; \left( G \pa W - 2 \pa G W -\frac{1}{21} \sqrt{\frac{2}{3}} \pa^3 U \right)(w)
= -\frac{1}{(z-w)^4} \left[ \frac{4(18+261k+29k^2)}{7(3+k)(6+k)} \right]
 W(w) \nonu \\
&& + \frac{1}{(z-w)^3} \left[- (-\frac{1}{6}) 
\left[\frac{4(18+261k+29k^2)}{7(3+k)(6+k)}\right] \pa W -5\sqrt{\frac{2}{3}} \left( G U -\frac{\sqrt{6}}{3} \pa W 
\right)\right](w) + {\cal O}((z-w)^{-2}),    
\nonu \\
&& G(z) \; \left( G O_{4''} -\frac{2}{9} \pa O_{\frac{9}{2}'} \right)(w)
=\frac{1}{(z-w)^3} \left[\frac{16(3+2k)(15+2k)}{9(3+k)(6+k)}  \right]O_{4''}(w) +  {\cal O}((z-w)^{-2}),
\nonu \\
&& G(z) \; \left( G \pa^2 U - 4 \pa G \pa U +\frac{5}{2} \pa^2 G U -\frac{1}{2\sqrt{6}} \pa^3 W
 \right)(w) =\frac{1}{(z-w)^5} \left[ \frac{5(18+153k+17k^2)}{(3+k)(6+k)} \right] U(w)
\nonu \\
&& +\frac{1}{(z-w)^4} (-\frac{2}{5}) \left[ \frac{5(18+153k+17k^2)}{(3+k)(6+k)} \right]  \pa U(w) \nonu \\
&& +\frac{1}{(z-w)^3} 
\left[  (\frac{1}{30})  \left[ \frac{5(18+153k+17k^2)}{(3+k)(6+k)} \right] \pa^2 U - 2\sqrt{6} 
\left( G W -\frac{1}{6\sqrt{6}} \pa^2 U \right) \right. \nonu \\
&& \left. + 10 \left( T U -\frac{1}{4} \pa^2 U  
\right)  \right](w) + {\cal O}((z-w)^{-2}),
\nonu \\
&& G(z) \; \left( T \pa W -\frac{3}{2} \pa T W -\frac{1}{8} \pa^3 W \right)(w)
=-\frac{1}{(z-w)^5} \, \frac{5}{2} \sqrt{\frac{3}{2}} \, U(w) 
-\frac{1}{(z-w)^4} (-\frac{2}{5}) \frac{5}{2} \sqrt{\frac{3}{2}} \pa U(w) 
\nonu \\
&& +  \frac{1}{(z-w)^3} \left[- (\frac{1}{30}) 
\frac{5}{2} \sqrt{\frac{3}{2}} \pa^2 U  -\frac{9}{2} \left( G W -\frac{1}{6\sqrt{6}} \pa^2 U \right)
+5\sqrt{\frac{2}{3}} \left( T U -\frac{1}{4} \pa^2 U  \right)
\right] (w) 
\nonu \\
&& + {\cal O}((z-w)^{-2}),
\nonu \\
&& G(z) \; \left( T O_{4''} -\frac{1}{6} \pa^2 O_{4''} \right)(w)
= \frac{1}{(z-w)^3} \, \frac{13}{6} O_{\frac{9}{2}'}(w) + {\cal O}((z-w)^{-2}),
\nonu \\
&& G(z) \; \left( G O_{\frac{9}{2}'} -\frac{1}{9} \pa^2 O_{4''} \right)(w)
= \frac{1}{(z-w)^3} \, \frac{(1422+927k+103k^2)}{9(3+k)(6+k)} 
O_{\frac{9}{2}'}(w) + {\cal O}((z-w)^{-2}).
\label{Finalexpexp}
\eea
Therefore, in (\ref{Finalexpexp}),  
the $k$-dependence for the quasi-primary fields contain $G(w)$,
whereas the $k$-dependence is not observed in 
the remaining quasi-primary fields.
This is obvious because the OPE between $G(z)$ and $G(w)$ has an explicit
$c$-dependence from (\ref{gg}). 
Steps can be taken to
check that the coefficients (\ref{CoeffA1}) hold in this case.


\end{document}